\documentclass[cernpreprint,atlasdraft=false,coverpage=false,texlive=2016,UKenglish,texmf]{atlasdoc}
 
\usepackage[full]{atlaspackage}
\usepackage{atlasbiblatex}
\usepackage{soul}
\usepackage{ulem}
\usepackage[process,jetetmiss,other]{atlasphysics}
 
\graphicspath{{logos/}{figures/}}
 
\usepackage{ANA-TOPQ-2018-08-PAPER-defs}
\usepackage{multirow}
\usepackage{changepage}
\usepackage{array}
\usepackage{makecell}
\usepackage{cellspace}
\usepackage{amsfonts}
\newcolumntype{P}[1]{>{\centering\arraybackslash}p{#1}}
 
\AtlasTitle{Measurements of the inclusive and differential production cross sections of a top-quark--antiquark
pair in association with a $Z$~boson at $\sqrt{s} = 13$~TeV with the ATLAS detector}
 
\AtlasVersion{2.3}
\AtlasJournal{Eur. Phys. J. C}
\AtlasJournalRef{Eur. Phys. J. C 81 (2021) 737}
\AtlasDOI{10.1140/epjc/s10052-021-09439-4}
 
\AtlasAbstract{
Measurements of both the inclusive and differential production cross sections of a top-quark--antiquark pair in association with a $Z$~boson~($t\bar{t}Z$) are
presented. The measurements are performed by targeting final states with three or four isolated leptons (electrons or muons) and are based on $\sqrt{s} = 13$~TeV
proton--proton collision data with an integrated luminosity of 139~fb$^{-1}$, recorded from 2015 to 2018 with the ATLAS detector at the CERN Large Hadron
Collider. The inclusive cross section is measured to be $\sigma_{t\bar{t}Z} = 0.99 \pm 0.05$~(stat.)~$\pm \, 0.08$~(syst.)~pb, in agreement with the most
precise theoretical predictions. The differential measurements are presented as a function of a number of kinematic variables which probe the kinematics
of the $t\bar{t}Z$ system. Both absolute and normalised differential cross-section measurements are performed at particle and parton levels for specific
fiducial volumes and are compared with theoretical predictions at different levels of precision, based on a $\chi^{2}/$ndf and $p$-value
computation. Overall, good agreement is observed between the unfolded data and the predictions.
}
 
\author{The ATLAS Collaboration}
 
\AtlasRefCode{TOPQ-2018-08}
 
\PreprintIdNumber{CERN-EP-2021-011}

\hypersetup{pdftitle={ATLAS document},pdfauthor={The ATLAS Collaboration}}
 
\begin{document}
 
\maketitle
\tableofcontents
\clearpage
 
\section{Introduction}
Precise measurements of the properties of the top quark, the heaviest known elementary particle, are an important check of the internal consistency of the Standard Model~(SM)
of particle physics and could provide hints of possible new physics beyond the SM~(BSM). The production cross sections of top-quark--antiquark pairs~(\ttbar), single top quarks,
as well as the top-quark mass, have been measured with a great level of precision~\cite{TOPQ-2018-15,CMS-TOP-18-004,TOPQ-2017-16,TOPQ-2016-03}. The large centre-of-mass energy and
luminosity of the Large Hadron Collider~(LHC) enable precise and differential cross-section measurements for SM processes with small production rates, such as the associated production
of a \ttbar pair and a \Zboson boson~(\ttZ).
 
The \ttZ production process is particularly interesting, as it provides direct access to the neutral coupling of the top quark to the electroweak~(EW) gauge
bosons~\cite{Buckley2016,Dror2016}. Deviations of the coupling strength of the top quark to the \Zboson boson~($t$--\Zboson coupling) from its SM value
might imply the existence of new effects in the EW symmetry breaking mechanism which could be probed in the context of effective field theory~(EFT)~\cite{EFT:Bylund}.
Various BSM models predict large deviations of the top quark's EW couplings from the SM value, which were probed by the previous generation of lepton
colliders~\cite{ALEPH:2005ab,Richard:2014upa}. Precise measurements of the inclusive and differential cross sections of the \ttZ process are, thus, of particular
interest. Differential cross-section measurements can also offer sensitivity to differences among the predictions from various Monte Carlo~(MC) generators and can,
therefore, serve as an important input to the tuning of MC parameter values~(MC tunes). Furthermore, the \ttZ process is an irreducible background in several searches for BSM
phenomena~\cite{EXOT-2016-16,SUSY-2018-09}, as well as in measurements of important SM processes, such as \ttbar production in association with a Higgs boson~\cite{HIGG-2018-13}
or single top-quark production in association with a \Zboson boson~\cite{TOPQ-2018-01}. The ATLAS Collaboration measured the inclusive \ttZ cross section using a subset of the
LHC \RunTwo data, collected in 2015 and 2016~\cite{TOPQ-2016-11} and a first differential measurement of the \ttZ process was carried out by the CMS Collaboration using the
2016 and 2017 data sets~\cite{CMS:2019too}.
 
Theoretical predictions of the \ttZ cross section exist at next-to-leading order~(NLO) with the resummation of soft gluon corrections computed at
next-to-next-to-leading-logarithm~(NNLL) precision~\cite{Broggio:2017kzi,Kulesza:2019} in perturbative quantum chromodynamics (QCD) with added EW corrections.
Recently they have been matched to the complete set of NLO corrections of both QCD and EW origin~\cite{Broggio:2019ewu,Kulesza:2020nfh} using the \MGNLO~(\MGNLOs)
framework~\cite{Frederix:2018nkq}.
 
This paper presents measurements of the inclusive and differential \ttZ production cross section in final states with three or four isolated leptons~(electrons or muons)
with the ATLAS detector~\cite{PERF-2007-01} at the LHC. The measurements were performed with $\rts = \SI{13}{\TeV}$ proton--proton~(\pp) collision data collected during
\RunTwo of the LHC~(2015--2018) and corresponding to an integrated luminosity of \SI{139}{\ifb}. The \Zboson boson is identified by targeting events featuring
an oppositely charged electron~($e$) or muon~($\mu$) pair. The detector signatures resulting from the hadronisation of final-state quarks from the decay of the
\ttbar system, in particular those from bottom (anti)quarks, are exploited by constructing target regions with different jet and $b$-jet multiplicities.
The inclusive measurement follows an analysis strategy similar to the previous ATLAS \ttZ measurement~\cite{TOPQ-2016-11}. The production cross
section is extracted by performing a simultaneous maximum-likelihood fit in the targeted analysis regions with the signal normalisation as the parameter of interest.
In addition, a set of normalised and absolute differential measurements are presented as a function of different variables which probe the SM predictions for the
kinematics of the \ttZ system. Some of these variables are found to be sensitive to potential EFT signals~\cite{EFT:Bylund}, while others are more interesting in
the context of MC tuning~\cite{ATL-PHYS-PUB-2017-007,ATL-PHYS-PUB-2018-009}. The differential measurements were performed at both particle and parton level in different
fiducial volumes in order to correct for various acceptance effects. For the first time, the \ttZ cross section is measured with the full \RunTwo data and includes
differential measurements performed as functions of variables related to the \ttbar system.
 
This paper is structured as follows. Section~\ref{sec:detector} provides a brief description of the ATLAS detector. In Section~\ref{sec:samples}, the simulation
of signal and background processes is discussed, followed by the definitions of the final-state objects at reconstruction, particle and parton levels in
Section~\ref{sec:reconstruction}. The event selection, both online during data taking and offline after data taking, and the background estimation are
discussed in Sections~\ref{sec:signalregions} and~\ref{sec:background}, respectively. The sources of systematic uncertainties that affect the measurements are
discussed in Section~\ref{sec:systematics}. The result of the inclusive cross-section measurement is presented in Section~\ref{sec:results_inclusive}.
Section~\ref{sec:differential} describes the differential variables and the unfolding procedure, followed by the results of the differential measurements.
Conclusions are drawn in Section~\ref{sec:conclusion}.
 
\section{ATLAS detector}
\label{sec:detector}
The ATLAS detector~\cite{PERF-2007-01} at the LHC covers nearly the entire solid angle around the collision point.\footnote{\AtlasCoordFootnote}
It consists of an inner tracking detector surrounded by a thin superconducting solenoid, electromagnetic and hadronic calorimeters, and a muon
spectrometer incorporating three sets of large superconducting toroidal magnets, each consisting of eight separate coils. The inner-detector system~(ID)
is immersed in a \SI{2}{\tesla} axial magnetic field and provides charged-particle tracking in the range $\abseta < 2.5$.
 
The high-granularity silicon pixel detector covers the vertex region and typically provides four measurements per track, with the first hit typically being
detected in the insertable B-layer~(IBL) installed before \RunTwo~\cite{ATLAS-TDR-2010-19,PIX-2018-001}. It is followed by the silicon microstrip tracker~(SCT),
which usually provides eight measurements per track. These silicon detectors are complemented by the transition radiation tracker~(TRT), which enables radially
extended track reconstruction up to $\abseta = 2.5$. The TRT also provides electron identification information based on the fraction of hits above a higher
energy-deposit threshold corresponding to transition radiation. Typically, around \num{30} TRT hits are measured in total.
 
The calorimeter system covers the pseudorapidity range $\abseta < 4.9$. Within the region $\abseta < 3.2$, electromagnetic calorimetry is provided by barrel and
endcap high-granularity lead/liquid-argon~(LAr) calorimeters, with an additional thin LAr presampler covering $\abseta < 1.8$ to correct for energy loss
in material upstream of the calorimeters. Hadronic calorimetry is provided by the steel/scintillator-tile calorimeter, segmented into three barrel structures
within $\abseta < 1.7$, and two copper/LAr hadronic endcap calorimeters. The solid angle coverage is extended with forward copper/LAr and tungsten/LAr calorimeter
modules optimised for electromagnetic and hadronic measurements respectively.
 
The muon spectrometer~(MS) comprises separate trigger and high-precision tracking chambers measuring the deflection of muons in a magnetic field generated
by superconducting air-core toroids. The field integral of the toroids ranges between \num{2.0} and \SI{6.0}{Tm} across most of the detector.
A set of precision chambers covers the region $\abseta < 2.7$ with three layers of monitored drift tubes, complemented by cathode-strip chambers in the
forward region, where the background rates are highest. The muon trigger system covers the range $\abseta < 2.4$ with resistive-plate chambers in the barrel,
and thin-gap chambers in the endcap regions.
 
Interesting events are selected to be recorded by the first-level trigger system implemented in custom hardware, followed by selections made by algorithms
implemented in software in the high-level trigger~\cite{TRIG-2016-01}. The first-level trigger accepts events from the \SI{40}{\MHz} bunch crossings at a rate below
\SI{100}{\kHz}, which the high-level trigger reduces in order to record events to disk at about \SI{1}{\kHz}.

 
\section{Data and simulated event samples}
\label{sec:samples}
The analysis is performed on data from \pp collisions at $\rts = \SI{13}{\TeV}$ delivered by the LHC and recorded by the ATLAS detector between 2015 and 2018. The
bunch spacing for this data-taking period was \SI{25}{ns} with an average number of \pp interactions per bunch crossing~(`pile-up') which varies by year
and LHC beam conditions and was in the range from \num{10} to \num{70} for almost all events. After requirements on the stability
of the beams, the operational status of all ATLAS detector components, and the quality of the recorded data, the total integrated luminosity of the data set
corresponds to \SI{139}{\ifb}. This value is derived from the calibration of the luminosity scale using $x$--$y$ beam-separation scans, following a methodology
similar to that detailed in Ref.~\cite{DAPR-2013-01}, and using the LUCID-2 detector~\cite{Avoni:2018} for the baseline luminosity measurements.
 
The data were collected using a combination of single-electron and single-muon triggers, with requirements on the identification, isolation, and \pt of
the leptons to maintain efficiency across the full momentum range while controlling the trigger rates~\cite{TRIG-2016-01}. For electrons the trigger thresholds
were $\pt = 26$, \num{60} and \SI{140}{\GeV}, whereas for muons, the thresholds were $\pt = 26$ and \SI{50}{\GeV}.\footnote{Lower \pt thresholds of \SI{24}{\GeV}
for electrons and \SI{20}{\GeV} for muons were applied for 2015 data.} Identification and isolation requirements were applied to the triggers with the lower \pt
thresholds~\cite{ATL-DAQ-PUB-2016-001,ATL-DAQ-PUB-2017-001,ATL-DAQ-PUB-2018-002}.
 
Signal and background processes considered in this analysis were modelled using simulated MC samples. The effect of pile-up interactions was modelled by overlaying
the hard-scattering event with simulated minimum-bias events generated with \PYTHIA[8.186]~\cite{Sjostrand:2007gs} using the \NNPDF[2.3lo] set of parton distribution
functions~(PDFs)~\cite{Ball:2012cx} and the A3 set of tuned MC parameters~\cite{ATL-PHYS-PUB-2016-017}. The simulated events were reweighted to match the pile-up
conditions observed in the measured data. For processes featuring \Wboson boson, \Zboson boson or top-quark production, the \Wboson, \Zboson and top-quark masses were set to
\SI{80.4}{\GeV}, \SI{91.2}{\GeV}~\cite{PDG} and \SI{172.5}{\GeV}, respectively. The decays of bottom and charm hadrons were simulated using the \EVTGEN program~\cite{Lange:2001uf}.
The MC samples were either processed through a full simulation of the ATLAS detector based on \GEANT~\cite{SOFT-2010-01,Agostinelli:2002hh} or a fast simulation~(\textsc{AtlFast2})
relying on parameterised showers in the calorimeter~\cite{ATL-PHYS-PUB-2010-013,ATL-SOFT-PUB-2018-002}.
 
The \ttZ signal process was modelled using the \MGNLOsV{2.3.3}~\cite{Alwall:2014hca} generator together with \EVTGEN~{1.2.0}, which provided a matrix element~(ME) calculation
at NLO in the strong coupling constant~(\alphas) with the \NNPDF[3.0nlo]~\cite{Ball:2014uwa} PDF set. The functional form of the renormalisation and factorisation scales~(\muR and \muF)
was set to $\mu_{\mathrm{r,f}} = 0.5 \times \sum_i \sqrt{m^2_{\mathrm{i}}+p^2_{\mathrm{T,i}}}$, where $i$ runs over all final-state particles generated from the ME calculation.
The $\ttbar\gamma^{*}$ contribution and the $Z/\gamma^{*}$ interference were included with dilepton invariant masses (\mll) down to \SI{5}{\GeV}. Top-quark decays were simulated at
leading order~(LO) using \MADSPIN~\cite{Frixione:2007zp,Artoisenet:2012st} to preserve all spin correlations. The events were interfaced with \PYTHIA[8.210]~\cite{Sjostrand:2014zea}
for the parton shower and hadronisation, using the A14 set of tuned parameters~\cite{ATL-PHYS-PUB-2014-021} and the \NNPDF[2.3lo] PDF set.
 
The SM theoretical prediction of the production cross section for the \ttZ process, including all \Zboson boson decay modes and taking into account the $\ttbar\gamma^{*}$ contribution and the
$Z/\gamma^{*}$ interference, is $\sigma_{\ttZ} = 0.88^{+0.09}_{-0.10}\,$pb and includes NLO QCD+EW corrections~\cite{ATL-PHYS-PUB-2016-005}. This value is an off-shell extension
of a cross-section calculation of $\sigma_{\ttZ} = 0.84^{+0.09}_{-0.10}\,$pb, which was reported in Ref.~\cite{deFlorian:2016spz}~(based on Ref.~\cite{Frixione:2015zaa}). The
uncertainties are due to the QCD scales, the proton PDFs, and \alphas.
 
The measured differential cross sections are compared with theoretical expectations obtained with different generators. Alternative \ttZ samples were simulated with
\SHERPA[2.2.1]~\cite{Bothmann:2019yzt} generator at NLO QCD accuracy, using both inclusive and multi-leg set-ups. In both cases, dynamic \muR and \muF scales were used as in
the nominal samples. The default \SHERPA[2.2.1] parton shower was used together with the \NNPDF[3.0nnlo] PDF set~\cite{Ball:2014uwa}. The multi-leg sample was simulated using the
\MEPSatNLO prescription~\cite{Hoeche:2011fd,Hoeche:2012yf,Catani:2001cc,Hoeche:2009rj} with up to one additional parton at NLO and with a merging scale of \SI{30}{\GeV}. Another
sample was generated with the same \MGNLOs and \EVTGEN versions as the nominal sample but using a different MC program for the modelling of the parton-shower and hadronisation:
\HERWIG[7]~\cite{Bahr:2008pv,Bellm:2015jjp} instead of \PYTHIA[8]. In addition, two alternative samples with the same settings as the nominal sample, but using a set of
variations of the A14 tune's parameters (A14 eigentune variation Var3c~\cite{ATL-PHYS-PUB-2014-021}), were employed to evaluate the uncertainty associated with the amount of
initial-state radiation~(ISR).
 
The production of a single top quark or antiquark in association with a \Zboson boson and one extra parton~(\tZq) was simulated using the \MGNLOsV{2.3.3} generator at NLO
QCD with the \NNPDF[3.0nnlo] PDF set. The events were interfaced with \PYTHIA[8.230] using the A14 tune and the \NNPDF[2.3lo] PDF set. The \tZq sample also includes off-shell
\Zboson decays to dilepton pairs with invariant masses in the range $\mll > \SI{30}{\GeV}$. Single top-quark or top-antiquark production in association with both a \Wboson and
a \Zboson boson~(\tWZ) was simulated at NLO with \MGNLOsV{2.2.2} and the \NNPDF[3.0nnlo] PDF set, using \PYTHIA[8.235] for the parton-shower simulation. The interference between
\ttZ and \tWZ was removed following a diagram removal approach referred to as the DR1 scheme~\cite{Demartin:2016axk}.
 
Events featuring the production of a \ttbar pair in association with a \Wboson or Higgs boson~(\ttW and \ttH) were generated using NLO QCD MEs in \MGNLOsV{2.3.3}~(for \ttW)
or~{2.6.0}~(for \ttH) with the \NNPDF[3.0nlo] PDF set and showered with \PYTHIA[8.210] or {8.230} using the A14 tune. MC samples featuring Higgs production in association with a
\Wboson or \Zboson boson~(\HWZ) were generated at LO with \PYTHIA[8.186] using the A14 tune and the \NNPDF[2.3lo] PDF set.
 
Diboson processes featuring the production of three charged leptons and one neutrino or four charged leptons~(\WZjets or \ZZjets, respectively)
were simulated using the \SHERPA[2.2.2] generator. In this set-up, multiple MEs were matched and merged with the \SHERPA parton shower based on the Catani--Seymour
dipole factorisation scheme~\cite{Gleisberg:2008fv,Schumann:2007mg} using the \MEPSatNLO prescription~\cite{Hoeche:2011fd,Hoeche:2012yf,Catani:2001cc,Hoeche:2009rj}.
The virtual QCD corrections for MEs at NLO accuracy were provided by the \OPENLOOPS library~\cite{Cascioli:2011va,Denner:2016kdg}. Samples were generated
using the  \NNPDF[3.0nnlo] PDF set, along with the dedicated set of tuned parton-shower parameters developed by the \SHERPA authors. The \VVjets events with up to one
additional parton were simulated at NLO, whereas events with two or three partons were simulated at LO precision.
 
The production of three or four top quarks~(\ttt and \tttt) and the production of a \ttbar pair with two \Wboson bosons~(\ttWW) were simulated at LO using
\MGNLOsV{2.2.2} interfaced to \PYTHIA[8.186] with the A14 tune and the \NNPDF[2.3lo] PDF set. Fully leptonic triboson processes~($WWW$, $WWZ$, $WZZ$ and $ZZZ$)
with up to six leptons in the final states were simulated with \SHERPA[2.2.2] and the \NNPDF[3.0nnlo] PDF set. Final states with no additional partons were calculated
at NLO, whereas final states with one, two or three additional partons, were calculated at leading order.
 
\section{Object reconstruction}
\label{sec:reconstruction}
The following subsections describe the definitions of final-state objects at reconstruction~(detector), particle, and parton levels, which are used to characterise the
final-state event topologies and to define the phase-space regions for the cross-section measurements.
 
\subsection{Reconstruction of detector-level objects}
 
Electron candidates are reconstructed from clusters of energy deposits in the electromagnetic calorimeter that are matched to a track in the ID. They are required to
satisfy $\pt > \SI{7}{\GeV}$, $\abseta < 2.47$ and a `Medium' likelihood-based identification requirement~\cite{PERF-2017-01,EGAM-2018-01}. Electron candidates
are excluded if their calorimeter clusters lie within the transition region between the barrel and the endcap of the electromagnetic calorimeter, $1.37 < \abseta < 1.52$.
The track associated with the electron must pass the requirements $|\zzsth| < \SI{0.5}{mm}$ and $|d_{0}| / \sigma(d_{0}) < 5$, where $z_{0}$ describes the longitudinal
impact parameter relative to the reconstructed primary vertex,\footnote{The primary vertex is defined as the vertex with the highest scalar sum of the squared transverse
momenta of associated tracks with $\pt > \SI{400}{\MeV}$.} \dzero is the transverse impact parameter relative to the beam axis, and $\sigma(d_{0})$ is the uncertainty
on \dzero.
 
Muon candidates are reconstructed from MS tracks matched to ID tracks in the pseudorapidity range of $\abseta < 2.5$. They must satisfy $\pt > \SI{7}{\GeV}$ along with the
`Medium' identification requirements defined in Ref.~\cite{PERF-2015-10,MUON-2018-03}. This criterion defines requirements on the number of hits in the different ID and MS subsystems
and on the significance of the charge-to-momentum ratio $q/p$. In addition, the track associated with the muon candidate must have $|\zzsth| < \SI{0.5}{mm}$ and
$|d_{0}| / \sigma(d_{0}) < 3$.
 
Isolation criteria are applied to the selected electrons and muons. For electrons, the scalar sum of the \pt of tracks within a variable-size cone around the electron,
excluding tracks originating from the electron itself, must be less than 6\% of the electron \pt. The track isolation cone radius $\DeltaR = \sqrt{(\Delta \eta)^2+(\Delta \phi)^2}$
is given by the smaller of $\DeltaR = \SI{10}{\GeV}/\pt$ and $\DeltaR = 0.2$. In addition, the sum of the transverse energy of the calorimeter \topos\footnote{\AtlasTopoClusterFootnote}
in a cone of $\DeltaR = 0.2$ around the electron is required to be less than 6\% of the electron \pt, excluding clusters originating from the electron itself. For muons, the scalar sum of
the \pt of tracks within a variable-size cone around the muon, excluding its own track, must be less than 6\% of the muon \pt, with the track isolation cone radius being given
by the minimum of $\DeltaR = \SI{10}{\GeV}/\pt$ and $\DeltaR = 0.3$.
 
Jets are reconstructed from \topos, using the \antikt jet clustering algorithm~\cite{Cacciari:2008gp} as implemented in the \Fastjet package~\cite{Fastjet}, with a radius parameter
of $R = 0.4$. They are calibrated through the application of a jet energy scale derived from \SI{13}{\TeV} data and simulation~\cite{PERF-2016-04}. Only jet candidates with
$\pT > \SI{25}{\GeV}$ and $\abseta < 2.5$ are considered in this analysis. To mitigate the impact of jets arising from additional \pp collisions in a given bunch crossing,
an additional selection criterion using a likelihood-based `jet-vertex-tagging'~(JVT) discriminant is applied to jets with $\pT < \SI{120}{\GeV}$ and $\abseta < 2.5$~\cite{PERF-2014-03}.
 
Jets containing $b$-hadrons~(`$b$-jets') are identified~(tagged) by the MV2c10 $b$-tagging algorithm~\cite{FTAG-2018-01}. The algorithm uses a multivariate discriminant with quantities such
as the impact parameters of associated tracks, and well-reconstructed secondary vertices. For the differential measurements, a selection that provides an 85\% efficiency for identifying
$b$-jets in simulated \ttbar events, with rejection factors against light-flavour jets and $c$-jets of \num{28} and \num{2}, respectively is used. Different calibrated $b$-tagging working
points~(WPs), corresponding to different $b$-jet selection efficiencies are used for the inclusive cross-section measurement. A method where exclusive bins in the $b$-tagging discriminant
corresponding to different identification efficiencies is employed. In the following, this approach is referred to as pseudo-continuous $b$-tagging~(PCBT).
 
Scale factors are applied as weights to MC events to correct for the mismodelling of efficiencies associated with the reconstruction, identification and trigger selection
of electrons and muons, as well as the JVT and $b$-tagging requirements for jets. The $b$-tagging scale factors are derived from a pseudo-continuous calibration as outlined above.
 
The missing transverse momentum is defined as the negative vector sum of the transverse momenta of all selected and calibrated physics objects~(electrons, photons, muons and jets).
Low-momentum tracks from the primary vertex that are not associated with any of the reconstructed physics objects described previously are also included as a `soft term' in the
calculation~\cite{PERF-2016-07}. The magnitude of the missing transverse momentum vector is denoted as \met.
 
Ambiguities can arise from the independent reconstruction of electron, muon and jet candidates in the detector. A sequential procedure~(overlap removal) is applied to
resolve these ambiguities and, thus, avoids a double counting of physics objects.\footnote{The lepton candidates considered for the overlap removed are electrons
selected with the `Loose' identification~\cite{PERF-2017-01,EGAM-2018-01}  and muons selected with the `Medium' identification requirement, but before placing isolation
requirements on the leptons.} It is applied as follows. If an electron candidate and a and muon candidate share a track, the electron candidate is removed. Jet candidates
within a distance of $\DeltaR_{y,\phi} =\sqrt{(\Delta y)^2+(\Delta \phi)^2} = 0.2$ from a remaining electron candidate are discarded. If multiple jets are found in this area, only the
closest jet is removed. If the electron-jet distance is between \num{0.2} and \num{0.4}, the electron candidate is removed. If the $\DeltaR_{y,\phi}$ between any
remaining jet and a muon candidate is less than \num{0.4}, the muon candidate is removed if the jet has more than two associated tracks, otherwise the jet is discarded.
 
\subsection{Particle- and parton-level objects and definitions of fiducial regions}
\label{subsec:pl_definitions}
 
In the measurements of differential \ttZ cross sections, the measured spectra are corrected for detector effects to so-called particle and parton levels using an unfolding procedure.
 
Parton-level objects were obtained from the MC record of the \ttZ event. The top quarks~(antiquarks) and \Zboson bosons were selected after final-state radiation and just
before their corresponding decay, \tWb or \Zll, respectively. The leptons originating from \Wboson and \Zboson bosons were selected directly from the decay vertex of the parent bosons.
 
The parton-level fiducial volumes for final states with three or four leptons were defined as follows: the \Zboson boson was required to decay into leptons,
whereas the \ttbar pair was required to decay via $\ttbar \to \Wboson^{+}b \Wboson^{-}\bar{b}$, with either one or both \Wboson bosons subsequently decaying leptonically.
 
The particular decay chains of interest are therefore:
 
\begin{eqnarray*}
&  & \Zboson \to \ell^{+}\ell^{-}, \ \ t \to \Wplus( \to \ell^{+}\nu_{\ell})+b, \ \ \bar{t}\to \Wminus(\to q\bar{q})+\bar{b} \\
\textrm{or} &  & \Zboson \to \ell^{+}\ell^{-}, \ \ t\to \Wplus(\to q\bar{q})+b, \ \ \ \ \ \bar{t}\to \Wminus(\to\ell^{-}\bar{\nu}_{\ell})+\bar{b} \\
\end{eqnarray*}
 
for a three-lepton final state, and
 
\begin{eqnarray*}
&  & \Zboson \to \ell^{+}\ell^{-}, \ \ t\to \Wplus(\to\ell^{+}\nu_{\ell})+b, \ \ \bar{t}\to \Wminus(\to\ell^{-}\bar{\nu}_{\ell})+\bar{b} \\
\end{eqnarray*}
 
for a four-lepton final state. The two decay chains for the three-lepton final state differ only in terms of which of the top quark or antiquark decayed hadronically.
 
The invariant mass of the lepton pair originating from the \Zboson boson has to be within a range of $\pm \SI{15}{\GeV}$ around the nominal \Zboson boson
mass~(\SI{91.2}{\GeV})~\cite{PDG} to be sensitive to on-shell \Zboson decays. Prompt $\tau$-leptons from \Zboson or \Wboson boson decays were not included in
the parton-level fiducial volume, regardless of their subsequent decay into leptons or hadrons. No kinematic requirements were applied to the parton-level objects in order
that the unfolded differential results at parton level can be more easily compared with fixed-order predictions.
 
Particle-level objects in simulated events were defined using quasi-stable particles with a mean lifetime greater than \SI{30}{ps} originating from \pp collisions. They were selected
after hadronisation but before the interaction of these particles with the detector components or consideration of pile-up effects. Electrons and muons were required to not to have
originated from a hadron in the MC generator event record, whether directly or through a $\tau$-lepton decay. This ensures that they originated from the \Zboson boson or the \Wboson
bosons from top-quark decays, without requiring a direct match with the parent boson. The four-momenta of the bare leptons were modified~(`dressed') by adding the four-momenta of all
radiated photons within a cone of size $\DeltaR = 0.1$, excluding photons from hadron decays, to take into account final-state photon radiation. Particle-level jets were reconstructed
with the \antikt algorithm with a radius parameter of $R = 0.4$ applied to all stable particles, but excluding both the neutrinos originating from the \Zboson boson or top quarks and the
selected electrons, muons and photons used in the definition of the charged leptons. If $b$-hadrons with $\pt > \SI{5}{\GeV}$ were found in the MC event record, they were clustered in the
stable-particle jets with their energies set to a negligible positive value~(`ghost-matching')~\cite{Cacciari:2008gn}. Particle-level jets containing one or more of these $b$-hadrons were
considered to originate from a $b$-quark. The particle-level missing transverse momentum was defined as the vector sum of the transverse momenta of all neutrinos found in the simulation
history of the event, excluding those originating from hadron decay.
 
The particle-level fiducial volume for final states with three or four leptons was defined by applying the same \pt and \abseta requirements as those summarised for the detector-level
selection in Tables~\ref{tab:sr-3l-def_main} and \ref{tab:sr-4l-def_main}, respectively.  In addition, the same requirements were placed on the number of jets and $b$-jets, and the same
requirements were  placed on the opposite-sign--same-flavour~(OSSF) lepton pair, along with the same invariant mass requirement for the \Zboson-mass window as that used in the detector-level
selection described in the following section~($|\mll - \mZ| < \SI{10}{\GeV}$).\footnote{It should be noted that this invariant mass requirement is more stringent than the on-shell \Zboson
requirement  used in defining the parton-level fiducial volumes in Section~\ref{subsec:pl_definitions} and provides better background rejection. The requirement was therefore tightened
for the  detector-level selection, and the same range was then adopted for the particle-level definitions.} For the particle-level fiducial volume for four-lepton final states, only
one OSSF lepton pair was required within the \Zboson-mass window --- the remaining lepton pair was required only to be opposite-sign. Only one of the jets is required to have originated
from a $b$-quark.
 
\section{Event selection and signal regions}
\label{sec:signalregions}
Only final states with exactly three or four isolated leptons~(electrons or muons) and at least two jets, as defined in Section~\ref{sec:reconstruction}, are considered.
All selected events are required to pass a single-electron or single-muon trigger. In addition, at least one reconstructed lepton with $\pt > \SI{27}{\GeV}$ is required
to be matched to the lepton reconstructed by the trigger algorithm and to be of the same flavour. Different signal regions are defined and optimised to achieve the best
sensitivity to \ttZ production with one or both top quarks decaying via $t \rightarrow b\Wboson \to b\ell\nu_{\ell}$. Furthermore, the regions are designed to contain a
sufficient number of signal events in order to reduce the statistical uncertainties of the differential \ttZ cross-section measurements. The signal regions are referred
to as `trilepton'~($3\ell$) and `tetralepton'~($4\ell$) signal regions, depending on the number of reconstructed leptons, and are meant to target events with one or two
prompt leptons, respectively from the \ttbar decay.
 
\subsection{Trilepton signal regions}
 
A summary of the definitions of the trilepton signal regions is provided in Table~\ref{tab:sr-3l-def_main}. The requirement on the minimum transverse momentum of the leading, sub-leading
and third lepton is \num{27}, \num{20} and \SI{20}{\GeV}, respectively. The sum of the three lepton charges is required to be~$\pm 1$. The OSSF lepton pair with the invariant mass
closest to the \Zboson boson mass is considered to originate from the \Zboson decay and its invariant mass~(labelled as \mZone) is required to be compatible with the mass of the \Zboson
boson~($|\mZone - \mZ| < \SI{10}{\GeV}$).  Furthermore, all OSSF lepton combinations are required to have $\mOSSF > \SI{10}{\GeV}$ to remove contributions arising from low-mass resonances.
Additional requirements are imposed on the total number of reconstructed jets~($N_{\mathrm{jets}}$) and $b$-tagged jets~($N_{b\textrm{-jets}}$) in the event.
 
Different $b$-jet requirements are used for the inclusive and differential cross-section measurements. For the inclusive measurement, a combination of two orthogonal regions which use
different $b$-tagging WPs~(PCBT, see Section~\ref{sec:reconstruction}) with 60\% and 70\% efficiency is employed. The tighter b-tagging WPs are used to suppress the \WZjets background
and reduce its impact on the overall precision of the measurement.  The two regions, referred to as $3\ell$-$Z$-$1b4j$-PCBT and $3\ell$-$Z$-$2b3j$-PCBT, are kept distinct from one another
during the fitting procedure used to perform the cross-section measurement. For the differential measurements, a looser, fixed WP corresponding  to an 85\% efficiency is used in order
to increase the data statistics.
 
\begin{table}[!htb]
\footnotesize
\caption[]{The definitions of the trilepton signal regions: for the inclusive measurement, a combination of the regions with pseudo-continuous
$b$-tagging $3\ell$-$Z$-$1b4j$-PCBT and $3\ell$-$Z$-$2b3j$-PCBT is used, whereas for the differential measurement only the region $3\ell$-$Z$-$2b3j$
with a fixed $b$-tagging WP is employed.}
\label{tab:sr-3l-def_main}
\def\arraystretch{1.4}
\begin{center}
\begin{tabular}{l | P{0.18\textwidth} P{0.18\textwidth} P{0.18\textwidth}}
\hline
\hline
Variable & $3\ell$-$Z$-$1b4j$-PCBT & $3\ell$-$Z$-$2b3j$-PCBT & $3\ell$-$Z$-$2b3j$       \\
&      inclusive          &  inclusive              &   differential           \\
\hline
$N_{\ell}\,(\ell=e,\mu)$  &              \multicolumn{3}{c}{$=3$}                       \\
&              \multicolumn{3}{c}{$\geq$1~OSSF lepton pair with $|\mZone - \mZ| < \SI{10}{\GeV}$} \\
&              \multicolumn{3}{c}{for all OSSF combinations: $\mOSSF > \SI{10}{\GeV}$}     \\
\pT$(\ell_{1},\,\ell_{2},\,\ell_{3})$ &  \multicolumn{3}{c}{$> 27,\,20,\,\SI{20}{\GeV}$}      \\
\hline
$N_{\mathrm{jets}}$          & $\geq 4$& $\geq 3$ & $\geq 3$          \\
$N_{b\textrm{-jets}}$     & $= 1$@60\%         & $\geq 2$@70\%      & $\geq 2$@85\%     \\
& veto add. $b$-jets@70\%           &     &                   \\
\hline
\hline
\end{tabular}
\end{center}
\end{table}
 
\FloatBarrier
 
\subsection{Tetralepton signal regions}
 
The definitions of the tetralepton signal regions are summarised in Table~\ref{tab:sr-4l-def_main}. The requirement on the transverse momentum of the four leading leptons
in all regions is \num{27}, \num{20}, \num{10} and \SI{7}{\GeV}, respectively. As in the case of the trilepton signal regions, all events are required to have at least one
OSSF lepton pair with an invariant mass satisfying $|\mZone - \mZ| < \SI{10}{\GeV}$. Furthermore, the remaining leptons which are not associated with the \Zboson boson~(non-\Zboson)
are required to have opposite charges, such that the sum of the four lepton charges is zero. As in the trilepton selection, a requirement that all OSSF lepton combinations satisfy
$\mOSSF > \SI{10}{\GeV}$ in order to suppress background contributions from low-mass resonances is applied.
 
The tetralepton signal regions are separated into different-flavour~(DF) and same-flavour~(SF) signal regions, according to the $b$-jet multiplicities and the flavour composition of
the non-\Zboson lepton pair. The \ZZjets background is suppressed by setting requirements on the jet and $b$-jet multiplicities, as well as by applying cuts on \met and the
invariant mass of the non-\Zboson lepton pair~(\mZtwo) in the case of the SF regions. In the SF regions, events with \mZtwo close to the \Zboson mass are accepted, but the \met
requirement is increased to reduce the \ZZjets background. If \mZtwo is not close to the \Zboson mass, the \met cut is relaxed. For the inclusive cross-section measurement, the
four tetralepton regions are included as separate bins in the fit, whereas for the differential measurements all the events are combined. Unlike the trilepton signal regions,
the $b$-jets are all selected using a fixed 85\% $b$-tagging efficiency WP. The tetralepton signal region selections are identical for the inclusive and differential measurements.
 
\begin{table}[!htb]
\footnotesize
\caption[]{The definitions of the four tetralepton signal regions. The regions are defined to target different $b$-jet multiplicities
and flavour combinations of the non-\Zboson leptons~($\ell\ell^{\,\textrm{non-}\Zboson}$).}
\label{tab:sr-4l-def_main}
\def\arraystretch{1.4}
\begin{center}
\begin{tabular}{l | P{0.222\textwidth} P{0.222\textwidth} P{0.122\textwidth} P{0.122\textwidth}}
\hline
\hline
Variable & $4\ell$-SF-$1b$ & $4\ell$-SF-$2b$ & $4\ell$-DF-$1b$ & $4\ell$-DF-$2b$                     \\
\hline
$\begin{array}{l} N_{\ell}(\ell=e,\mu) \end{array}$ & \multicolumn{4}{c}{$=4$}                      \\
& \multicolumn{4}{c}{$\geq$1~OSSF lepton pair with $|\mZone-\mZ| < \SI{10}{\GeV}$} \\
& \multicolumn{4}{c}{for all OSSF combinations: $\mOSSF > \SI{10}{\GeV}$}   \\
$\begin{array}{l} \pT\,(\ell_{1},\,\ell_{2},\,\ell_{3},\,\ell_{4}) \end{array}$  & \multicolumn{4}{c}{$> 27,\,20,\,10,\,\SI{7}{\GeV}$} \\
\hline
$\begin{array}{l} \ell\ell^{\,\textrm{non-}\Zboson} \end{array}$ & $e^{+}e^{-}$~~or~~$\mu^{+}\mu^{-}$ & $e^{+}e^{-}$~~or~~$\mu^{+}\mu^{-}$ & $e^{\pm}\,\mu^{\mp}$ & $e^{\pm}\,\mu^{\mp}$ \\
$\begin{array}{l} \met \end{array}$ & $> \SI{100}{\GeV}$, if~$|\mZtwo-\mZ| \leq \SI{10}{\GeV}$ & $> \SI{50}{\GeV} ,$ if~$|\mZtwo-\mZ| \leq \SI{10}{\GeV}$ & -- & -- \\
& $> \SI{50}{\GeV} ,$ if~$|\mZtwo-\mZ| > \SI{10}{\GeV}$ & -- & & \\
$\begin{array}{l} N_{\mathrm{jets}} \end{array}$ & $\geq2$ & $\geq2$ & $\geq2$ & $\geq2$ \\
$\begin{array}{l} N_{b\textrm{-jets}}@85\% \end{array}$ & $= 1$ & $\geq 2$ & $= 1$ & $\geq 2$ \\
\hline
\hline
\end{tabular}
\end{center}
\end{table}
 
\FloatBarrier
 
\section{Background estimation}
\label{sec:background}
Several processes can lead to background contaminations in the signal regions. The contributions from SM processes featuring the production of three or four prompt
leptons\footnote{The term `prompt' refers to leptons which are directly produced by the hard-scatter process or by the decays of heavy resonances such as \Wboson, \Zboson
or Higgs bosons.} is discussed in Section~\ref{subsec:prompt}, whereas the estimation of processes where at least one of the reconstructed leptons is a fake lepton
is explained in Section~\ref{subsec:fakes}.
 
\subsection{Prompt lepton background}
\label{subsec:prompt}
 
The dominant SM background processes in the trilepton and tetralepton regions are \VVjets production with $WZ \rightarrow \ell\ell\ell\nu$ and $ZZ \rightarrow \ell\ell\ell\ell$
decays, respectively. The normalisations of these processes are obtained from data and measured in dedicated \WZjets and \ZZjets control regions~(CRs) as defined in Table~\ref{tab:cr-def}.
The CRs are common to both the inclusive and the differential cross-section measurements. Invariant mass requirements on the OSSF lepton pairs are applied to select the \Zboson bosons
expected in both regions. A $b$-jet veto is applied in $3\ell$-$WZ$-CR to suppress the \ttZ contribution and to ensure orthogonality with the trilepton signal region. In $4\ell$-$ZZ$-CR,
no requirements are placed on the number of jets or $b$-jets. The invariant mass requirements on the two OSSF lepton pairs are sufficient to yield a very high \ZZjets purity in this region.
Orthogonality with the tetralepton signal regions is ensured through the use of an \met requirement~($\SI{20}{\GeV} < \met < \SI{40}{\GeV}$), where the lower bound is set so that the
selected events are more similar kinematically to those in the signal regions. The \WZjets purity in $3\ell$-$WZ$-CR is approximately 80\%, while the \ZZjets purity in $4\ell$-$ZZ$-CR
is approximately 97\%.
 
\begin{figure}[htb!]
\subfloat[]{\includegraphics[width=0.49\textwidth]{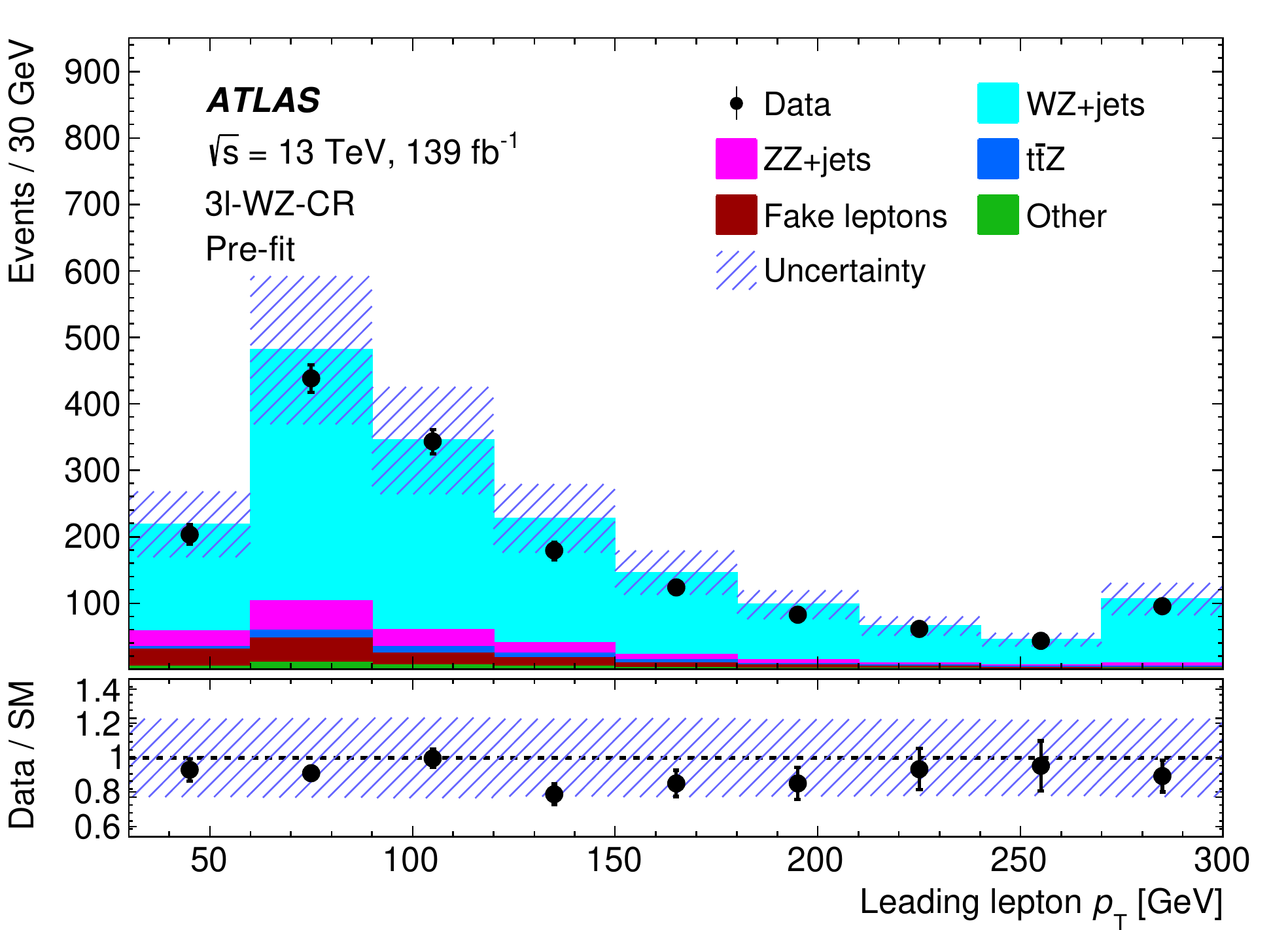}  \label{fig:cr-wz-a}}
\subfloat[]{\includegraphics[width=0.49\textwidth]{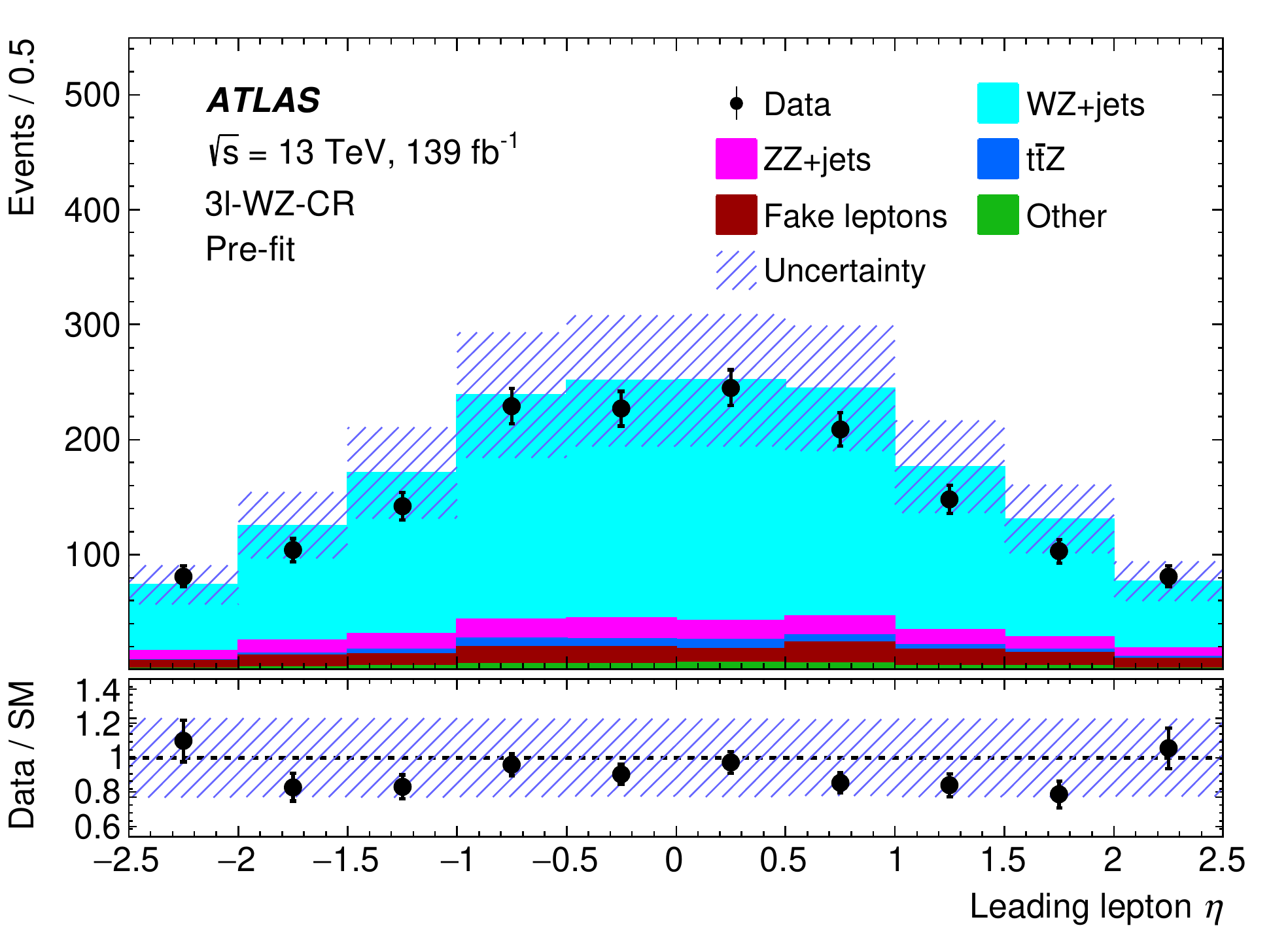} \label{fig:cr-wz-b}}
\caption{Distribution of the (a)~\pt and (b)~$\eta$ of the leading lepton in the \WZjets control region. The shaded band corresponds to the total uncertainty
(systematic and statistical) of the total SM prediction. The lower panel shows the ratio of the data to the SM prediction. The results and uncertainties are shown
before the fit to data is performed. The category `other' contains all processes mentioned in Section~\ref{sec:samples} which are not listed separately.
Events with a leading lepton \pt above \SI{300}{\GeV} are included in the uppermost bin of~(a).}
\end{figure}

\begin{figure}[htb!]
\subfloat[]{\includegraphics[width=0.49\textwidth]{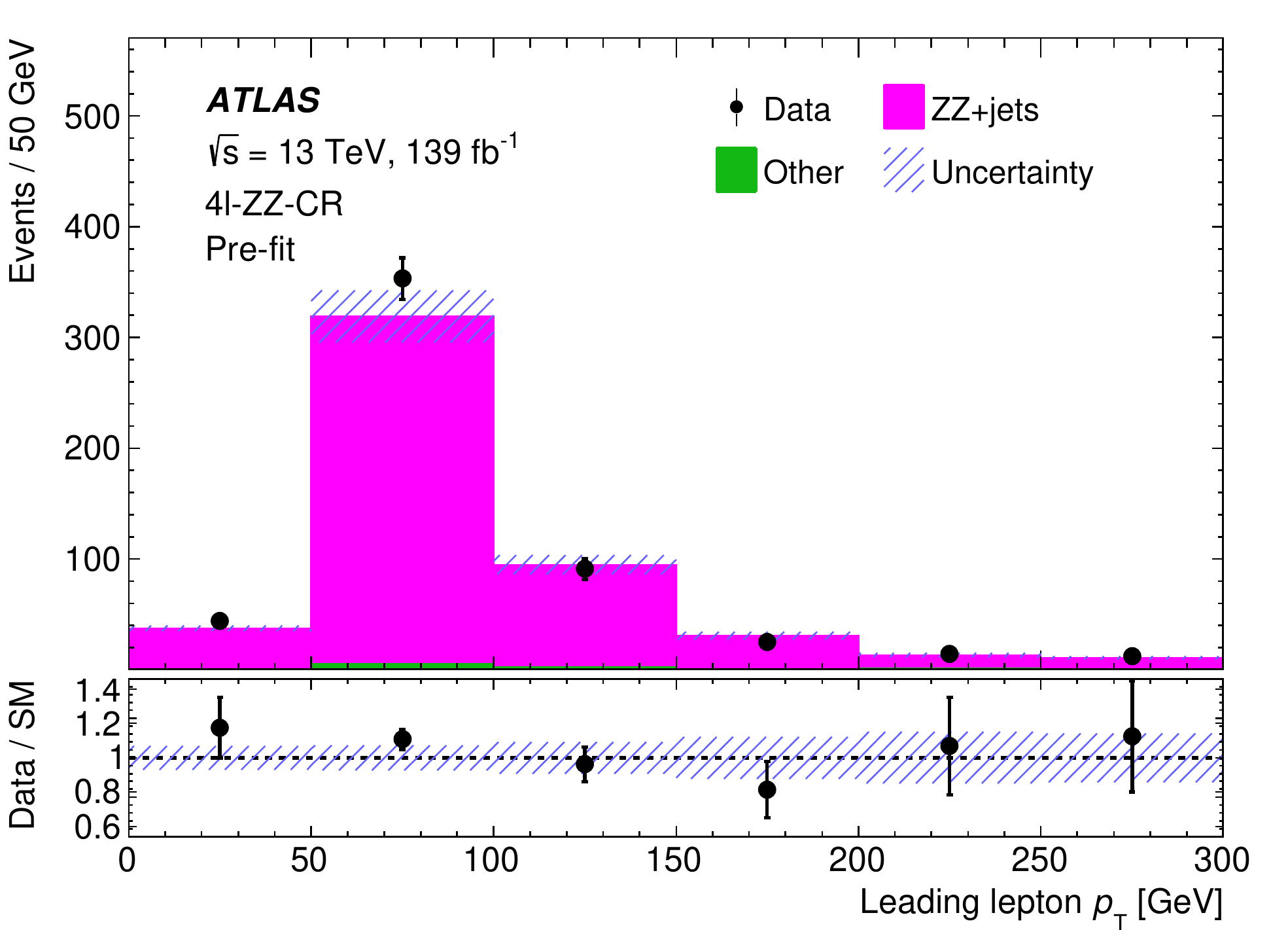} \label{fig:cr-zz-a}}
\subfloat[]{\includegraphics[width=0.49\textwidth]{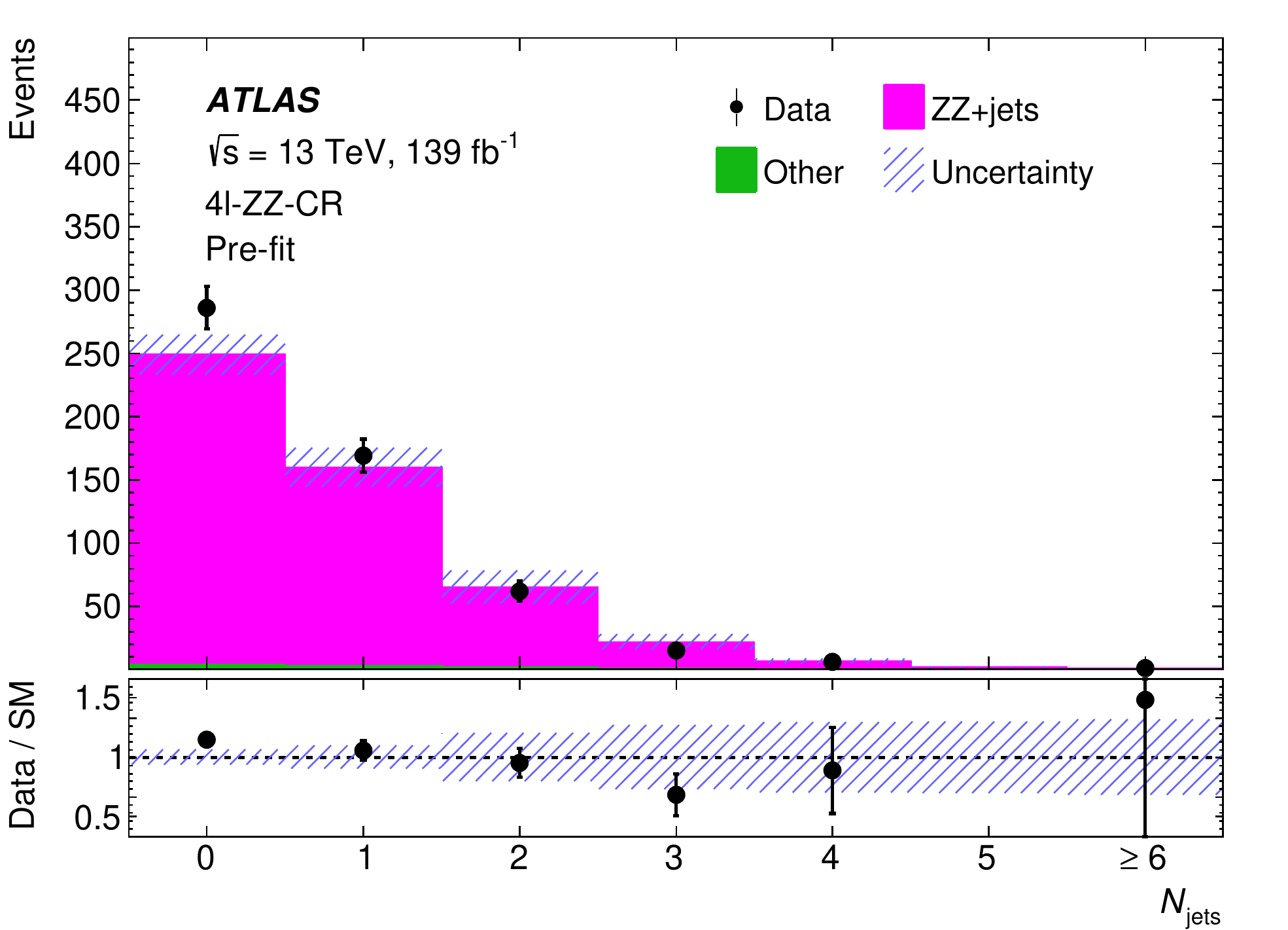}  \label{fig:cr-zz-b}}
\caption{Distribution of (a)~the \pt of the leading lepton and (b)~the number of selected jets in the \ZZjets control region. The shaded band corresponds to the total
uncertainty~(systematic and statistical) of the total SM prediction. The lower panel shows the ratio of the data to the SM prediction. The results and uncertainties
are shown before the fit to data is performed. The category `other' contains all processes mentioned in Section~\ref{sec:samples} which are not listed separately.
Events with a leading lepton \pt above \SI{300}{\GeV} are included in the uppermost bin of~(a).}
\end{figure}

The event yields in these control regions are extrapolated to the signal regions in accord with simulation. As the control regions are mostly populated by $WZ/ZZ$ plus light-flavour
jet events, only the predictions from these light-flavour components in the signal regions are constrained by the observed data yields in the control regions. The $WZ/ZZ + b$- and
$c$-jet\footnote{These backgrounds are defined by separating \VVjets events into light-flavour jet, $b$-jet and $c$-jet components, depending on whether a $b$- or $c$-hadron is found
in the MC event record of any of the selected jets.} backgrounds are constrained to their MC predictions, but with additional normalisation uncertainties assigned~(more details are provided
in Section~\ref{subsec:sys-bkg}). Figures~\ref{fig:cr-wz-a} and~\ref{fig:cr-wz-b} show, respectively, the \pt and $\eta$ distributions of the leading lepton for the \WZjets control region.
The \pt distribution and the number of selected jets in the \ZZjets control region are shown in Figures~\ref{fig:cr-zz-a} and~\ref{fig:cr-zz-b}. All distributions in the control regions
are shown before the simultaneous fit to data is applied~(pre-fit).
 
Another important background in the signal regions is \tWZ production, which can lead to final states very similar to those of the \ttZ signal. A relevant background process
in the trilepton regions is \tZq production, which contributes more for lower jet multiplicities. Other background processes, such as \ttWH, \ttWW, three/four top-quark production,
\HWZ or triboson production can also contribute to the signal regions, but are significantly smaller than the other processes mentioned above.
 
The MC samples used to simulate these processes are described in Section~\ref{sec:samples}. Besides the $WZ/ZZ$ plus light-flavour jets background, for which control
regions are employed to obtain the normalisation, the contributions from all SM processes leading to three or four prompt leptons are estimated entirely from MC simulation
and normalised to their theoretical cross-section predictions.
 
\begin{table}[!htb]
\footnotesize
\caption[]{Definitions of the control regions targeting the \WZjets, $WZ \rightarrow \ell\ell\ell\nu$~(left) and \ZZjets, $ZZ \rightarrow \ell\ell\ell\ell$
processes~(right): the control regions are used to obtain normalisations of the light-flavour components of the \VVjets backgrounds from data.}
\label{tab:cr-def}
\def\arraystretch{1.4}
\begin{center}
\begin{tabular}{l | c | c}
\hline
\hline
Variable & $3\ell$-$WZ$-CR & $4\ell$-$ZZ$-CR \\
\hline
$N_{\ell}\,(\ell=e,\mu)$ & $=3$ & $=4$            \\
& 1~OSSF lepton pair with & 2~OSSF lepton pairs with \\
& $|m_{\ell\ell}-\mZ| < \SI{10}{\GeV}$ & $|m_{\ell\ell}-\mZ| < \SI{10}{\GeV}$ \\
 
$\pt(\ell_{1},\,\ell_{2},\,\ell_{3},\,\ell_{4})$ & $> 27,\,20,\,\SI{20}{\GeV}$ & $> 27,\,20,\,10,\,\SI{7}{\GeV}$ \\
$N_{\mathrm{jets}}$ & $\geq 3$ & -- \\
$N_{b\textrm{-jets}}$@85\% & $=0$ & -- \\
\met & -- & $\SI{20}{\GeV} < \met < \SI{40}{\GeV}$ \\
\hline
\hline
\end{tabular}
\end{center}
\end{table}
 
\subsection{Fake lepton background}
\label{subsec:fakes}
 
Different types of objects, which are misidentified as leptons, are referred to as `fake leptons' throughout the rest of the document. In the signal regions of
this analysis, this background arises mainly from dileptonic \ttbar decays where additional non-prompt leptons arise from heavy-flavour hadron decays.
 
To estimate the contribution of fake leptons in the signal regions, a fully data-driven method, called the `matrix method' is employed. Descriptions of
this technique can be found in Refs.~\cite{ATLAS-CONF-2014-058,SUSY-2013-09}. It relies on the prompt and fake leptons having different probabilities of passing the identification,
isolation and impact parameters requirements. The method uses data events selected with the same criteria as in the signal regions, but with looser lepton
selections\footnote{Whereas `Medium' electrons are used for the nominal selection, the identification WP is relaxed to `LooseAndBLayerLH'~\cite{PERF-2017-01}
for the looser electron selection. No isolation requirements are applied for electrons and muons in the looser selection.} than the ones defined in
Section~\ref{sec:reconstruction}.
 
An alternative version of the matrix method is described in Ref.~\cite{Varnes:2016nrb}. It evaluates the total number of fake electrons and muons entering the
signal regions via the maximisation of a likelihood function. The likelihood function is constructed from a product of Poisson probability functions that
represent the numbers of leptons passing different quality criteria for the signal regions. The observed number of leptons selected with the looser criteria and the
probabilities~(efficiencies) for fake or prompt leptons to satisfy the nominal lepton requirements are fixed, while the expectation values of the Poisson
functions --- the numbers of fake leptons in the signal regions --- are obtained from the likelihood maximisation. This method offers a more robust fake-lepton
estimation for statistically limited regions. For the differential measurements, the estimations are performed separately for each bin of the measured variables.
 
The probabilities of fake leptons to satisfy the nominal lepton requirements~(fake-lepton efficiencies) are obtained from data. They are measured separately for
electrons and muons in events with exactly two leptons with the same charge~(same-sign) and at least one $b$-jet identified at the 85\% efficiency WP. The measurements
are performed after subtracting the contributions, estimated from MC events, of charge-misidentified electrons and prompt leptons in the same-sign region. It has been
checked that the dominant fake lepton source in this region are heavy-flavour hadron decays, as in the signal regions. The fake-lepton efficiencies are approximately
10\% for electrons and 15\% for muons, with an increase to around 20\% for muons with  $\pt < \SI{12}{\GeV}$ and $> \SI{35}{\GeV}$. The equivalent probabilities for
prompt leptons~(prompt-lepton efficiencies) are obtained from \Zll simulation and the respective scale factors for electrons or muons. The prompt-lepton efficiencies
are in most cases higher than 90\% for both the electrons and muons. They increase for larger lepton \pt values and reach $> \SI{98}{\%}$ for $\pt > \SI{35}{\GeV}$.
Both the fake- and prompt-lepton efficiencies are parameterised as a function of \pt and \abseta of the respective lepton.
 
Systematic uncertainties are assigned to the fake-lepton estimates to account for differences in the relative contributions of the various fake-lepton sources between the
signal regions and the regions used for the efficiency measurements. Further uncertainties arise from the subtraction of prompt and charge-misidentified leptons, as well as
from the dependencies of the fake-lepton efficiencies on the number of jets/$b$-jets in the events. The method is also affected by statistical uncertainties arising from the
limited number of events in the data sample used to evaluate the fake-lepton yields, as well as the statistical limitations of the efficiency measurements. Similarly to the
nominal values, the uncertainties of the fake- and prompt-lepton efficiencies are binned in \pt and \abseta of the leptons and propagated to the fake-lepton estimation in the
signal regions. The overall uncertainties are approximately 50\% for both electrons and muons, but they can fluctuate for the differential measurements, depending on the variable
and the kinematic region.

\subsection{Pre-fit event yields in signal and control regions}
 
To validate the SM background modelling explained in the previous sections, Table~\ref{tab:yields_inclusive_prefit} presents a comparison between the total expected
background prediction and the observed data events in the trilepton and tetralepton signal regions, as well as in the \WZjets and \ZZjets control regions. The event yields
and uncertainties are shown before applying the fitting procedure. The statistical and all systematic uncertainties as explained in Section~\ref{sec:systematics} are
considered except the normalisation uncertainties of the processes which are free parameters in the fit. Within the uncertainties, agreement between data and the
SM predictions is observed in nearly all regions.
 
\begin{table}[!htb]
\footnotesize
\caption[]{The observed and expected numbers of events in the trilepton and tetralepton signal regions, as well as in the \VVjets control regions. The predictions are
shown before the fit to data. The $WZ$ and \ZZjets backgrounds are listed separately for their light-flavour~($l$), $b$-jet and $c$-jet components. The category `fake leptons'
refers to the contributions from fake and non-prompt leptons. `Other' includes the contributions from \HWZ, \ttWW, three/four top-quark production and triboson processes.
Background categories with event yields shown as `--' do not contribute significantly to a region. The indicated uncertainties consider statistical errors as well as all
experimental and theoretical systematic uncertainties, except the normalisation uncertainties of the fitted background components.}
\label{tab:yields_inclusive_prefit}
\def\arraystretch{1.4}
\begin{center}
\resizebox{\textwidth}{!}{
\begin{tabular}{l | c c c c c c | c c}
\hline
\hline
Region & $3\ell$-$Z$-$1b4j$ & $3\ell$-$Z$-$2b3j$ & $4\ell$-SF-$1b$ & $4\ell$-SF-$2b$ & $4\ell$-DF-$1b$ & $4\ell$-DF-$2b$ & $3\ell$-$WZ$-CR & $4\ell$-$ZZ$-CR \\
& -PCBT & -PCBT & & & & & & \\
\hline
\ttZ            & $164 \pm 14$~~   & $210 \pm 9$~~~~    & $13.0 \pm 1.6$~~ & $23.3 \pm 1.8$~~ & $16.8 \pm 1.4$~~ & $22.5\pm 1.0$~~ & $44 \pm 11$    & $0.7 \pm 0.1$  \\
$WZ + l$        & $3.1 \pm 2.3$  & $0.3\pm 0.3$     & --             & --             & --             & --            & $1160 \pm 370$~~ & -- \\
$WZ + b$      	& $30 \pm 18$    & $16.0 \pm 9.4$~~ & --             & --             & --             & --            & $17 \pm 10$      & -- \\
$WZ + c$      	& $12.5 \pm 5.9$~~ & $2.1 \pm 1.0$  & --             & --             & --             & --            & $230 \pm 100$    & -- \\
$ZZ + l$        & $0.3 \pm 0.2$  & $0.03 \pm 0.03$ & $1.8 \pm 0.9$ & $1.0 \pm 0.5$  & $0.5 \pm 0.2$ & $0.02 \pm 0.01$ & $116 \pm 17$~~    & $461 \pm 37$~~   \\
$ZZ + b$        & $3.5 \pm 1.8$  & $2.2 \pm 1.1$   & $1.1 \pm 0.7$ & $2.8 \pm 1.7$  & $0.2 \pm 0.1$ & $0.08 \pm 0.05$ & $2.0 \pm 1.0$     & $14.3 \pm 7.6$~~ \\
$ZZ + c$        & $0.8 \pm 0.3$  & $0.2 \pm 0.1$   & $1.0 \pm 0.5$ & $1.2 \pm 0.6$  & $0.2 \pm 0.1$ & $0.02 \pm 0.01$ & $13.7 \pm 4.4$~~  & $20.7 \pm 7.5$~~ \\
\tWZ            & $23.7 \pm 4.2$~~ & $19.3 \pm 7.1$~~ & $2.7 \pm 0.5$  & $2.0 \pm 0.8$  & $3.6 \pm 1.1$  & $2.1 \pm 0.9$ & $13.3 \pm 1.2$~~ & $0.2 \pm 0.1$    \\
\tZq            & $11.7 \pm 5.0$~~ & $29.3 \pm 9.3$~~ & --             & --             & --             & --            & $9.1 \pm 3.5$ & --              \\
\ttWH           & $5.8 \pm 0.9$  & $10.1 \pm 2.2$~~ & $0.5 \pm 0.1$  & $0.9 \pm 0.1$  & $0.6 \pm 0.1$  & $0.8 \pm 0.1$ & $1.8 \pm 0.4$  & $0.01 \pm 0.01$ \\
Fake leptons    & $30 \pm 15$    & $14.6 \pm 7.4$~~ & $0.7 \pm 0.6$  & $0.9 \pm 0.8$  & $0.9 \pm 0.8$  & $0.3 \pm 0.3$ & $86 \pm 43$    & $7.7 \pm 3.6$   \\
Other           & $0.7 \pm 0.4$  & $1.5 \pm 0.7$  & $0.7 \pm 0.3$  & $0.3 \pm 0.1$  & $0.7 \pm 0.4$  & $0.2 \pm 0.1$ & $12.4 \pm 6.4$~~ & $0.9 \pm 0.6$   \\
\hline
SM total        & $286 \pm 31$~~   & $306 \pm 21$~~   & $21.3 \pm 2.3$~~ & $32.4 \pm 3.0$~~ & $23.4 \pm 2.0$~~ & $26.0 \pm 1.5$~~ & $1710 \pm 390$~~ & $505 \pm 39$~~    \\
\hline
Data     & 272 & 343 & 19 & 33 & 33 & 32 & 1569 & 539 \\
\hline
\hline
\end{tabular}
}
\end{center}
\end{table}
 
\FloatBarrier
 
\section{Systematic uncertainties}
\label{sec:systematics}
The signal and background predictions in all signal regions are affected by several sources of experimental and theoretical systematic uncertainty.
These are considered for both inclusive and differential measurements presented in Sections~\ref{sec:results_inclusive} and~\ref{sec:differential}.
The uncertainties can be classified into the different categories which are described in the following subsections.
 
\subsection{Detector-related uncertainties}
\label{subsec:sys-det}
 
The uncertainty in the combined 2015--2018 integrated luminosity is 1.7\%~\cite{ATLAS-CONF-2019-021}, obtained using the LUCID-2 detector~\cite{Avoni:2018} for
the primary luminosity measurements. This systematic uncertainty affects all processes modelled using MC simulations apart from the light-flavour components of
the \VVjets backgrounds, whose normalisations are taken from data in control regions.
 
The uncertainty in the reweighting of the MC pile-up distribution to match the distribution in data is evaluated by varying the pile-up correction factors
used to perform the reweighting.
 
Uncertainties associated with the lepton selection arise from the trigger, reconstruction, identification and isolation efficiencies, and the lepton momentum scale
and resolution~\cite{PERF-2015-10,EGAM-2018-01,MUON-2018-03}. They are below 1\% for the individual sources and have a total impact of 2--2.5\% on the measurements.
Uncertainties associated with the jet selection arise from the jet energy scale~(JES), the JVT requirement and the jet energy resolution~(JER). The JES and its uncertainties
are derived by combining information from test-beam data, collision data and simulation~\cite{JETM-2018-05}. The uncertainties in the JER and JVT increase at lower jet \pt.
The overall effect of uncertainties related to jet selection and calibration is approximately 2\%.
 
The efficiency of the flavour-tagging algorithm is measured for each jet flavour using control samples in data and in simulation. From these
measurements, correction factors are derived to correct the tagging rates in the simulation. In the case of $b$-tagged jets, the correction factors and
their uncertainties are estimated from data using dileptonic \ttbar events~\cite{FTAG-2018-01}. In the case of $c$-jets, they are derived from
jets arising from \Wboson boson decays in \ttbar events~\cite{ATLAS-CONF-2018-001}. In the case of light-flavour jets, the correction factors are derived
using dijet events~\cite{ATLAS-CONF-2018-006}. Sources of uncertainty affecting the $b$- and $c$-tagging efficiencies are evaluated as a function of
jet \pt, including bin-to-bin correlations. The uncertainties in the efficiency for tagging light-flavour jets depend on the jet \pt and on $\eta$.
An additional uncertainty is assigned to account for the extrapolation of the $b$-tagging efficiency measurement from the \pt region used
to determine the correction factors to regions with higher \pt. The impact of flavour-tagging uncertainties on the measurements depends on the signal
regions and is 2--3\% in total.

\subsection{Signal modelling uncertainties}
 
Different sources of systematic uncertainty in the theoretical predictions of the \ttZ process are considered. To evaluate the effect of \muR and \muF uncertainties,
the scales used in the ME of the \MGNLOs~+~\PYTHIA[8] samples are varied simultaneously, as well as individually, by factors of \num{2.0} and \num{0.5} relative to their
nominal values. The uncertainty due to the ISR is estimated using a set of variations of the A14 tune's parameter values. Uncertainties associated with the choice of PDF set
are evaluated according to the \PDFforLHC prescription~\cite{Butterworth:2015oua} using eigenvector variations from multiple NLO PDF sets, the effects of which are added in
quadrature.
 
The systematic uncertainty due to the modelling of the parton shower, the hadronisation and the underlying event --- called the parton-shower uncertainty in the following --- is
quantified by employing an alternative \ttZ sample generated with \MGNLOs, but interfaced to \HERWIG[7] instead of \PYTHIA[8].

\subsection{Background modelling uncertainties}
\label{subsec:sys-bkg}
 
The normalisation of the \WZjets and \ZZjets backgrounds with light-flavour jets are obtained from data, as discussed in Section~\ref{subsec:prompt}. The \VVjets
components with $b$- or $c$-jets are constrained to their MC predictions and normalisation uncertainties of 50\%~($WZ/ZZ + b$) and 30\%~($WZ/ZZ + c$) are assigned to
them. These uncertainties are evaluated from data/MC comparisons in $\Zboson + b/c$ events~\cite{STDM-2017-38}, but also take into account differences in the heavy-flavour
jet fractions between \Zjets and \VVjets events. Modelling uncertainties of \VVjets related to the \muR and \muF scales and the PDF choice are obtained with the same prescription
as for the signal. Uncertainties attributed to the resummation scale and CKKW matching scale~\cite{Hoeche:2009rj,Hoeche:2012yf} are evaluated from alternative \VVjets
samples with variations of these scale choices.
 
Uncertainties related to the \muR and \muF scales and the PDF of the \tWZ background are evaluated in the same way as for the \ttZ and \VVjets samples. An additional uncertainty
is assigned to the \tWZ process to account for the interference between the \ttZ and \tWZ processes. It is evaluated by switching to an alternative diagram removal
scheme~(DR1 vs DR2)~\cite{Demartin:2016axk} and obtaining an uncertainty from the differences observed in the signal regions.
 
Scale and PDF uncertainties of the \tZq background are obtained in the same way as for the previously described samples. In addition, a normalisation uncertainty of 30\%
is assigned, motivated by the measurements of this process presented in Refs.~\cite{TOPQ-2016-14,CMS-TOP-16-020}. As for the \ttZ signal, the uncertainty due to the ISR
is evaluated using alternative samples with \PYTHIA[8] A14 Var3c eigentune variations.
 
For the fake-lepton background, statistical as well as systematic uncertainties are considered as explained in Section~\ref{sec:background}. They are evaluated for
each signal region independently and applied as normalisation uncertainties of the total fake-lepton background contribution in each region.
 
For the \ttH background, a normalisation uncertainty of approximately 10\% due to the choice of QCD scales and PDF is used~\cite{deFlorian:2016spz}. For processes giving
smaller backgrounds, namely \HWZ, \ttW, \ttWW, triboson and three/four top-quark production, a conservative overall normalisation uncertainty of 50\% is applied.
 
\section{Results of the inclusive cross-section measurement}
\label{sec:results_inclusive}
The ratio of the measured value of the inclusive \ttZ production cross section to its corresponding SM prediction~(\muttZ) is obtained from a simultaneous fit to the
numbers of events in the trilepton and tetralepton signal regions~(as defined in Tables~\ref{tab:sr-3l-def_main} and~\ref{tab:sr-4l-def_main}), as well as the \WZjets
and \ZZjets control regions (defined in Table~\ref{tab:cr-def}). For trilepton events, only the dedicated regions for the inclusive measurement are included in the fit.
The fit is based on the profile-likelihood technique, with a likelihood function as a product of Poisson probability functions given by the observed event yields in the
signal and control regions. The value of \muttZ as well as the normalisations of the light-flavour components of the \VVjets backgrounds are treated as free parameters in
the fit. The systematic uncertainties described in Section~\ref{sec:systematics} are included in the fit as nuisance parameters constrained by Gaussian functions. None of
the uncertainty parameters is found to be significantly constrained or pulled by the fit. The calculation of confidence intervals and hypothesis testing is performed using
a modified frequentist method as implemented in the \textsc{RooStats} framework~\cite{Cowan:2010js,Read:2002hq,RooFit}.
 
\begin{figure}[!htb]
\centering
\includegraphics[width=0.99\textwidth]{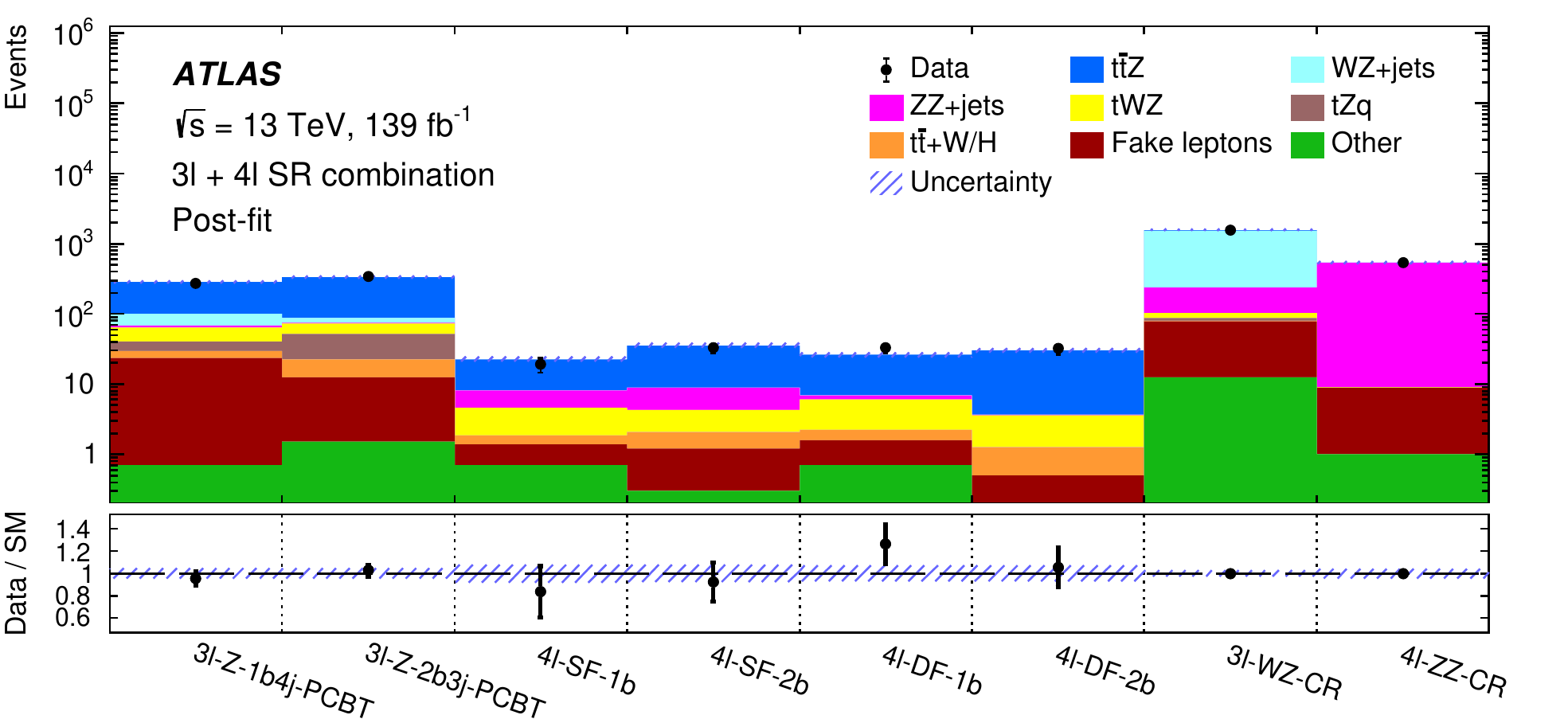}
\caption{The observed and expected event yields in the trilepton and tetralepton signal regions, as well as the \VVjets control regions, after the combined fit. The bottom
panel shows the ratio of data to the total SM prediction. The size of the combined statistical and systematic uncertainty in the sum of the fitted signal and backgrounds
is indicated by the blue hatched band.}
\label{fig:inclusive_postfit}
\end{figure}

Within their uncertainties, the fitted normalisations of the light-flavour components of the \WZjets and \ZZjets backgrounds are compatible with unity, but can vary by
up to 10\% from their initial value. The observed and expected total event yields in the signal regions and the \VVjets control regions after the combined
fit~(post-fit) are shown in Figure~\ref{fig:inclusive_postfit} and detailed in Table~\ref{tab:yields_inclusive_postfit}.
The strong anti-correlation between the $WZ + l$ and $WZ + c$ backgrounds results in a smaller total uncertainty of the fitted SM background expectation in $3\ell$-$WZ$-CR
compared with the uncertainties of the individual $WZ + l$ and $WZ + c$ components.
 
\begin{figure}[htb!]
\subfloat[]{\includegraphics[width=0.49\textwidth]{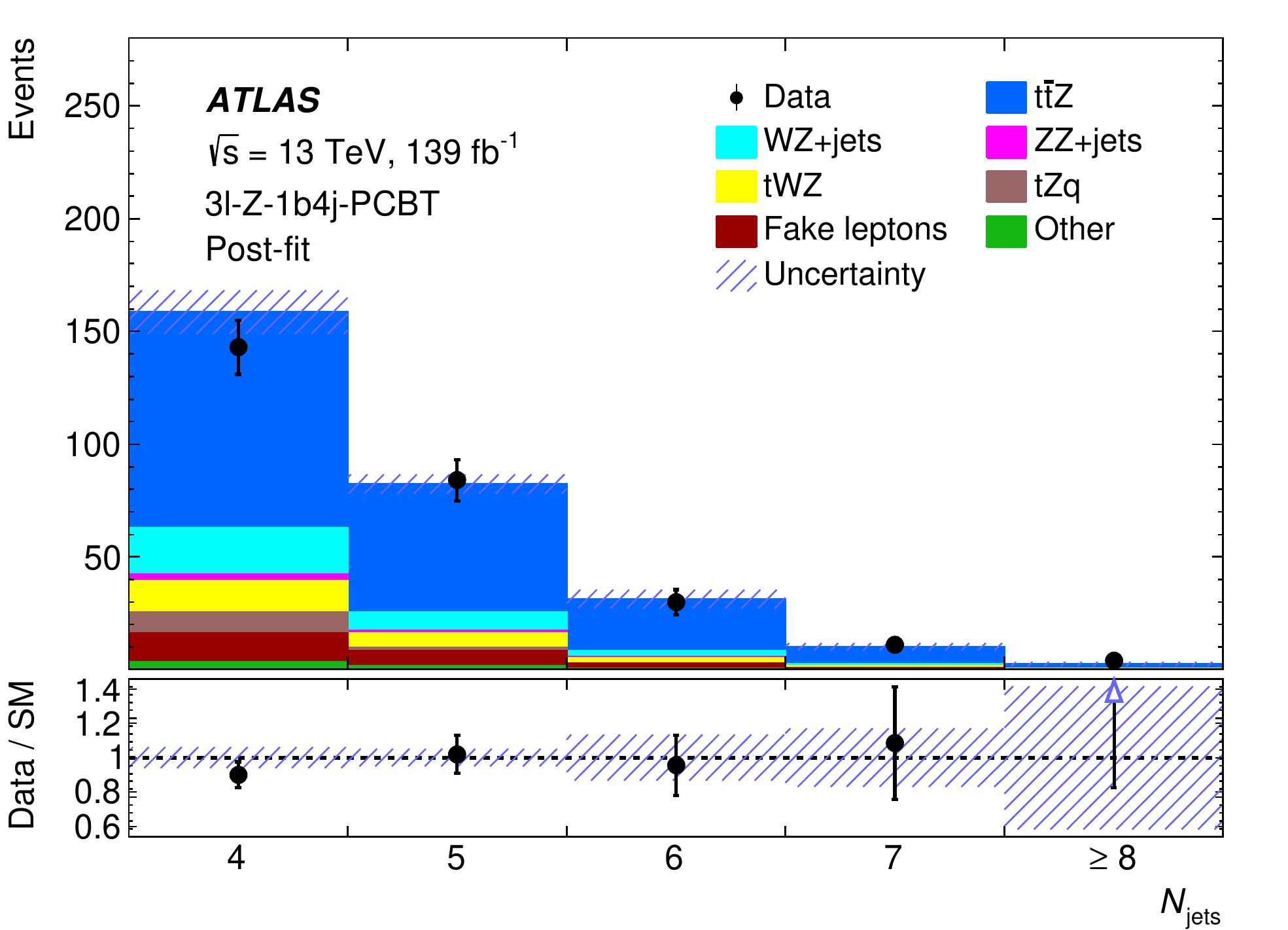} \label{fig:sr-3l-a}}
\subfloat[]{\includegraphics[width=0.49\textwidth]{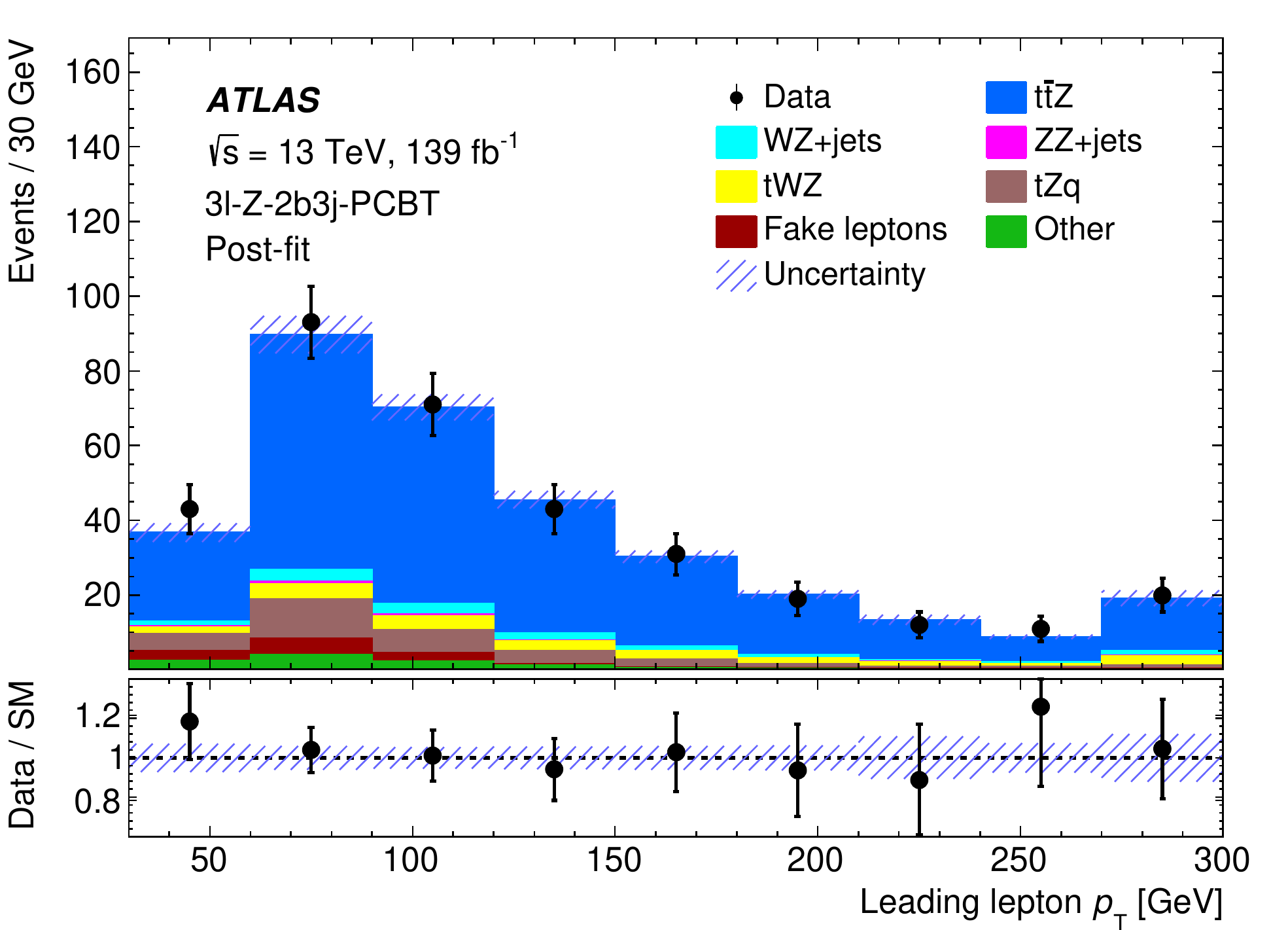}  \label{fig:sr-3l-b}}
\caption{Post-fit distributions of (a)~$N_{\mathrm{jets}}$ in $3\ell$-$Z$-$1b4j$-PCBT and (b)~the~\pt of the leading lepton in $3\ell$-$Z$-$2b3j$-PCBT. The shaded band includes all
sources of statistical and systematic uncertainty after the combined fit. The lower panel shows the ratio of data to the total SM prediction. The uppermost bins include all events
above the $x$-axis ranges. The blue triangular marker in the lower panel of~(a) points to the position of a data point which lies slightly beyond the $y$-axis range shown.}
\label{fig:sr-3l}
\end{figure}

\begin{figure}[htb!]
\subfloat[]{\includegraphics[width=0.49\textwidth]{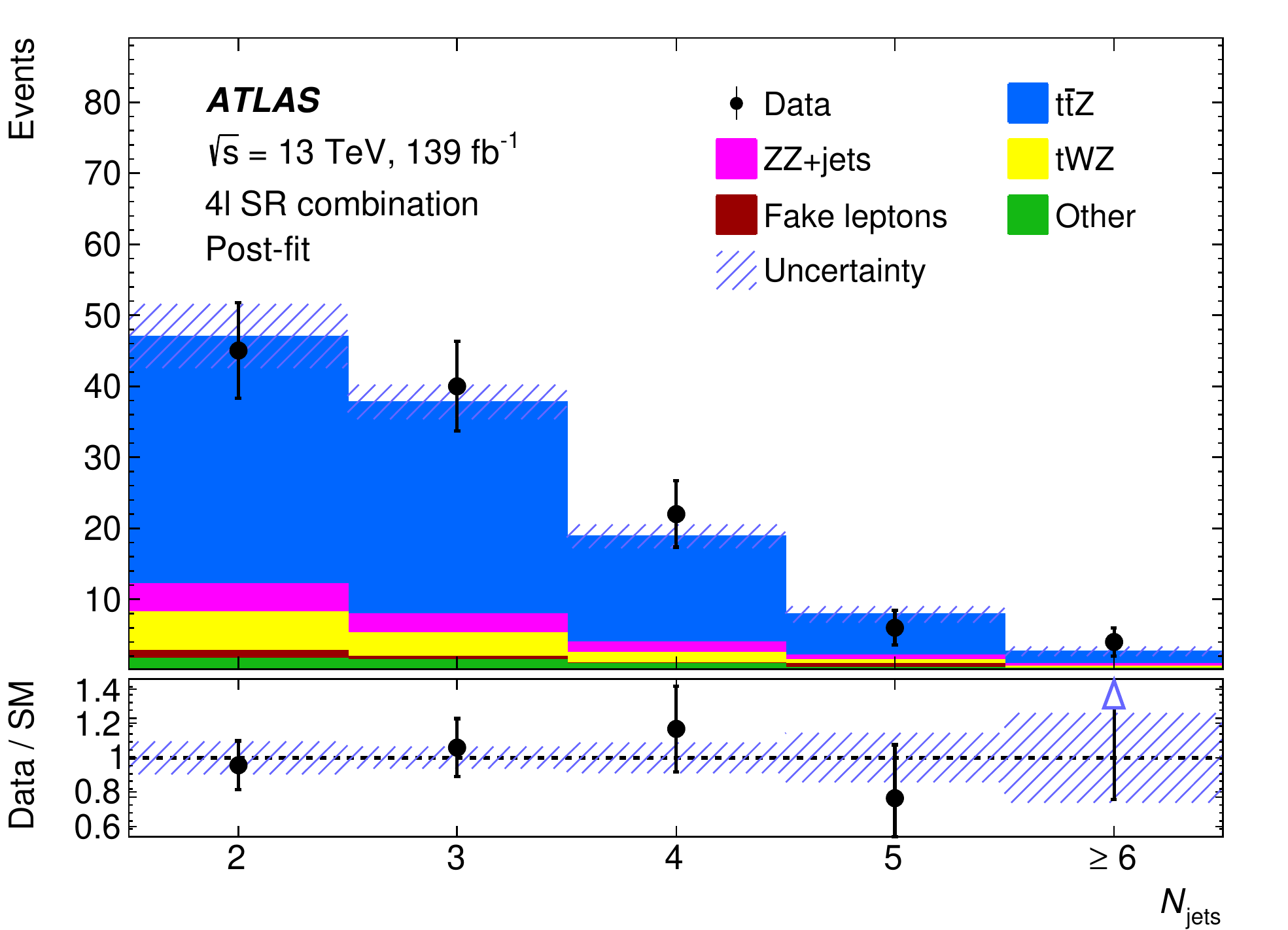} \label{fig:sr-4l-a}}
\subfloat[]{\includegraphics[width=0.49\textwidth]{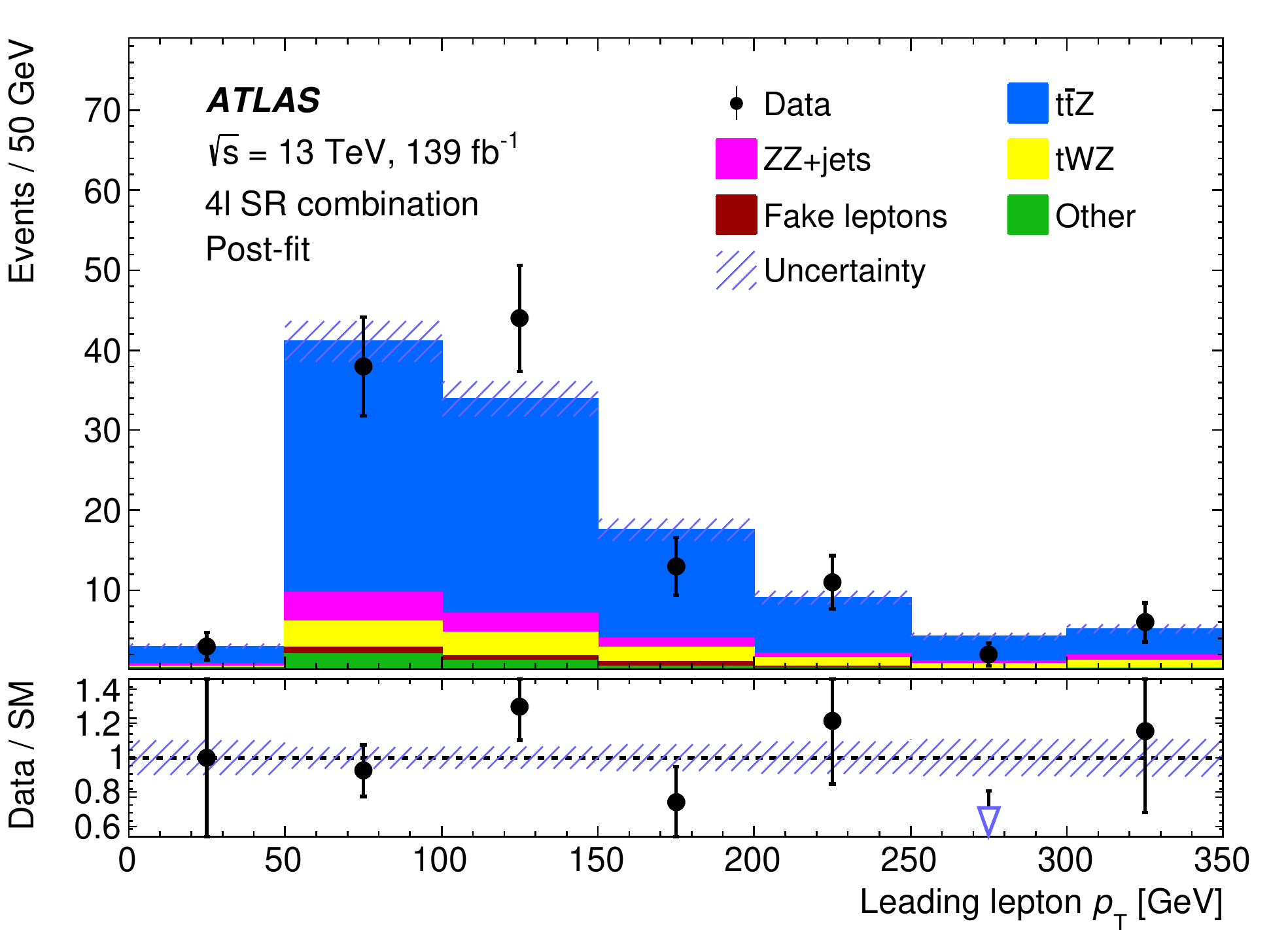}  \label{fig:sr-4l-b}}
\caption{Post-fit distributions of (a)~$N_{\mathrm{jets}}$ and (b)~the~\pt of the leading lepton in the combination of the tetralepton signal regions. The shaded band includes all
sources of statistical and systematic uncertainty after the combined fit. The lower panel shows the ratio of data to the total SM prediction. The uppermost bins include all events
above the $x$-axis ranges. The blue triangular markers in the lower panels point to the positions of data points which lie slightly beyond the $y$-axis range shown.}
\end{figure}

\begin{figure}[htb!]
\subfloat[]{\includegraphics[width=0.49\textwidth]{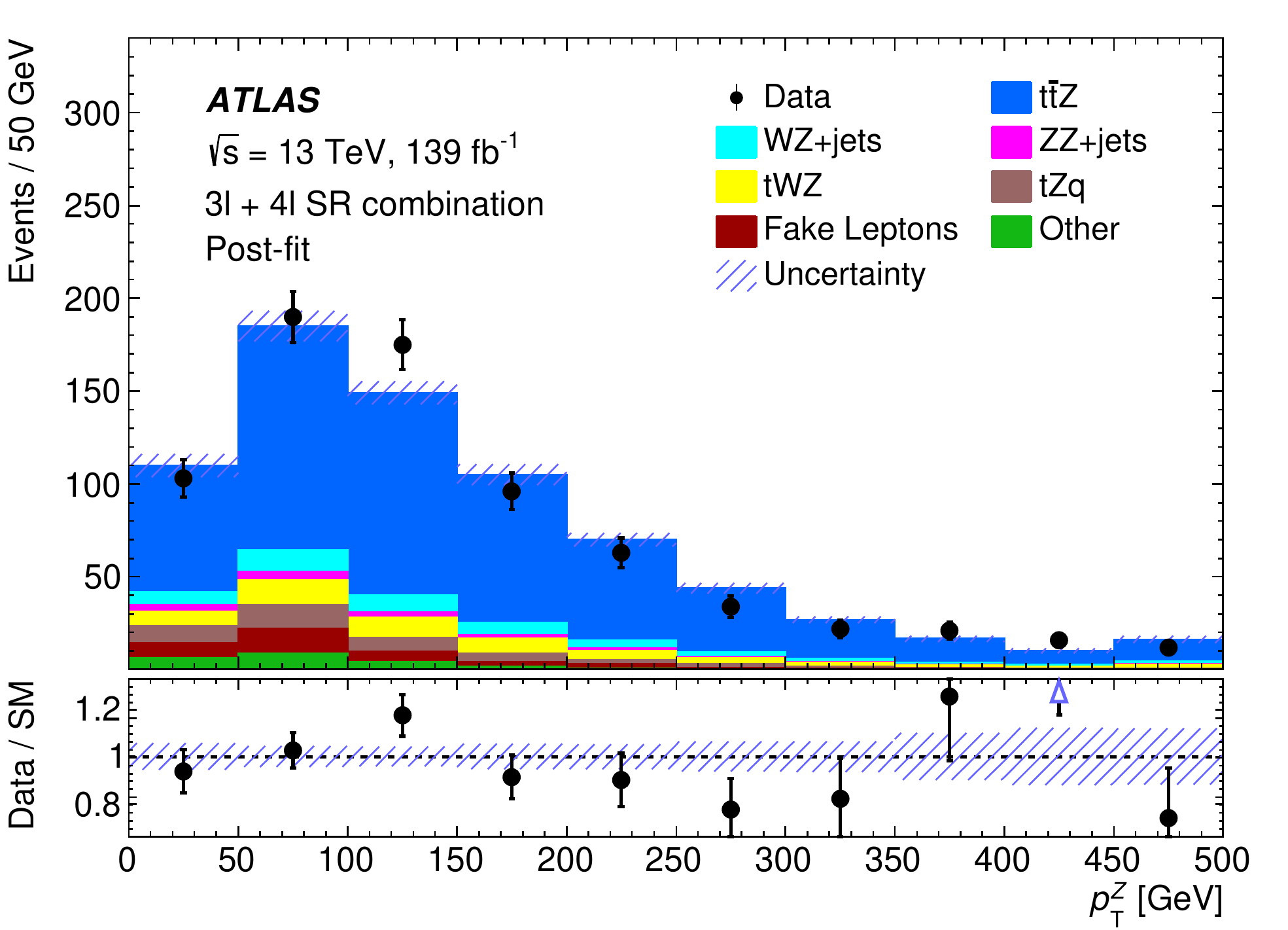}  \label{fig:sr-comb-a}}
\subfloat[]{\includegraphics[width=0.49\textwidth]{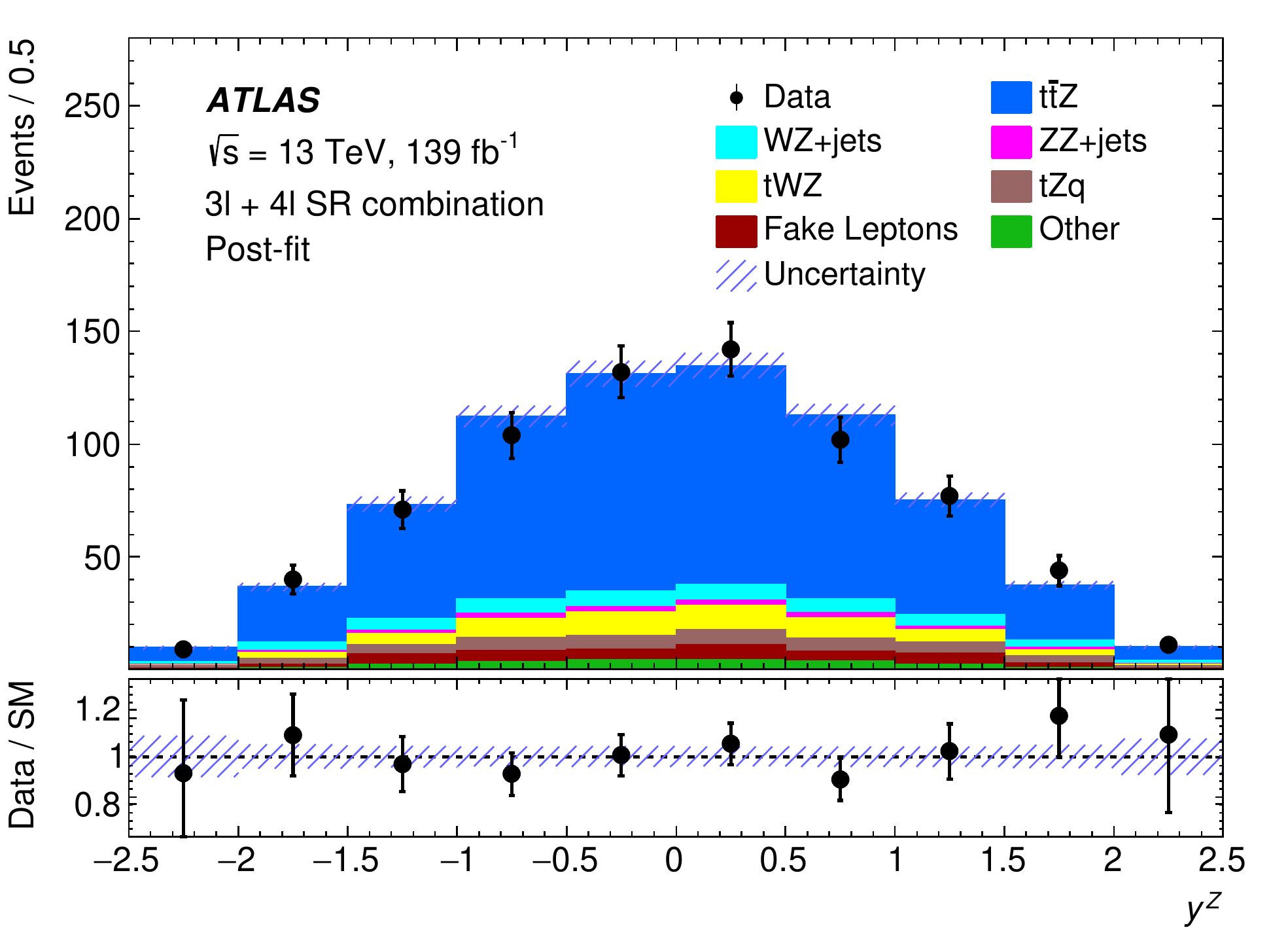} \label{fig:sr-comb-b}}
\caption{Post-fit distributions of the (a)~\pt and (b)~rapidity of the \Zboson boson in the combination of the trilepton and tetralepton regions. The shaded band includes all sources
of statistical and systematic uncertainty after the combined fit. The lower panel shows the ratio of data to the total SM prediction. The uppermost bins include all events
above the $x$-axis ranges. The blue triangular marker in the lower panel of~(a) points to the position of a data point which lies slightly beyond the $y$-axis range shown.}
\label{fig:sr-comb}
\end{figure}

Comparisons between data and the post-fit SM predictions for some selected variables which offer sensitivity to the quality of the background modelling in the
signal regions are also presented. The number of selected jets with $\pt > \SI{25}{\GeV}$ in signal region $3\ell$-$Z$-$1b4j$-PCBT is shown in Figure~\ref{fig:sr-3l-a}.
The \pt of the leading lepton in $3\ell$-$Z$-$2b3j$-PCBT is given in Figure~\ref{fig:sr-3l-b}. Figure~\ref{fig:sr-4l-a} depicts the number of selected jets
and Figure~\ref{fig:sr-4l-b} the \pt of the leading lepton in the combination of the four tetralepton regions. Figure~\ref{fig:sr-comb-a} shows the \pt and Figure~\ref{fig:sr-comb-b}
the rapidity~($y$) of the reconstructed \Zboson boson in the combination of the trilepton and tetralepton regions.
 
\begin{table}[!htb]
\footnotesize
\caption[]{The observed and expected numbers of events in the trilepton and tetralepton signal regions, as well as the \VVjets control regions, after the combined fit.
The definitions of the background categories are the same as in Table~\ref{tab:yields_inclusive_prefit}. Categories with event yields shown as `--' do not contribute
significantly to a region. The indicated uncertainties consider all experimental and theoretical systematic uncertainties as well as the statistical errors. As systematic uncertainties
might be correlated between different processes, the individual uncertainties do not necessarily add up in quadrature to the uncertainty of the total SM prediction.}
\label{tab:yields_inclusive_postfit}
\def\arraystretch{1.4}
\begin{center}
\resizebox{\textwidth}{!}{
\begin{tabular}{l | c c c c c c | c c}
\hline
\hline
Region & $3\ell$-$Z$-$1b4j$ & $3\ell$-$Z$-$2b3j$ & $4\ell$-SF-$1b$ & $4\ell$-SF-$2b$ & $4\ell$-DF-$1b$ & $4\ell$-DF-$2b$ & $3\ell$-$WZ$-CR & $4\ell$-$ZZ$-CR \\
& -PCBT & -PCBT & & & & & & \\
\hline
\ttZ            & $185 \pm 16$~~   & $247 \pm 20$~~   & $14.5 \pm 1.7$~~ & $26.9 \pm 2.5$~~ & $19.3 \pm 1.8$~~ & $26.7 \pm 2.3$~~ & $45 \pm 11$    & $0.8 \pm 0.1$ \\
$WZ + l$        & $2.4 \pm 1.8$  & $0.2\pm 0.3$   & --             & --             & --             & --             & $1068 \pm 110$~~ & -- \\
$WZ + b$        & $20 \pm 11$    & $10.8 \pm 6.1$~~ & --             & --             & --             & --             & $11.2 \pm 6.3$~~ & -- \\
$WZ + c$        & $10.8 \pm 4.8$~~ & $1.8 \pm 0.8$  & --             & --             & --             & --             & $207 \pm 87$~~   & -- \\
$ZZ + l$        & $0.3 \pm 0.2$  & $0.02 \pm 0.02$ & $1.7 \pm 0.7$ & $0.9 \pm 0.5$  & $0.5 \pm 0.1$ & $0.02 \pm 0.01$ & $121 \pm 15$~~    & $496 \pm 26$~~   \\
$ZZ + b$        & $3.0 \pm 1.6$  & $2.0 \pm 1.0$   & $0.9 \pm 0.6$ & $2.5 \pm 1.5$  & $0.2 \pm 0.1$ & $0.07 \pm 0.07$ & $1.8 \pm 0.9$     & $12.9 \pm 7.1$~~ \\
$ZZ + c$        & $0.7 \pm 0.2$  & $0.1 \pm 0.1$   & $0.9 \pm 0.5$ & $1.1 \pm 0.6$  & $0.2 \pm 0.1$ & $0.02 \pm 0.01$ & $13.0 \pm 4.1$~~  & $19.8 \pm 7.1$~~ \\
\tWZ            & $23.8 \pm 4.0$~~ & $20.5 \pm 7.0$~~ & $2.7 \pm 0.4$  & $2.2 \pm 0.8$  & $3.8 \pm 1.1$  & $2.3 \pm 0.9$  & $13.2 \pm 1.2$~~ & $0.2 \pm 0.1$   \\
\tZq            & $10.8 \pm 4.5$~~ & $29.7 \pm 9.0$~~ & --             & --             & --             & --             & $8.6 \pm 3.2$  & --              \\
\ttWH           & $5.8 \pm 0.9$  & $10.1 \pm 2.2$~~ & $0.5 \pm 0.1$  & $0.9 \pm 0.1$  & $0.6 \pm 0.1$  & $0.8 \pm 0.1$  & $1.8 \pm 0.4$  & $0.01 \pm 0.01$ \\
Fake leptons    & $23 \pm 11$    & $11.0 \pm 5.3$~~ & $0.7 \pm 0.3$  & $0.9 \pm 0.4$  & $0.9 \pm 0.5$  & $0.3 \pm 0.1$  & $65 \pm 31$    & $7.9 \pm 3.1$   \\
Other           & $0.7 \pm 0.4$  & $1.5 \pm 0.7$  & $0.7 \pm 0.3$  & $0.2 \pm 0.1$  & $0.7 \pm 0.4$  & $0.2 \pm 0.1$  & $12.4 \pm 6.3$~~ & $1.0 \pm 0.5$   \\
\hline
SM total        & $286 \pm 13$~~   & $334 \pm 15$~~   & $22.5 \pm 1.8$~~ & $35.6 \pm 2.7$~~ & $26.1 \pm 1.9$~~ & $30.3 \pm 2.2$~~ & $1569 \pm 43$~~~~  & $539 \pm 23$~~    \\
\hline
Data     & 272 & 343 & 19 & 33 & 33 & 32 & 1569 & 539 \\
\hline
\hline
\end{tabular}
}
\end{center}
\end{table}
 
Table~\ref{tab:mu_inclusive} summarises the measured \muttZ parameters obtained from the individual fits in the trilepton and tetralepton regions, as well as the value from the combined
$3\ell + 4\ell$ fit. The values obtained from the fit in the different regions are compatible within their uncertainties. The $3\ell$-channel events represent the dominant
contribution to the combined result, and the individual $3\ell$ result can be seen to differ only slightly from that using the combined selections. The total systematic uncertainties
in the $4\ell$ channel are smaller than those in the $3\ell$ channel, but the overall precision is poorer in the $4\ell$ channel due to the limited number of data events.
 
\begin{table}[!htb]
\small
\caption[]{Measured \muttZ parameters obtained from the fits in the different lepton channels. The uncertainties include statistical and systematic sources.
The uncertainty of the theoretical prediction of the \ttZ cross section~(see Section~\ref{sec:samples}) is not considered for the \muttZ values.}
\label{tab:mu_inclusive}
\def\arraystretch{1.5}
\begin{center}
\begin{tabular}{l | l}
Channel &  \hspace{1.3cm} \muttZ \\
\hline
Trilepton   & $1.17 \pm 0.07 \,\mathrm{(stat.)}\,^{+0.12}_{-0.11}\,\mathrm{(syst.)}$ \\
Tetralepton & $1.21 \pm 0.15 \,\mathrm{(stat.)}\,^{+0.11}_{-0.10}\,\mathrm{(syst.)}$ \\
Combination ($3\ell+4\ell$)    & $1.19 \pm 0.06 \,\mathrm{(stat.)}\,\pm 0.10 \,\mathrm{(syst.)}$ \\
\end{tabular}
\end{center}
\end{table}
 
The measured \muttZ value and its uncertainty based on the fit results from the combined trilepton and tetralepton channels are converted to a cross-section measurement.
The value corresponds to the phase-space region where the invariant mass of the decay products of the \Zboson boson lies between \num{70} and \SI{110}{\GeV}.
The cross section is measured to be
\begin{equation*}
\sigma(\pp \rightarrow \ttZ) = 0.99 \pm 0.05 \,\mathrm{(stat.)} \pm 0.08 \,\mathrm{(syst.)} \,\mathrm{pb}.
\end{equation*}
The result agrees with the SM prediction of $0.84^{+0.09}_{-0.10}\,$pb at NLO QCD and EW accuracy~\cite{deFlorian:2016spz,Frixione:2015zaa}
and more recent calculations including NNLL corrections or the complete set~(QCD and EW) of NLO corrections~\cite{Broggio:2017kzi,Kulesza:2019}.
 
The contributions from the relevant uncertainties of the measured cross section are summarised in Table~\ref{tab:syst_inclusive}. For this table, the uncertainties are
grouped into several type-related categories and are shown together with the total uncertainty. As none of the uncertainties show significant asymmetries, they are symmetrised.
The dominant uncertainty sources can be attributed to the \ttZ parton shower, the modelling of the \tWZ background, and jet flavour-tagging. It should be noted that the uncertainty
in the cross section due to the systematic uncertainty on the luminosity is larger than the 1.7\% mentioned in Section~\ref{subsec:sys-det}, as the luminosity affects both signal and
background normalisation.
 
\begin{table}[!htb]
\small
\caption[]{List of relative uncertainties of the measured inclusive \ttZ cross section from the combined fit. The uncertainties are symmetrised for presentation and grouped into
the categories described in the text. The quadrature sum of the individual uncertainties is not equal to the total uncertainty due to correlations introduced by the fit.}
\label{tab:syst_inclusive}
\def\arraystretch{1.1}
\begin{center}
\begin{tabular}{l | c}
Uncertainty & $\Delta\sigma_{\ttZ}/\sigma_{\ttZ}$~[\%] \\
\hline
\ttZ parton shower         & 3.1 \\
\tWZ modelling             & 2.9 \\
$b$-tagging                & 2.9 \\
\VVjets modelling          & 2.8 \\
\tZq modelling             & 2.6 \\
Lepton                     & 2.3 \\
Luminosity                 & 2.2 \\
Jets + \met                & 2.1 \\
Fake leptons               & 2.1 \\
\ttZ ISR                   & 1.6 \\
\ttZ \muF and \muR scales  & 0.9 \\
Other backgrounds          & 0.7 \\
Pile-up      	           & 0.7 \\
\ttZ PDF                   & 0.2 \\
\hline
Total systematic           & 8.4 \\
Data statistics            & 5.2 \\
\hline
Total                      & 10  \\
\end{tabular}
\end{center}
\end{table}
 
\FloatBarrier
 
\section{Differential cross-section measurements}
\label{sec:differential}
 
\subsection{Description of the observables and reconstruction of the $t\bar{t}$ system}
A set of ten observables were selected for the differential cross-section measurements which probe the kinematics of the \ttZ system. The definitions of these variables are
summarised in Table~\ref{tab:variables}.  With the exception of the number of reconstructed jets~($N_{\mathrm{jets}}$), which is unfolded to particle level only, all distributions
are unfolded to both particle and parton level. Two of the variables, namely the transverse momentum and the absolute value of the rapidity of the \Zboson boson~($\pT^{Z}$ and $|y^{Z}|$),
which are sensitive to \ttZ generator modelling and various BSM effects, are defined identically for the trilepton and tetralepton selections. The differential measurements for these
variables are therefore performed using an inclusive selection denoted by $3\ell + 4\ell$.
 
\begin{table}[!h]
\footnotesize
\caption[]{Summary of the variables used for the differential measurements. Some variables are considered for the trilepton or tetralepton signal regions only, as indicated.
The jet multiplicity is measured for the two topologies separately, whereas for the variables related only to the kinematics of the \Zboson boson ($\pT^{Z}$ and $|y^{Z}|$),
the trilepton and tetralepton regions are combined.}
\label{tab:variables}
\def\arraystretch{1.5}
\begin{center}
\resizebox{\textwidth}{!}{
\begin{tabular}{p{0.018\textwidth}|p{0.11\textwidth}|p{0.84\textwidth}}
\cline{2-3}
& \multicolumn{1}{c|}{Variable} 	                       &  Definition  \\ \hline \hline
\multirow{2}{*}{\rotatebox{90}{$3\ell+4\ell$}} & $\pT^{Z}$     &  Transverse momentum of the \Zboson boson  \\
& $|y^{Z}|$                                                    &  Absolute value of the rapidity of the \Zboson boson  \\ \hline
\multirow{4}{*}{\rotatebox{90}{$3\ell$}} & $N_{\mathrm{jets}}$ &  Number of selected jets with $\pt > \SI{25}{\GeV}$ and $\abseta < 2.5$      \\
& $\pT^{\ell,{\textrm{non-}}Z}$	                               &  Transverse momentum of the lepton which is not associated with the \Zboson boson  \\
& $|\Delta\phi(Z,t_{\mathrm{lep}})|$	                       &  Azimuthal separation between the \Zboson boson and the top quark (antiquark) featuring the \Wln decay \\
& $|\Delta y(Z,t_{\mathrm{lep}})|$	                       &  Absolute rapidity difference between the \Zboson boson and the top quark (antiquark) featuring the \Wln decay  \\ \hline
\multirow{4}{*}{\rotatebox{90}{$4\ell$}} & $N_{\mathrm{jets}}$ &  Number of selected jets with $\pt > \SI{25}{\GeV}$ and $\abseta < 2.5$      \\
& $|\Delta\phi(\ell^{+}_{t}, \ell^{-}_{\bar{t}})|$             &  Azimuthal separation between the two leptons from the \ttbar system \\
& $|\Delta\phi(\ttbar,Z)|$	                               &  Azimuthal separation between the \Zboson boson and the \ttbar system \\
& $\pT^{\ttbar}$				               &  Transverse momentum of the \ttbar system \\
\hline
\hline
\end{tabular}
}
\end{center}
\end{table}
 
The jet multiplicity is a natural variable to use to probe the modelling of QCD radiation and hadronisation in MC generators.  It is measured separately
for the trilepton and tetralepton selections due to the different number of final-state quarks from the decay of the \ttbar system in the two channels.
The transverse momentum of the lepton which is not associated with the \Zboson boson~($\pT^{\ell,{\textrm{non-}}Z}$) in the trilepton signal regions
provides a good test of the modelling of the \pt of the top quark (antiquark) and its decay products in the MC generator.
 
The absolute azimuthal separation and rapidity difference between the \Zboson boson and the leptonic top quark~($|\Delta\phi(Z,t_{\mathrm{lep}})|$ and $|\Delta y(Z,t_{\mathrm{lep}})|$)
in the trilepton signal regions, as well as the absolute azimuthal separation between the \Zboson boson and the \ttbar system~($|\Delta\phi(\ttbar,Z)|$) in the tetralepton regions,
provide direct probes of the \ttZ vertex. These variables therefore offer sensitivity to a number of BSM effects which could modify the coupling between the \Zboson boson and the top quark.
 
The absolute azimuthal separation between the two leptons associated with the top quarks~($|\Delta\phi(\ell^{+}_{t},\ell^{-}_{\bar{t}})|$) in tetralepton events provides
sensitivity to BSM effects modifying the spin correlations between the two top quarks. The transverse momentum of the \ttbar system~($\pt^{\ttbar}$) is sensitive to the MC
modelling of the hard-scattering process as well as the modelling of the QCD radiation in the parton shower.
 
In order to construct the $|\Delta\phi(Z,t_{\mathrm{lep}})|$ and $|\Delta y(Z,t_{\mathrm{lep}})|$ variables in the trilepton regions, the full four-vector of the leptonic top
quark from the \ttbar system~($t_{\mathrm{lep}}$) is required.\footnote{The leptonic top quark is defined as the top quark or antiquark which decays via \tWb, \Wlnu.} For both
detector- and particle-level quantities the reconstructed \met~(both its magnitude and azimuthal angle), is first attributed to the neutrino from the associated \Wboson boson decay.
The SM value of the \Wboson boson mass~\cite{PDG} is then used to determine the $z$-component of the neutrino momentum by analytically solving the corresponding quadratic equation.
In many cases the solution is ambiguous. For those, both real solutions are considered. For cases in which the discriminant of the quadratic equation is
negative, the \pt of the neutrino is set to the particular value which yields a single solution.  In order to form the final top-quark candidate, the reconstructed leptonically
decaying \Wboson boson candidate --- or candidates in the case of two neutrino solutions --- is added, via a four-vector sum, to the closer~(in \DeltaR) of the two reconstructed
jets in the event with the highest output from the $b$-tagging algorithm~(MV2c10). At particle level, the two jets which are ghost-matched to a $b$-hadron~(as described in
Section~\ref{subsec:pl_definitions}) are considered. In the case of only a single such ghost-matched jet, that jet is selected to form the top-quark candidate. Events with
two distinct neutrino solutions will have two possible top-quark candidates, so the one with an invariant mass of the \Wboson--$b$ system more consistent with a top-quark decay is chosen.
 
In the tetralepton channel the \ttbar system is reconstructed in the transverse plane only. The underlying assumption is that the two neutrinos from the \ttbar decay
represent the dominant source of missing transverse momentum in the event; the value of the reconstructed \met can, therefore, be taken to be a reasonable proxy for the vector sum
of the  neutrino momenta in the transverse plane. Such a partial reconstruction avoids the need to determine the full kinematics of both neutrinos separately, while still allowing
the reconstruction of the $\pt^{\ttbar}$ and $|\Delta\phi(\ttbar,Z)|$ variables for the differential measurements. The selection of the two $b$-tagged jets is performed analogously
to the trilepton case. At detector level, the two reconstructed jets with the highest $b$-tagging score are selected. At particle level, the two jets ghost-matched to a $b$-hadron
are selected; in the case of only one ghost-matched jet, the jet with the highest \pt of those remaining is selected as the second $b$-jet.
 
\subsection{Unfolding procedure}
To measure the differential cross-section distributions at particle and parton levels in the specific fiducial phase-spaces defined in Section~\ref{subsec:pl_definitions},
an iterative Bayesian unfolding procedure is used~\cite{DAgostini:1995}. It relies on the Bayesian probability formula starting from a given prior of the particle- or parton-level
distribution and iteratively updating it with the posterior distribution. The unfolding is performed using the \textsc{RooUnfold} package~\cite{Adye:2011gm}. The differential \ttZ
cross sections are calculated using the following equation:
\begin{equation*}
\dfrac{\mathrm{d}\sigma_{\ttZ}}{\mathrm{d}X^i} = \dfrac{1}{\mathcal{L}\cdot\mathcal{B}\cdot\Delta X^i\cdot \epsilon_\mathrm{\,eff}^i} \cdot \sum_j[\mathcal{M}^{-1}]_{ij} \cdot f_\mathrm{acc}^j\cdot\left(N_\mathrm{obs}^j-N_\mathrm{bkg}^j\right),
\end{equation*}
where $X$ denotes the variable used for the differential measurement (with the bin-width $\Delta X$), the index $i$ indicates the bin at particle (or parton) level and $j$ the
detector-level bin.
 
The migration matrix $\mathcal{M}$ quantifies the detector response and can be derived from the bin-to-bin migrations of events from particle or parton level to detector level
in the nominal \ttZ simulation for each of the considered differential variables. Its inverse, $\mathcal{M}^{-1}$, is determined through the iterative unfolding procedure.
For each $j$, $N_{\mathrm{obs}}^{j}$ denotes the number of observed data events, and $N_{\mathrm{bkg}}^{j}$ is the expected background contribution. The various background contributions
are estimated in the same way as for the inclusive measurement~(see Section~\ref{sec:background}). In this case, the $WZ/ZZ + l$ backgrounds are corrected by normalisation parameters
obtained from an inclusive fit based on the combined $3\ell + 4\ell$ channels. A statistics-only version of the fit was performed solely for the extraction of the normalisation parameters
in this case. The acceptance corrections, $f_{\mathrm{acc}}^{j}$, account for events that are generated outside the fiducial phase-space but pass the detector-level selection, whereas the
efficiency correction terms, $\epsilon_{\,\mathrm{eff}}^{i}$, correct for events that are in the fiducial phase-space but are not reconstructed in the detector. In either case, the term
`fiducial' refers to  the corresponding type of unfolding being performed --- either to parton or particle level. The integrated luminosity is denoted by $\mathcal{L}$. The branching ratio
$\mathcal{B}$ is that of the \ttZ system to final states with three or four charged electrons or muons, originating directly from either the \Zboson boson decay or the decay of the \Wboson
bosons from the \ttbar system, and is used to extrapolate the measurements to cover all \ttbar and \Zboson decays. The branching ratio correction is only applied for the parton-level
measurements  and corresponds to the decay channels applicable for the fiducial region based on the particular variable involved~(see Table~\ref{tab:variables}). The values for
$\mathcal{B}$, calculated using inputs taken from Ref.~\cite{PDG}, are \num{0.0193}~$(3\ell)$, \num{0.0030}~$(4\ell)$, and \num{0.0223}~$(3\ell + 4\ell)$.
 
\begin{figure}[h!]
\subfloat[]{\includegraphics[width=0.45\textwidth]{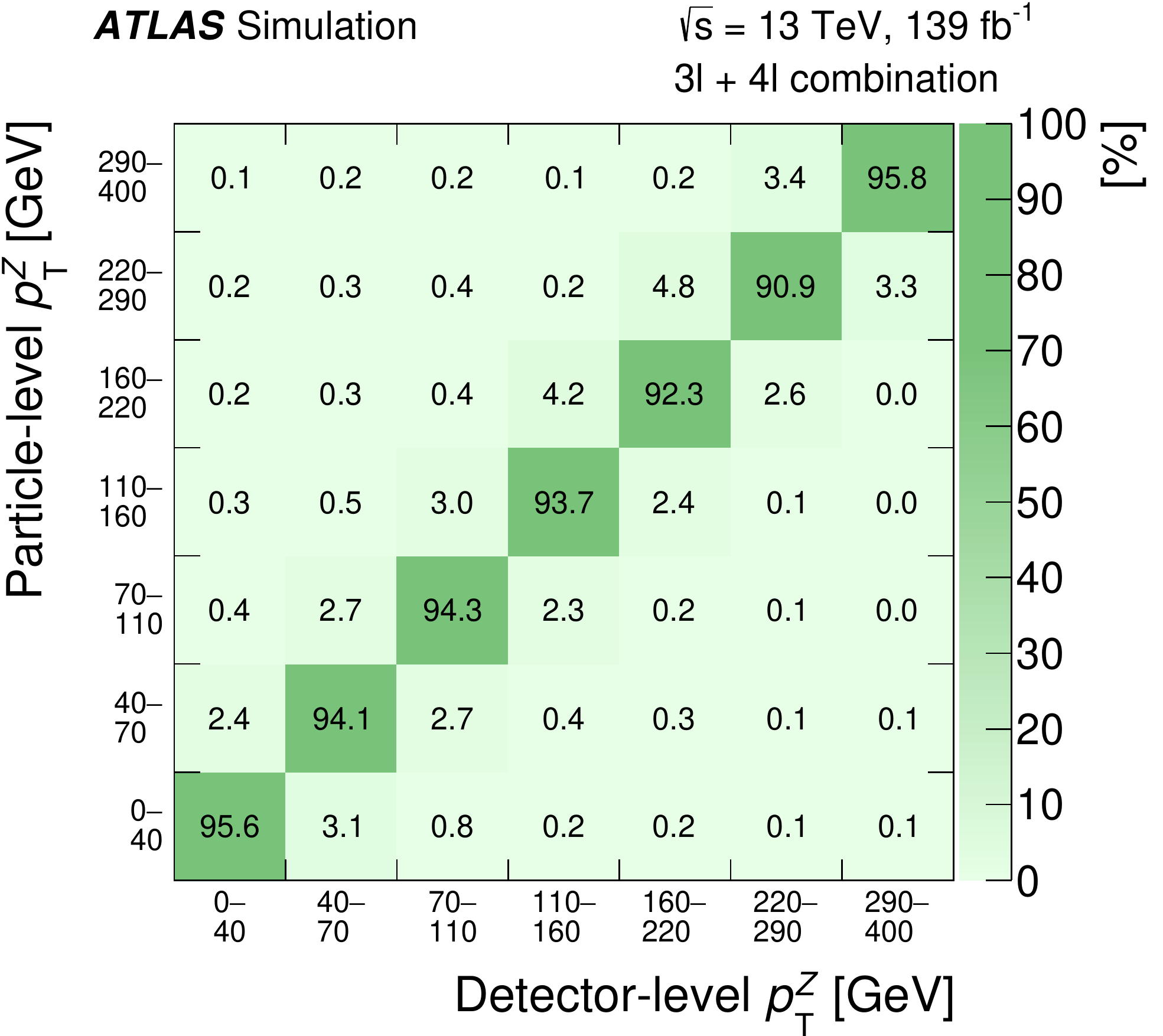}}
\hfill
\subfloat[]{\includegraphics[width=0.45\textwidth]{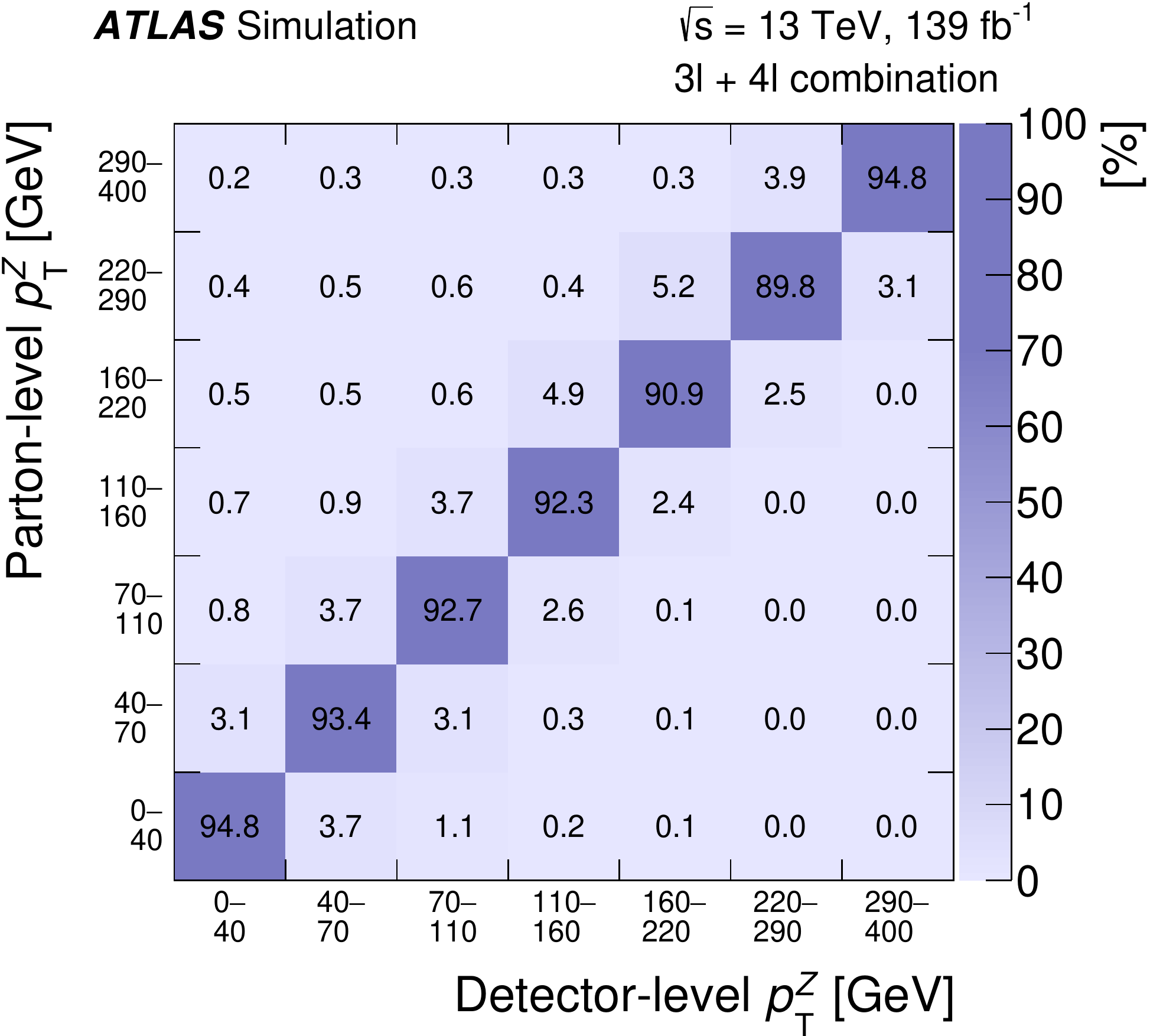}}
\caption{The migration matrices for the \pt of the \Zboson boson in the combination of the trilepton and tetralepton regions. The matrices quantify the migrations from (a)~particle or
(b)~parton level to detector level. The quoted values are expressed as a percentage. The matrices are normalised such that the sum of any given row is 100\%, although small differences
may be present due to rounding.}
\label{fig:mm-example}
\end{figure}

\begin{figure}[h!]
\subfloat[]{\includegraphics[width=0.49\textwidth]{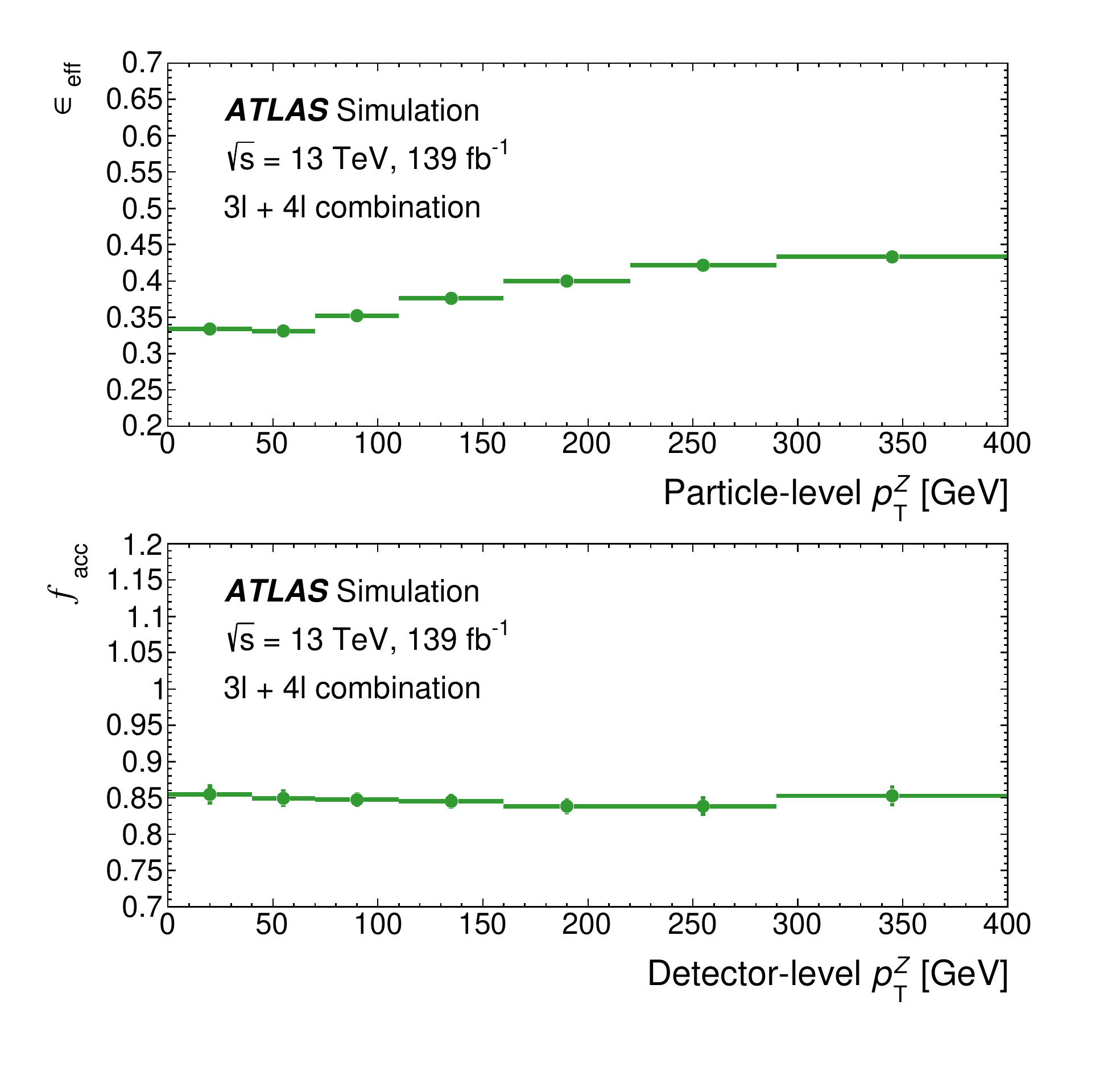}}
\hfill
\subfloat[]{\includegraphics[width=0.49\textwidth]{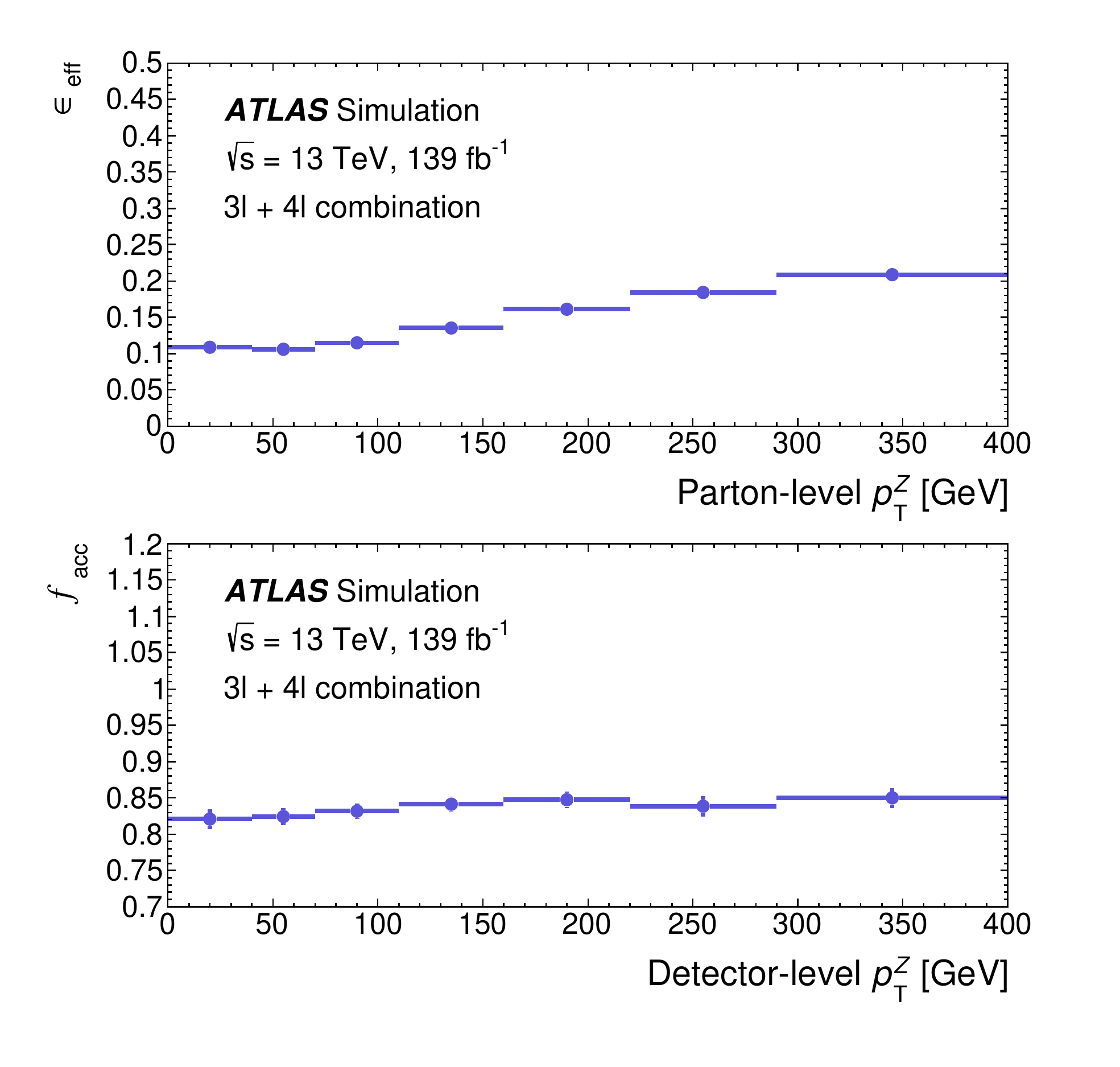}}
\caption{The efficiency ($\epsilon_{\,\mathrm{eff}}$) and acceptance ($f_{\mathrm{acc}}$) correction factors for the (a)~particle-level and (b)~parton-level measurements
as a function of the \pt of the \Zboson boson in the combination of the trilepton and tetralepton regions. The error bars represent the MC statistical uncertainties per bin
based on the nominal signal sample.}
\label{fig:effacc-example}
\end{figure}

Figure~\ref{fig:mm-example} shows the particle- and parton-level migration matrices that are used for the differential cross-section measurements depending on the \pt of
the \Zboson boson. The matrices are normalised such that the sum of entries in each row is equal to one. The entries in the matrices represent the fraction of events at
either particle or parton level in a $y$-axis bin that are reconstructed at detector level in an $x$-axis bin. Thus, the fraction of events in the diagonal elements shows
the quality of the resolution for a specific variable. In the case of $\pt^{Z}$, these fractions lie between 90\% and 96\% for both particle and parton level. For some of the
other variables which do not depend only on the \Zboson boson reconstruction~(e.g.~$N_{\mathrm{jets}}$, $\pt^{\ttbar}$), the migrations between bins can be significantly larger
and reach a level of 20--25\%. Figure~\ref{fig:effacc-example} depicts the corresponding correction factors as a function of $\pt^{Z}$: $\epsilon_{\,\mathrm{eff}}$ increases
for larger $\pt^{Z}$ due to higher lepton reconstruction efficiencies for increasing transverse momenta. It lies between 33\% and 43\% at particle level and between 10\% and 22\%
at parton level. The values of $f_{\mathrm{acc}}$ are in all bins higher than 80\% for both particle and parton level and show no notable dependence on $\pt^{Z}$.
 
The choice of binning is determined separately for each variable by performing a multi-dimensional scan in order to strike a reasonable balance between three partially competing aspects:
retaining a large number of bins; limiting the relative impact from statistical uncertainties of the measured data; and ensuring large values~($> \SI{50}{\%}$) for the diagonal elements of
the matrices associated with the bin migrations between particle/parton and detector level. As a result, the binning for the differential measurements differs from that shown in
Figures~\ref{fig:sr-3l}--\ref{fig:sr-comb}. The stability of the unfolding procedure is determined by constructing pseudo-data sets by randomly sampling events from
the nominal \ttZ MC sample, such that the pseudo-data sets contain approximately the same number of events as in the measured data. So-called `pull tests' are performed as part of
the binning optimisation to verify that the unfolding is stable for the selected number and range of bins. In addition, linear re-weightings are applied to the pseudo-data to test
the ability of the unfolding procedure to correct the pseudo-data back to their underlying true spectra, obtained from the MC event record. The number of iterations used in the
iterative Bayesian unfolding is also optimised with pseudo-experiments: for each iteration, a $\chi^{2}$ per degree-of-freedom~(ndf) is calculated by comparing the bin contents
of the unfolded pseudo-data with those from the previous iteration. In the case of the first iteration, the unfolded pseudo-data are instead compared with the corresponding
generator-level distribution. Iterations are performed until the $\chi^{2}/$ndf value of a given distribution stabilises at a constant value while the statistical uncertainty returned
from the unfolding procedure is kept as low as possible. For all variables, the number of iterations used lies between two and five. Systematic uncertainties are propagated to the unfolded
distributions by varying the detector-level distributions within the uncertainties and repeating the unfolding procedure.
 
The normalised differential cross sections are obtained by dividing the distributions by the integrated fiducial cross sections, which are computed by adding up the contributions
from all bins. The evaluation of systematic uncertainties is performed after the normalisation is done and it is on the same prescriptions employed for the absolute differential measurements.
 
\FloatBarrier
 
\subsection{Results of the differential measurements}
\begin{figure}[htb!]
\centering
\subfloat[]{\includegraphics[width=0.49\textwidth]{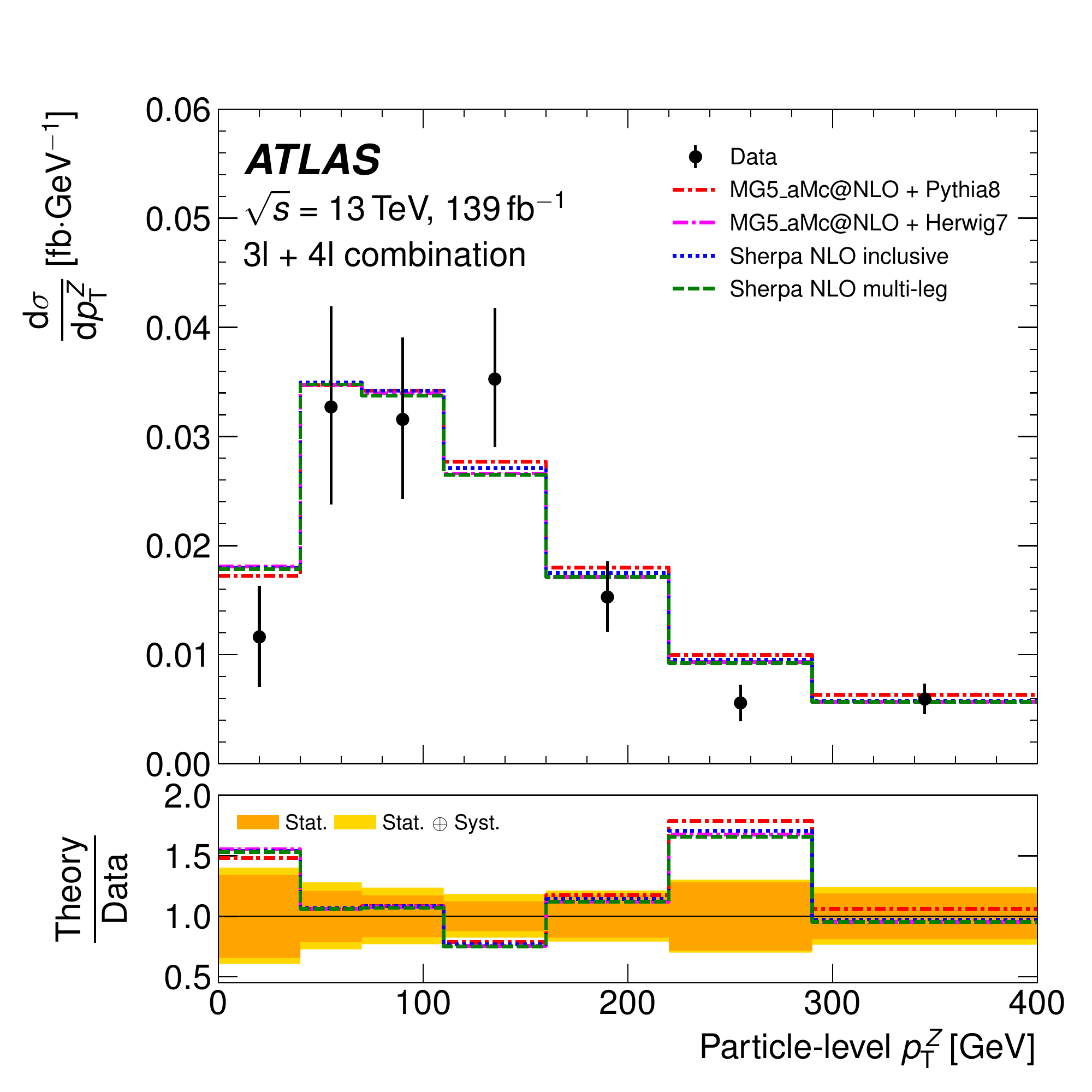} \label{fig:x-sec-results-1a}}
\subfloat[]{\includegraphics[width=0.49\textwidth]{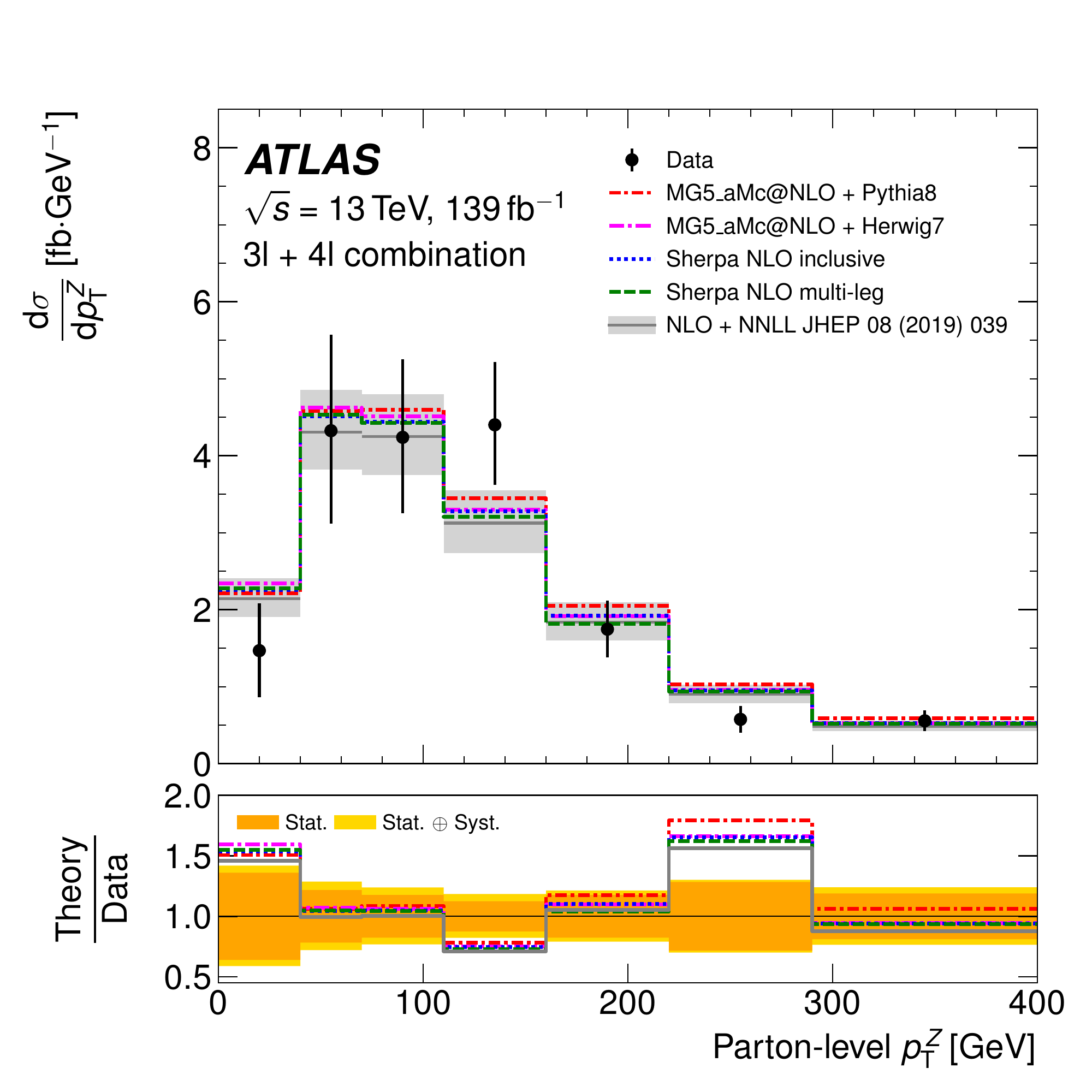}  \label{fig:x-sec-results-1b}}
 
\subfloat[]{\includegraphics[width=0.49\textwidth]{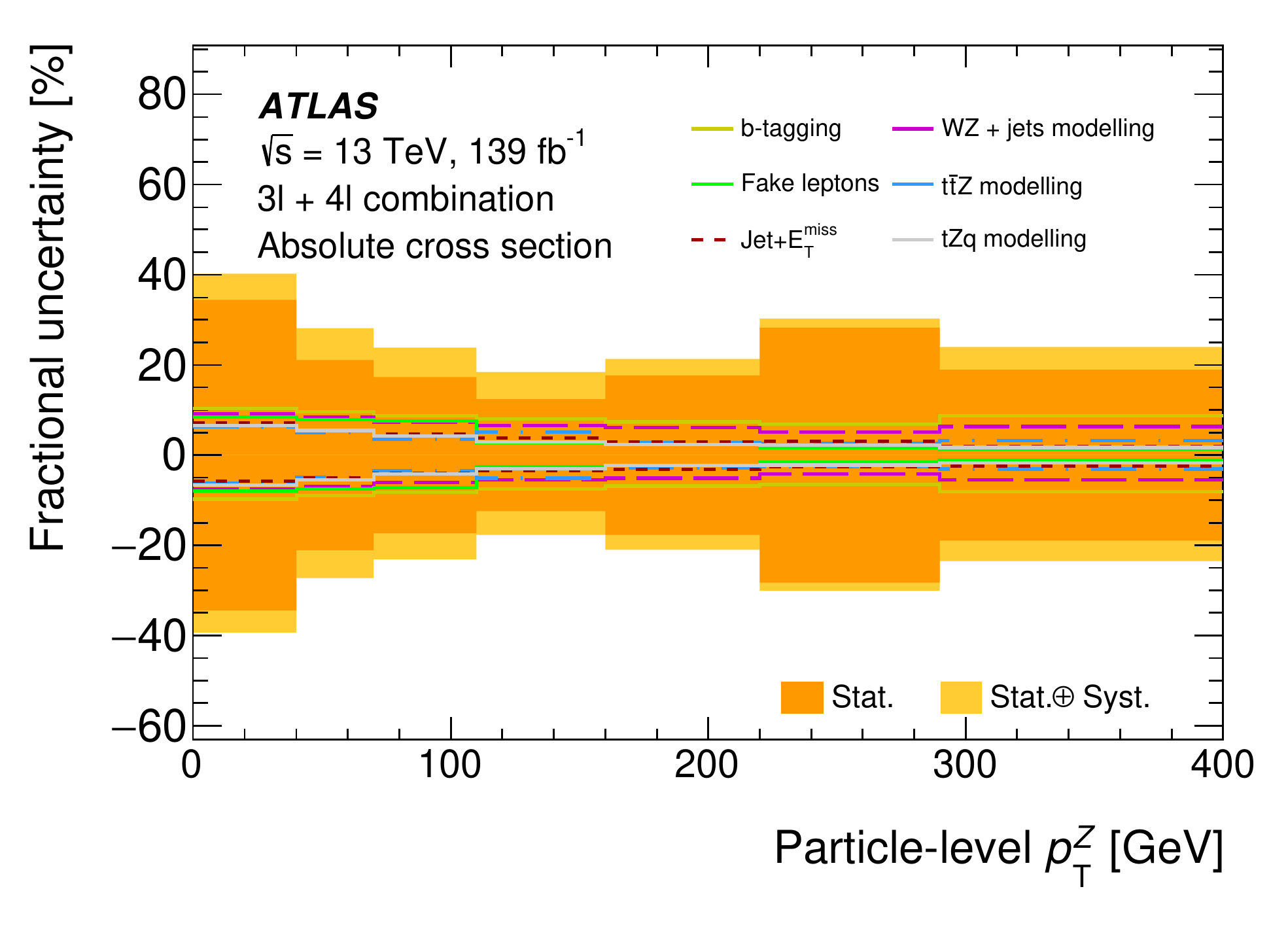}}
\subfloat[]{\includegraphics[width=0.49\textwidth]{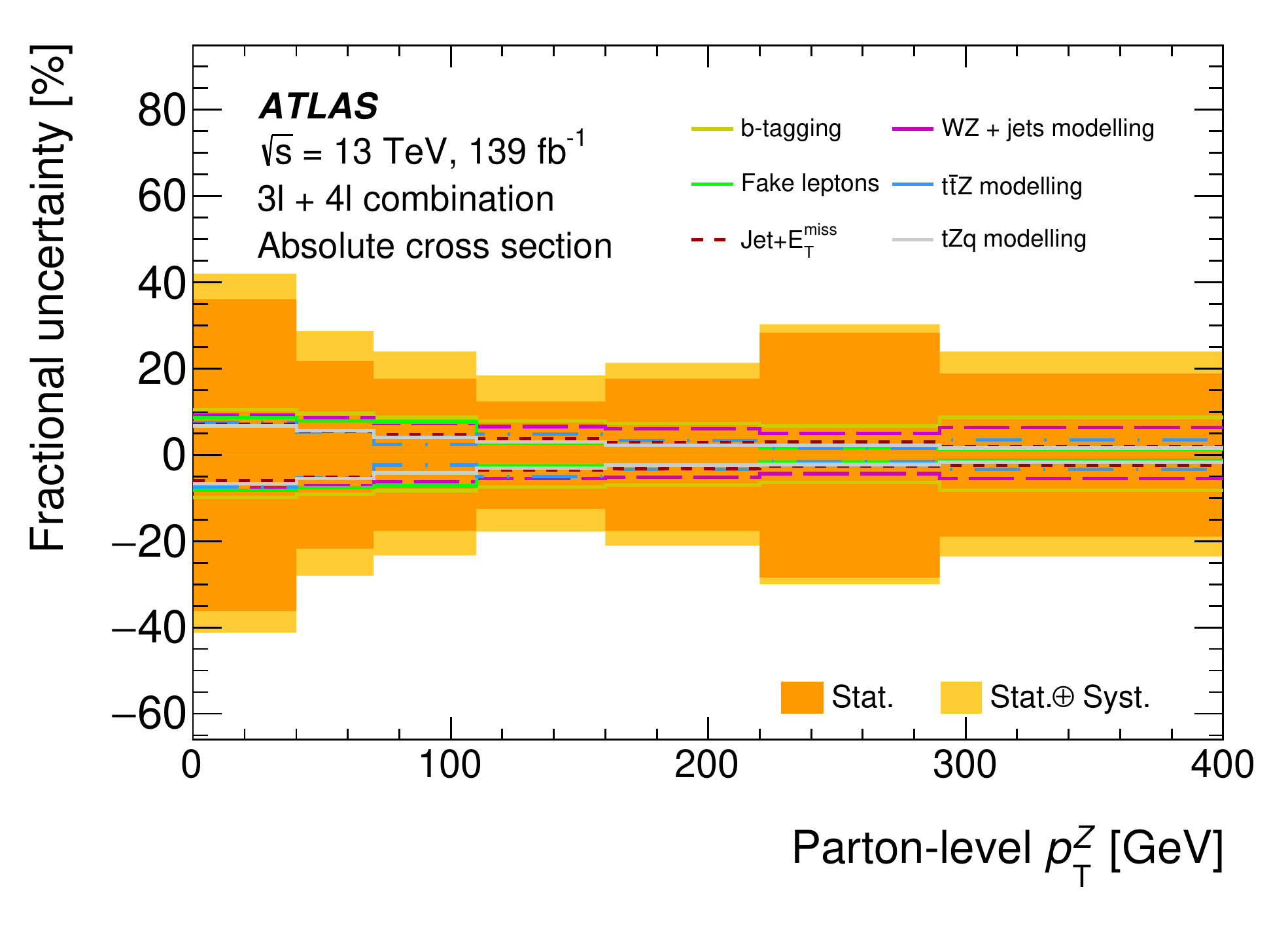}}
 
\caption{Absolute differential \ttZ cross sections measured at (a)~particle level and (b)~parton level as a function of the transverse momentum of the reconstructed \Zboson boson.
(c,d)~show the relative contributions from different categories of systematic uncertainties per bin. The large difference between the $y$-axis scales in (a) and (b) is a
result of different efficiency and acceptance corrections between the particle- and parton-level measurements, together with the branching ratio correction
of $\mathcal{B} = 0.0223$ for the combined $3\ell + 4\ell$ channels, which is applied only for the parton-level result.}
\label{fig:x-sec-results-1}
\end{figure}

The measured differential \ttZ cross sections unfolded to particle and parton levels for the \pt of the reconstructed \Zboson boson are presented in Figure~\ref{fig:x-sec-results-1}.
 
\begin{figure}[htb!]
\centering
\subfloat[]{\includegraphics[width=0.49\textwidth]{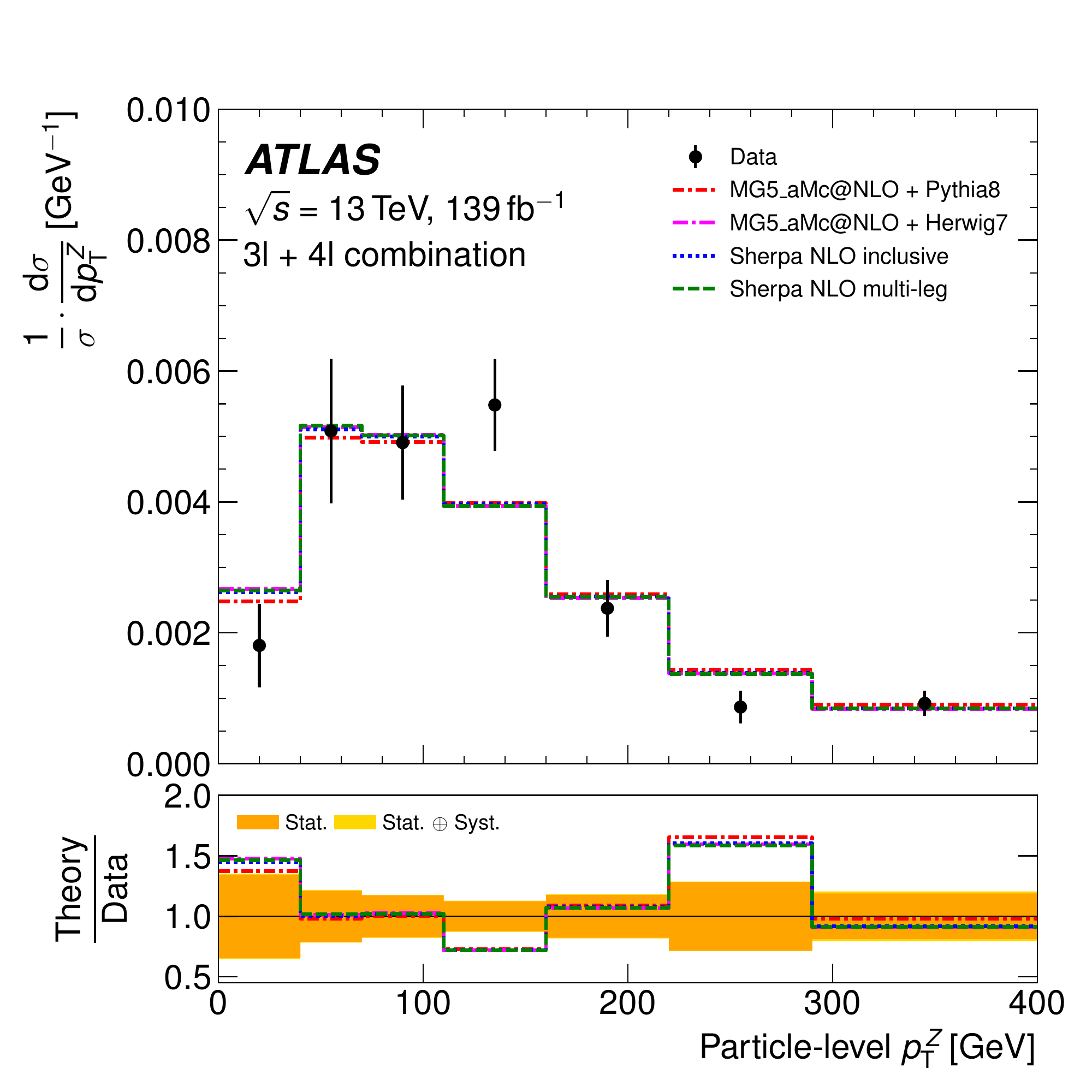} \label{fig:x-sec-results-2a}}
\subfloat[]{\includegraphics[width=0.49\textwidth]{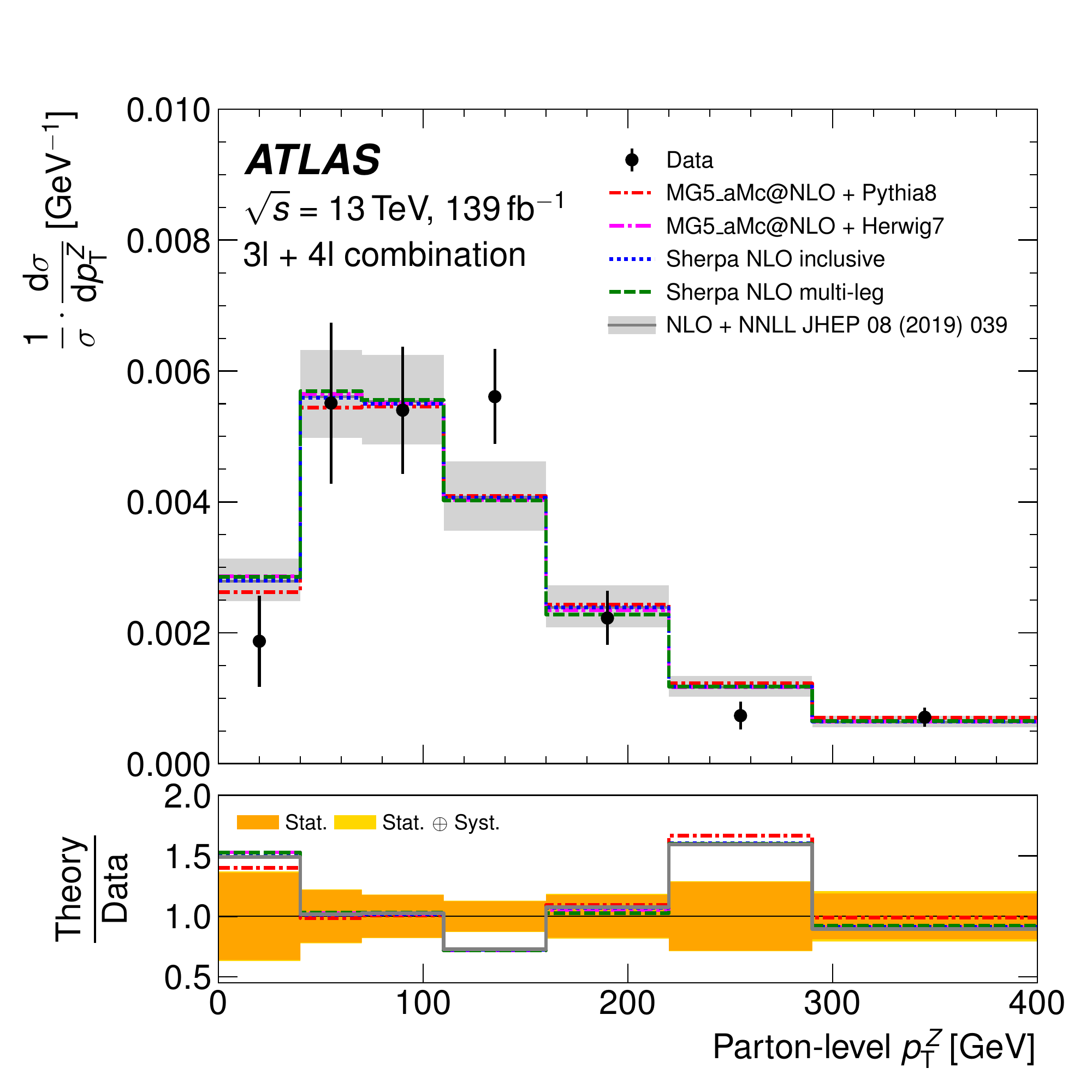} \label{fig:x-sec-results-2b}}
 
\subfloat[]{\includegraphics[width=0.49\textwidth]{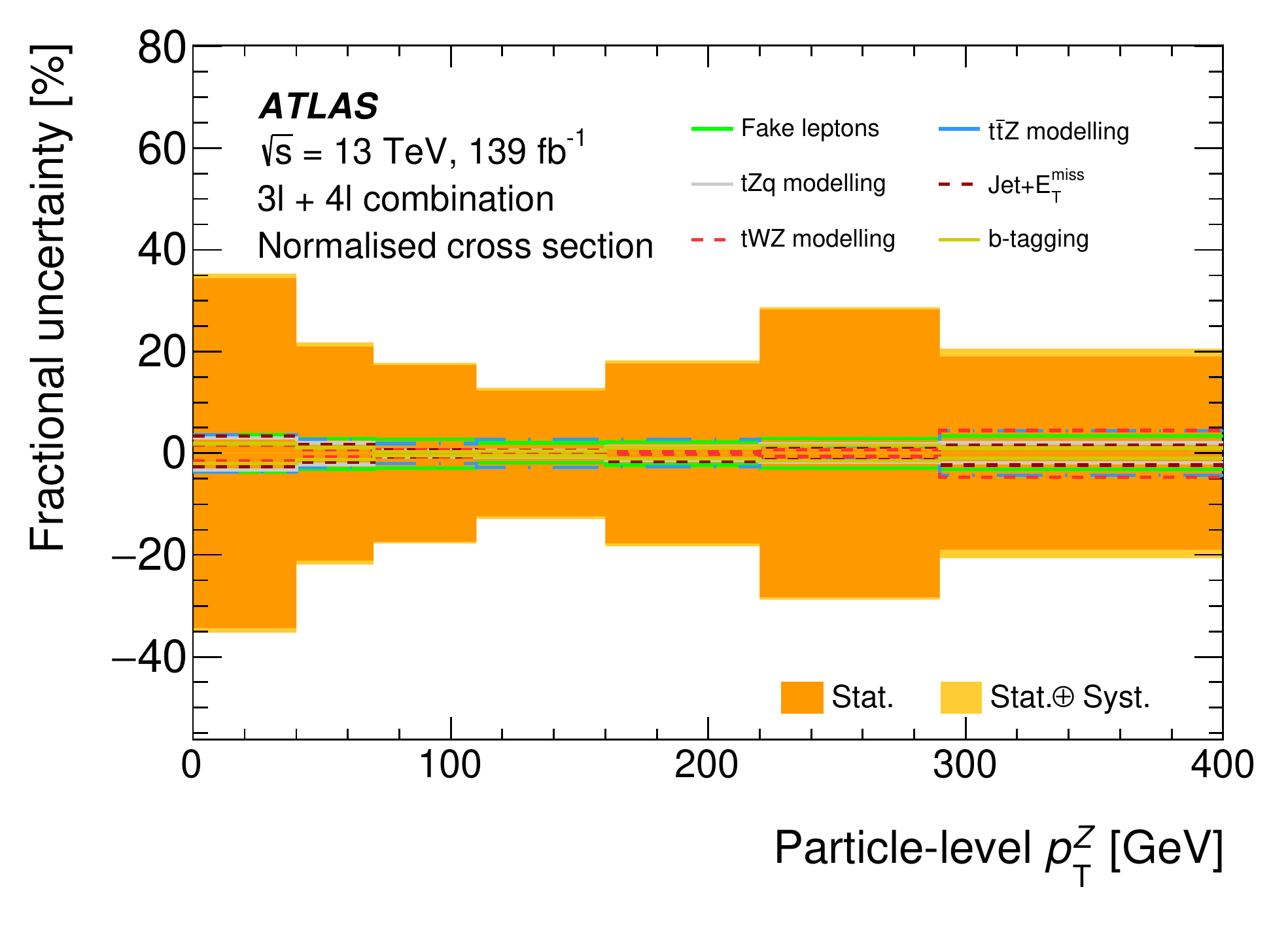}}
\subfloat[]{\includegraphics[width=0.49\textwidth]{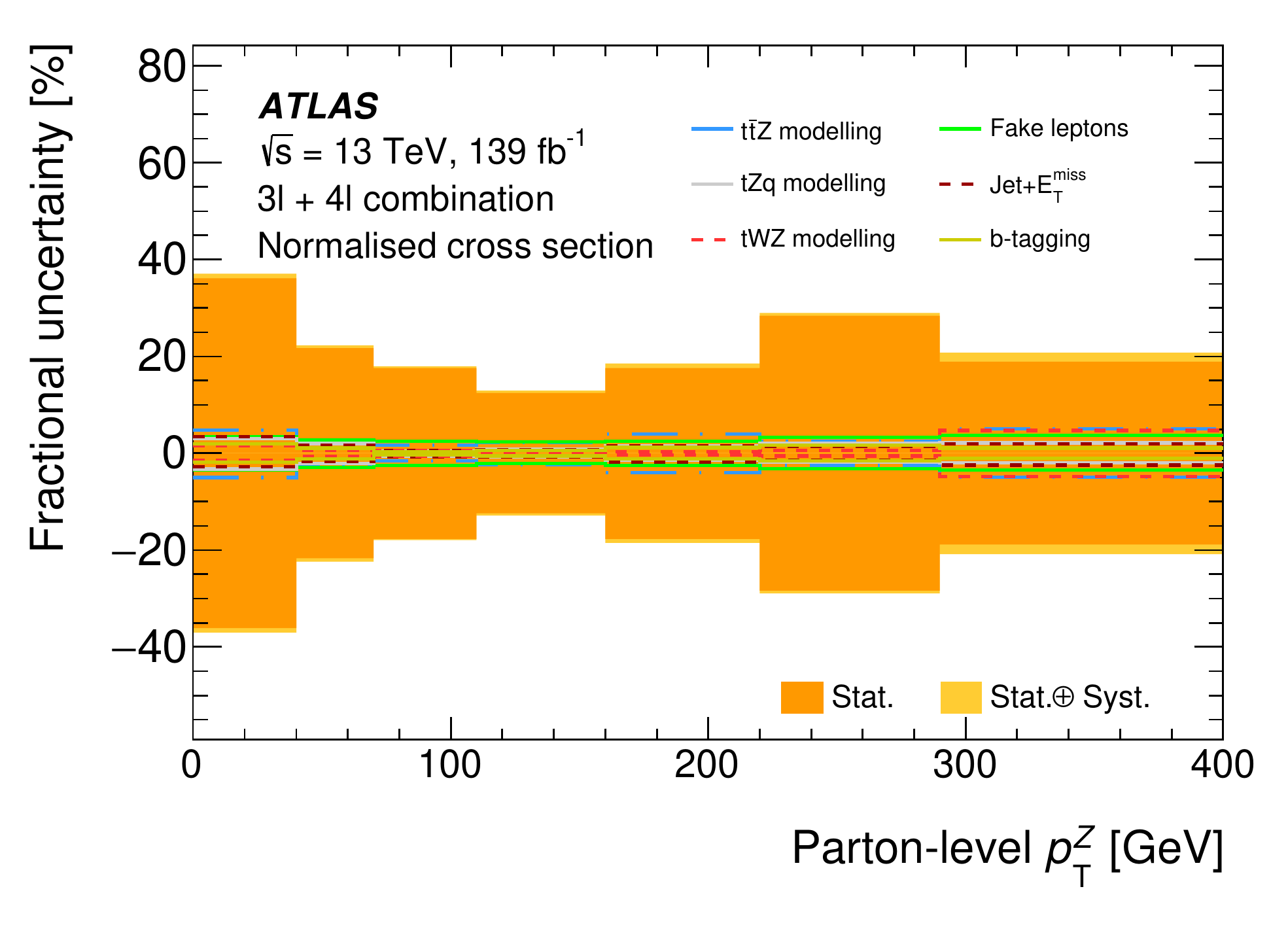}}
 
\caption{Normalised differential \ttZ cross sections measured at (a)~particle level and (b)~parton level as a function of the transverse momentum of the reconstructed \Zboson
boson. (c,d)~show the relative contributions from different categories of systematic uncertainties per bin.}
\label{fig:x-sec-results-2}
\end{figure}
 
The results are displayed in the seven \pt bin ranges used when performing the unfolding, with any additional contributions beyond \SI{400}{\GeV} included in the uppermost bin.
The relative contributions from statistical and systematic uncertainties in each bin are shown in the theory-to-data ratio panels of the upper figures, where the net effect corresponds
to a sum in quadrature of the two. In the lower figures, the same relative contributions are shown as well as a decomposition of the systematic uncertainties into various
categories.\footnote{Unlike the presentation of the uncertainties shown in Table~\ref{tab:syst_inclusive}, the category `\ttZ modelling' contains the sum in quadrature of all individual
modelling uncertainties associated with the \ttZ signal: parton shower, \muF, \muR scales, PDF and ISR.} The black data points in the upper figures correspond to the measured
unfolded data and error bars representing the sum in quadrature of statistical and systematic uncertainties. The total uncertainty of this measurement is between 20\% and 40\%, depending
on the bin, with the dominant uncertainty arising from the limited number of data events. Other significant sources of systematic uncertainty are associated with \ttZ modelling and
$b$-tagging. Figure~\ref{fig:x-sec-results-2} shows the same set of results for the normalised distributions for this variable. The uncertainties on the normalised cross sections are
notably smaller~(15--35\%) than those of the absolute cross sections because several systematic uncertainties cancel out. The differential cross sections measured in data are compared
with the NLO QCD predictions from different \ttZ generators, as described in Section~\ref{sec:samples}. The predictions are shown for \MGNLOs interfaced to \PYTHIA[8]~(red) or
\HERWIG[7]~(magenta), as well as for \SHERPA[2.2.1] inclusive~(blue) and multi-leg~(green). For the $\pt^{Z}$ measurement, the different generators provide very similar predictions.
 
\begin{figure}[htb!]
\centering
\subfloat[]{\includegraphics[width=0.49\textwidth]{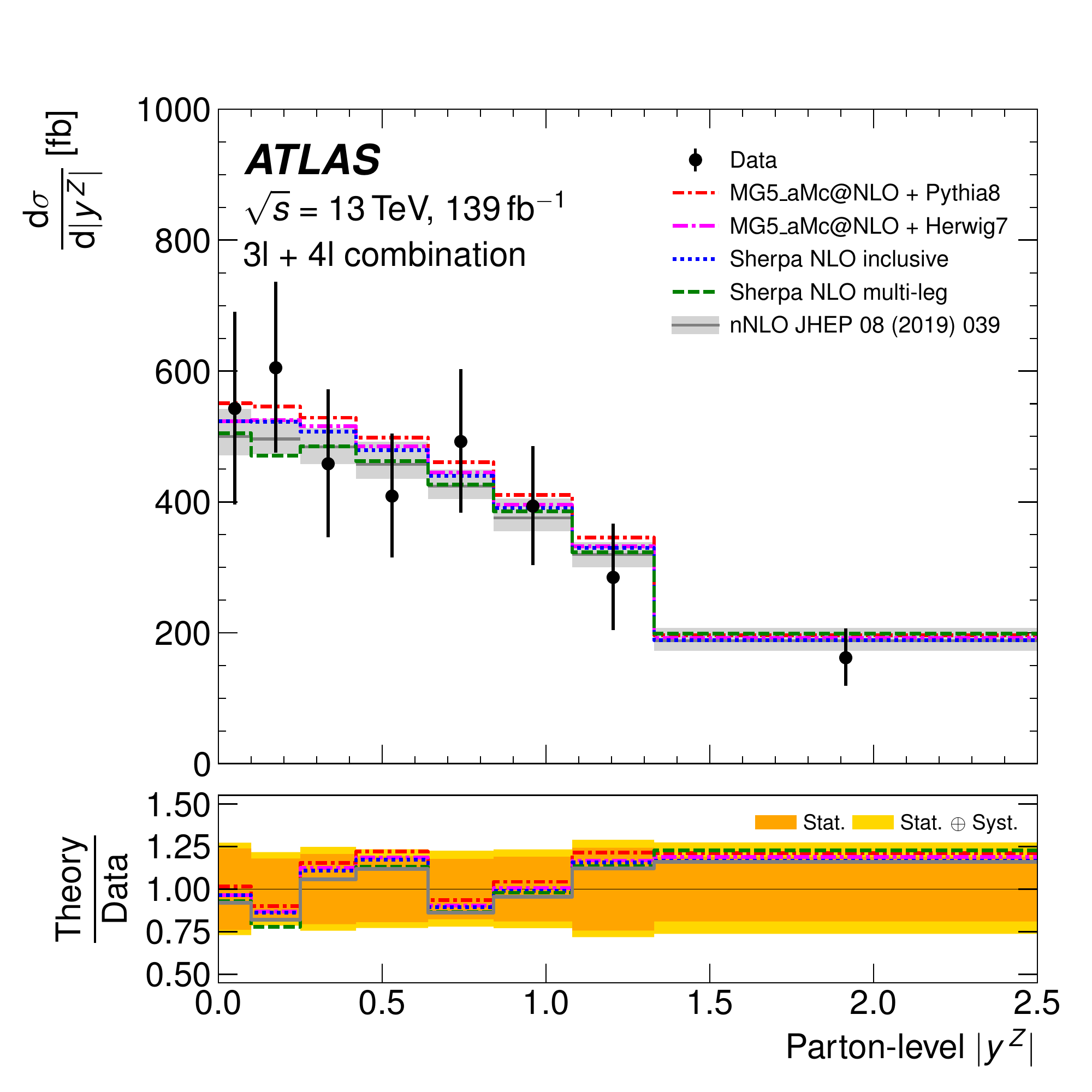} \label{fig:x-sec-results-3a}}
 
\subfloat[]{\includegraphics[width=0.49\textwidth]{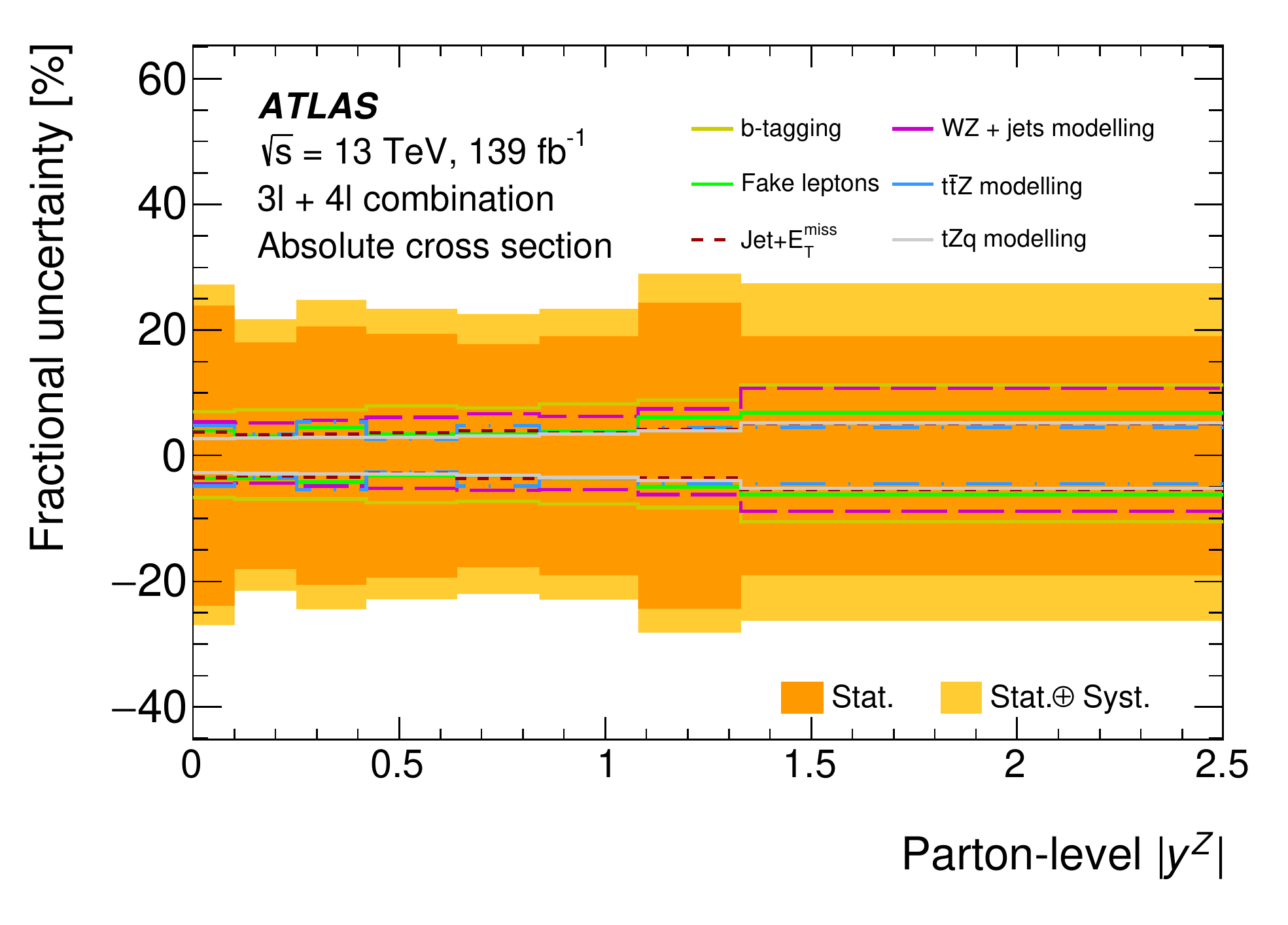}}
 
\caption{(a)~absolute differential \ttZ cross sections and (b)~the relative contributions from different categories of systematic uncertainties per bin measured at parton level
as a function of the absolute rapidity of the \Zboson boson.}
\label{fig:x-sec-results-3}
\end{figure}

\begin{figure}[htb!]
\centering
\subfloat[]{\includegraphics[width=0.49\textwidth]{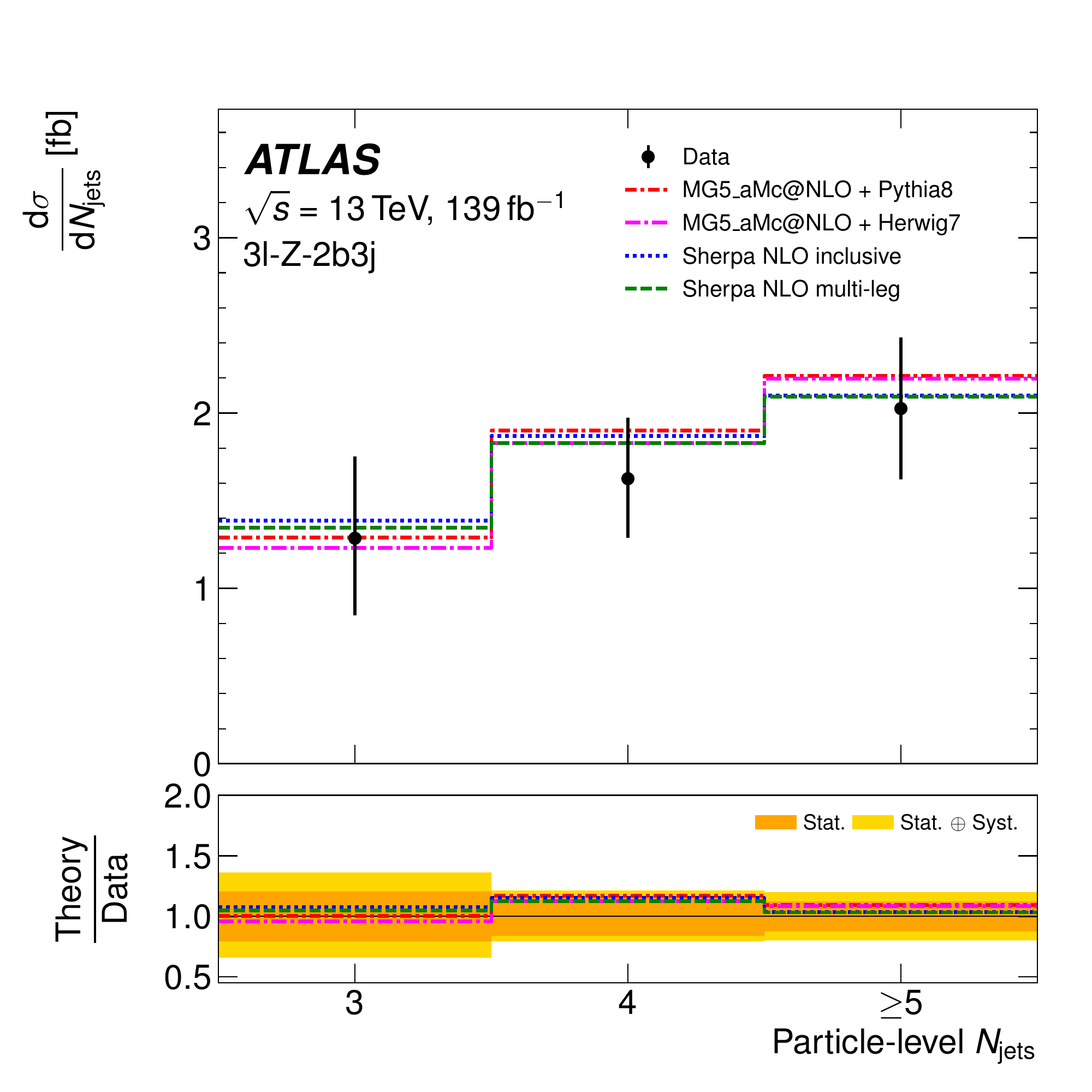} \label{fig:x-sec-results-4a}}
\subfloat[]{\includegraphics[width=0.49\textwidth]{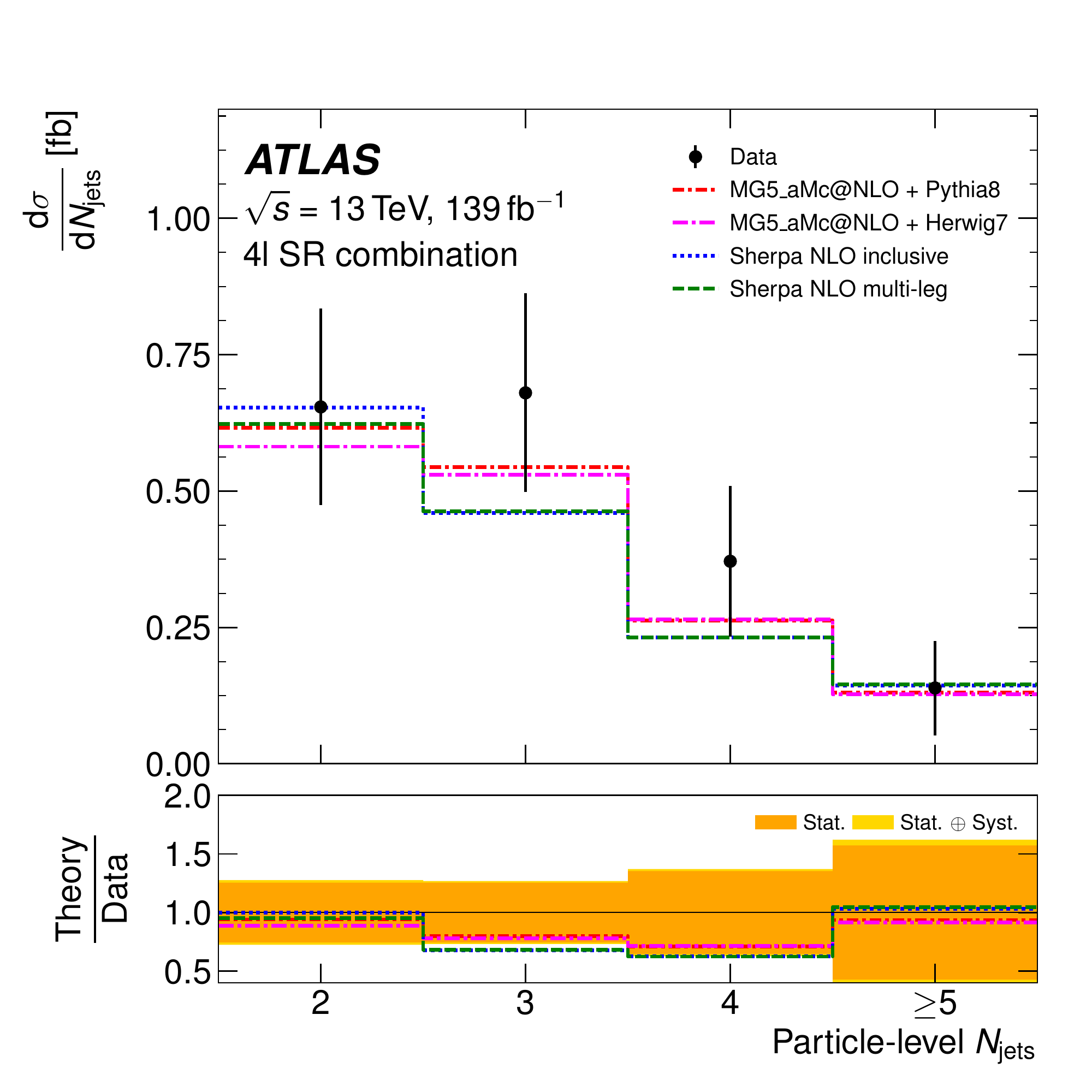} \label{fig:x-sec-results-4b}}
 
\subfloat[]{\includegraphics[width=0.49\textwidth]{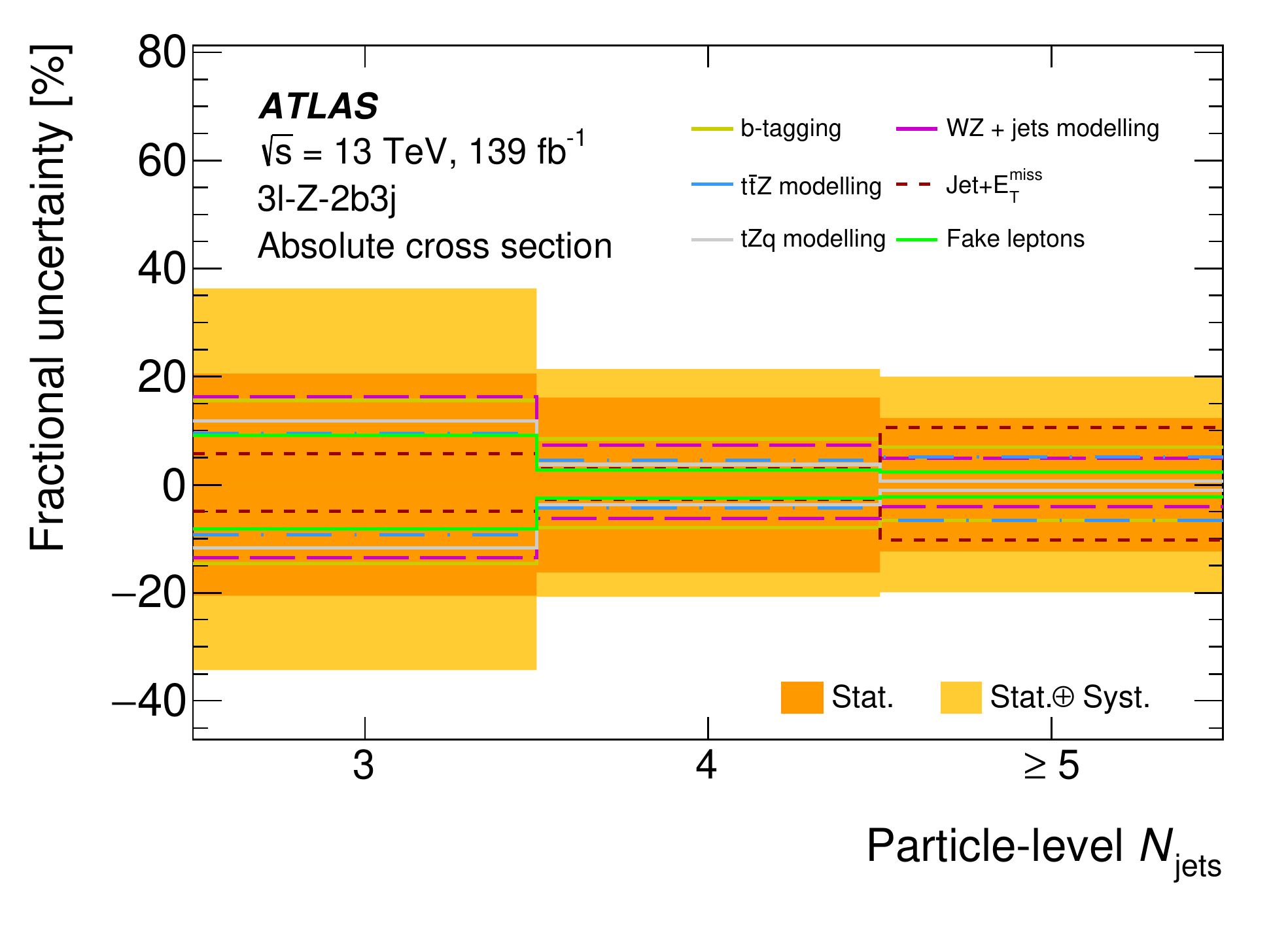}}
\subfloat[]{\includegraphics[width=0.49\textwidth]{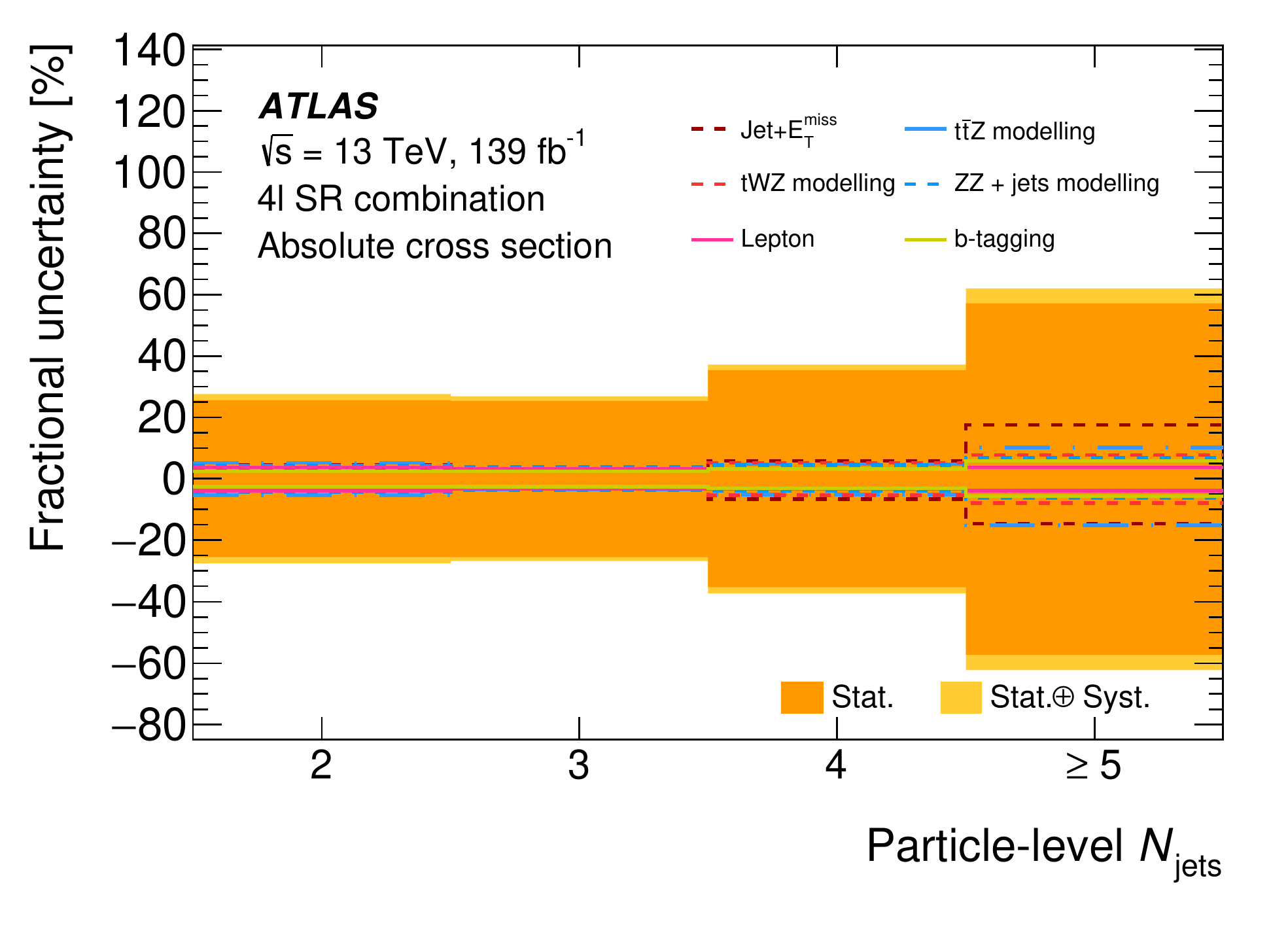}}
 
\caption{Absolute differential \ttZ cross section measured at particle level as a function of $N_{\mathrm{jets}}$ with $\pt > \SI{25}{\GeV}$ for (a)~the trilepton
and (b)~the tetralepton event selection. (c,d)~show the relative contributions from different categories of systematic uncertainties per bin.}
\end{figure}

\begin{figure}[htb!]
\centering
\subfloat[]{\includegraphics[width=0.49\textwidth]{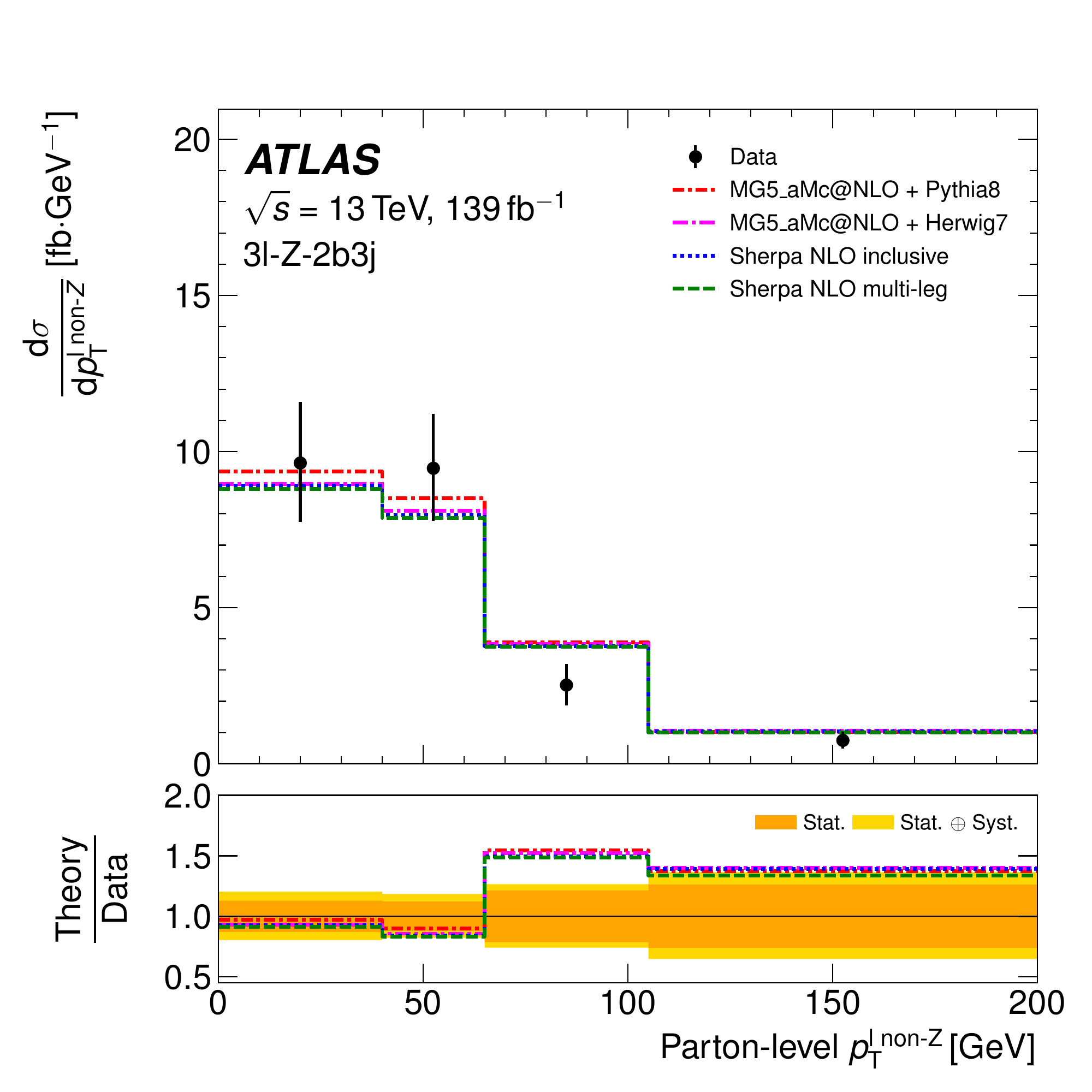} \label{fig:x-sec-results-5a}}
\subfloat[]{\includegraphics[width=0.49\textwidth]{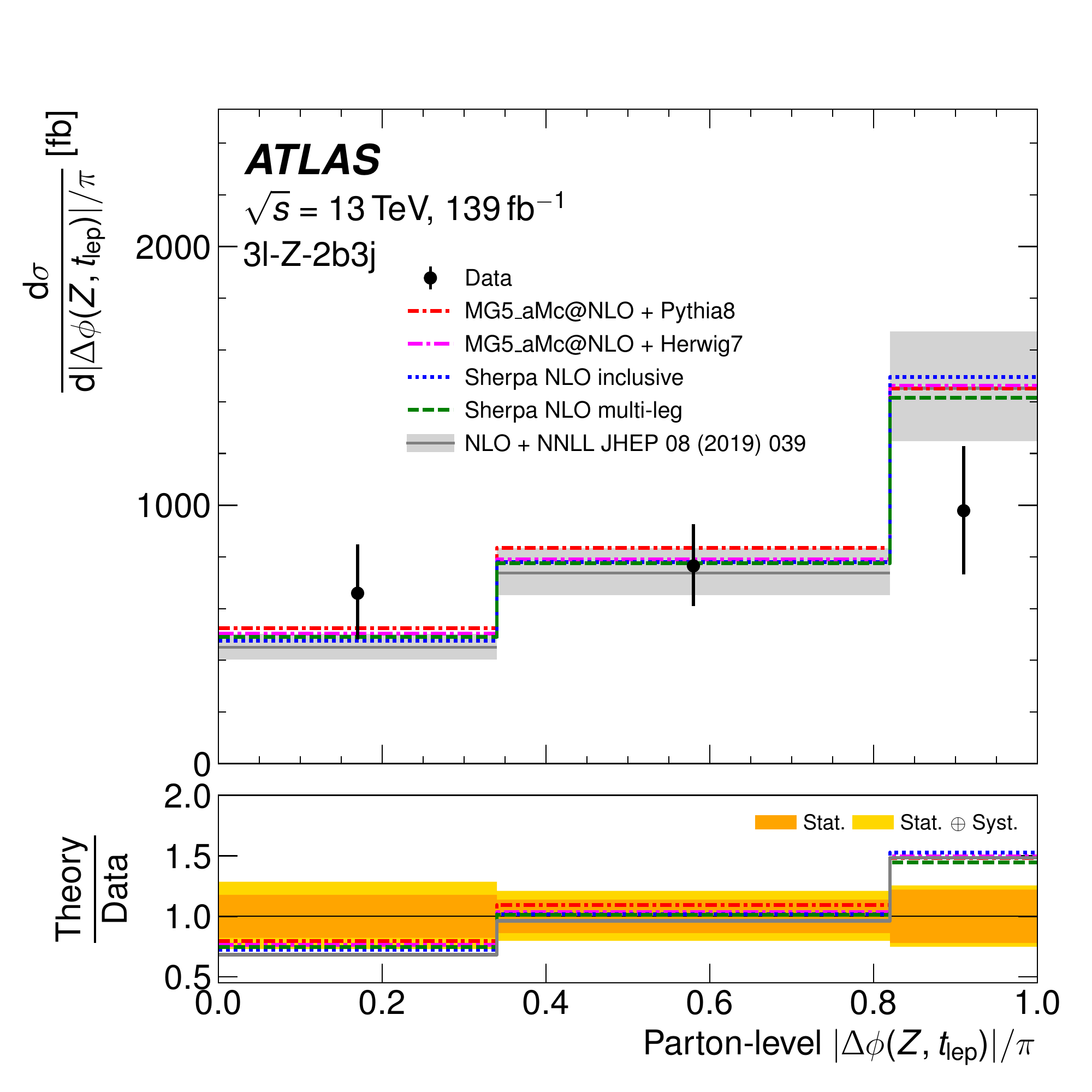} \label{fig:x-sec-results-5b}}
 
\subfloat[]{\includegraphics[width=0.49\textwidth]{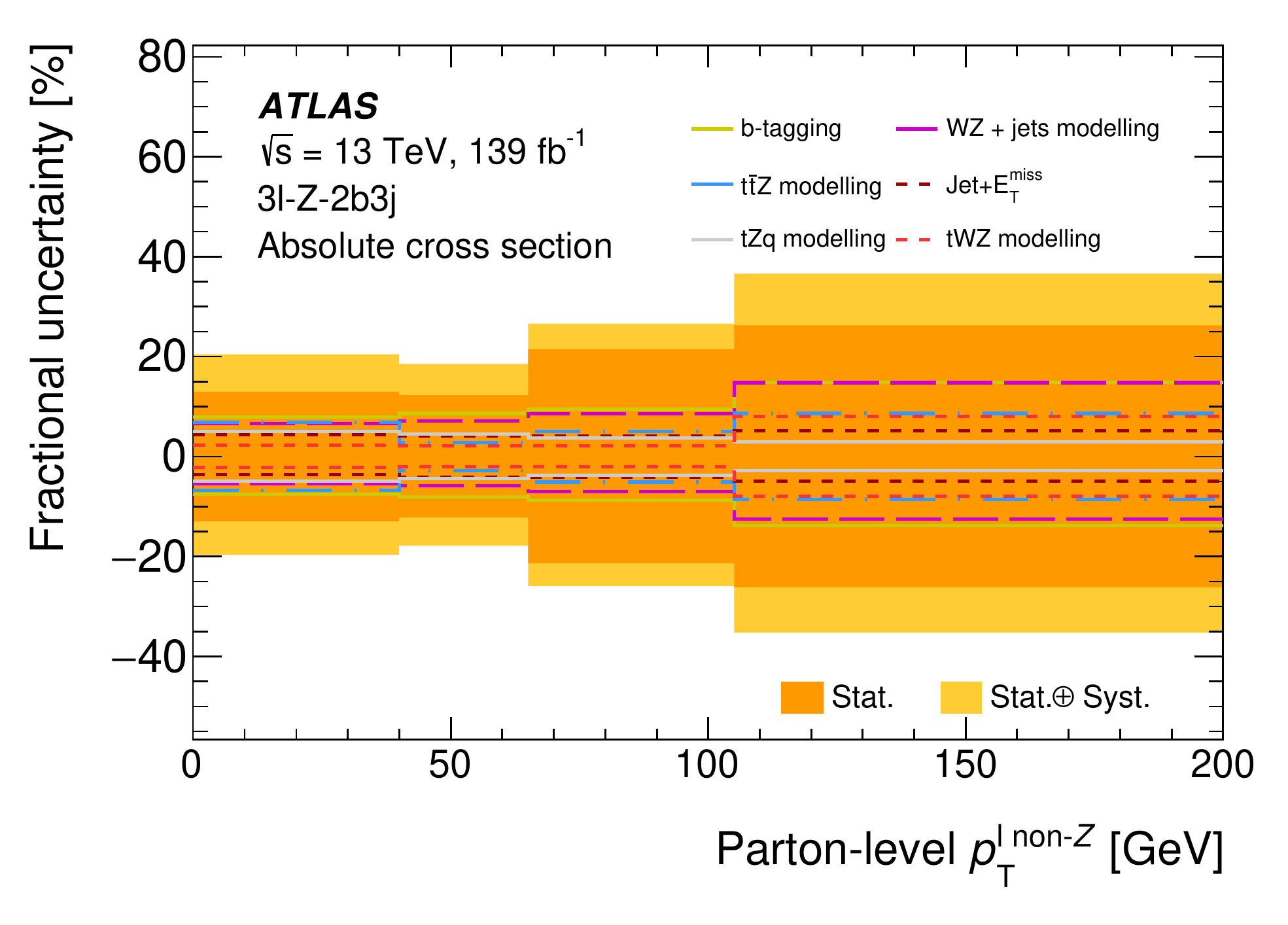}}
\subfloat[]{\includegraphics[width=0.49\textwidth]{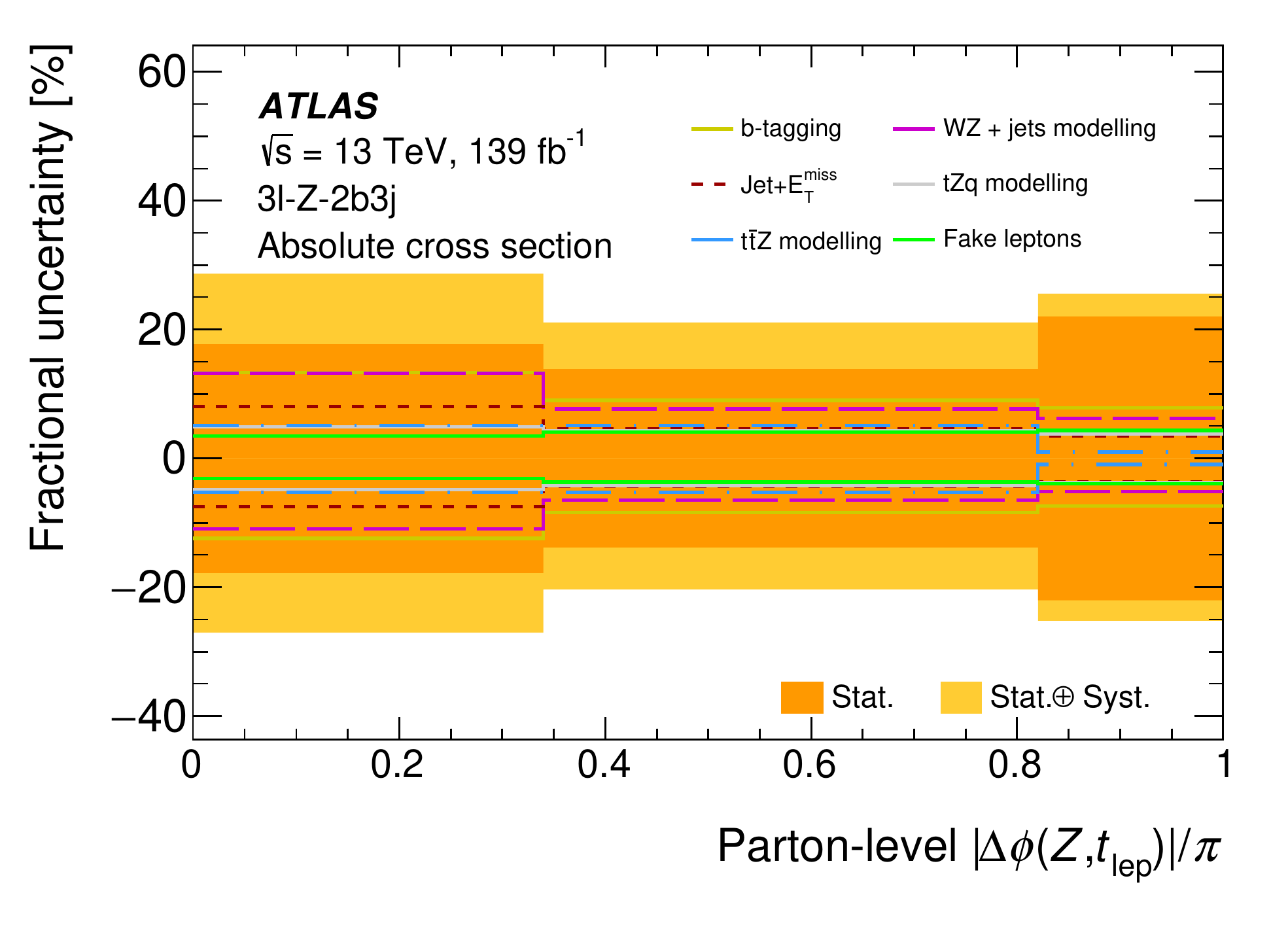}}
 
\caption{Absolute differential \ttZ cross sections measured at parton level as a function of (a)~$\pT^{\ell,{\textrm{non-}}Z}$ and (b)~the azimuthal separation~($\Delta\phi$)
between the \Zboson boson and the reconstructed top quark. (c,d)~show the relative contributions from different categories of systematic uncertainties per bin.}
\end{figure}

\begin{figure}[htb!]
\centering
\subfloat[]{\includegraphics[width=0.49\textwidth]{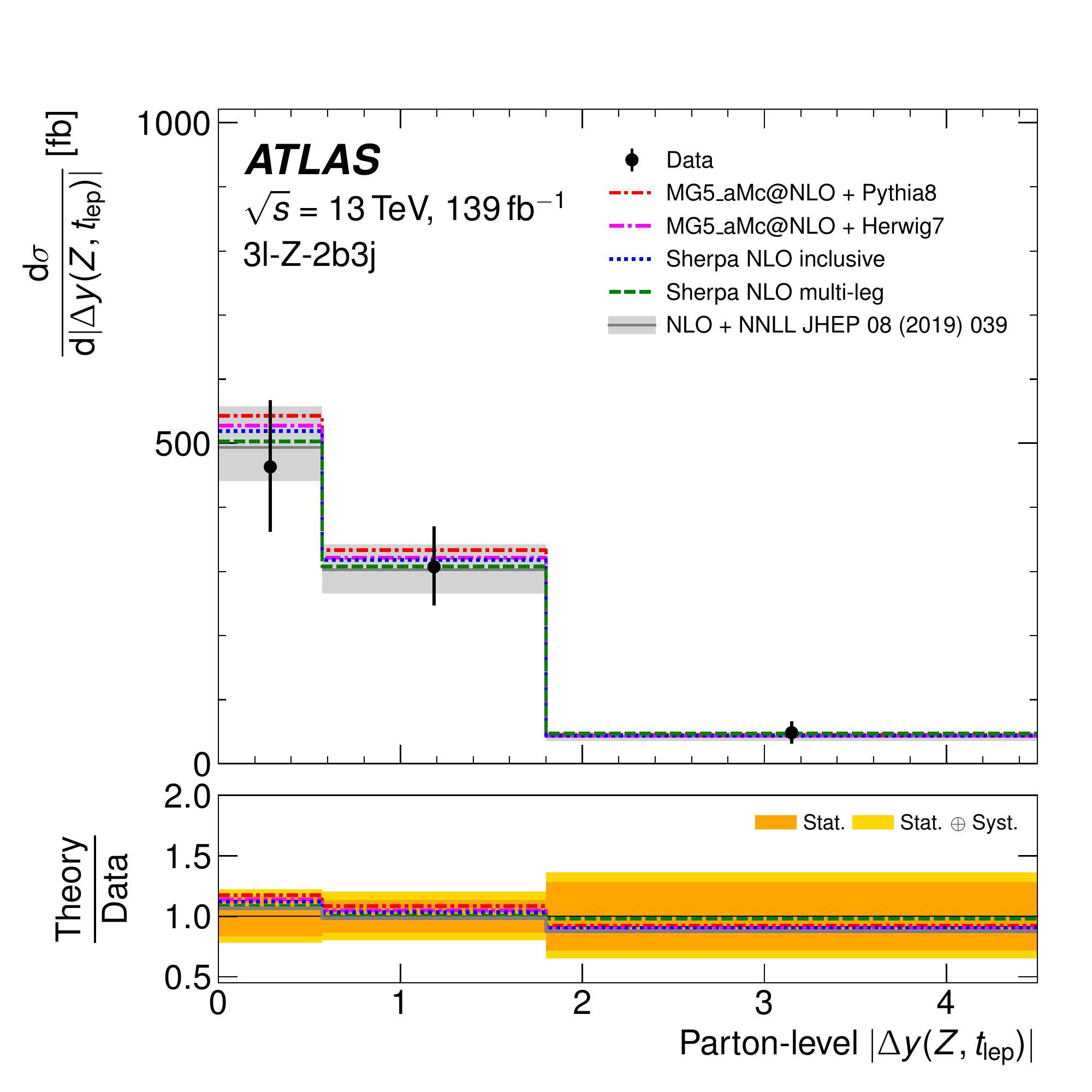} \label{fig:x-sec-results-6a}}
\subfloat[]{\includegraphics[width=0.49\textwidth]{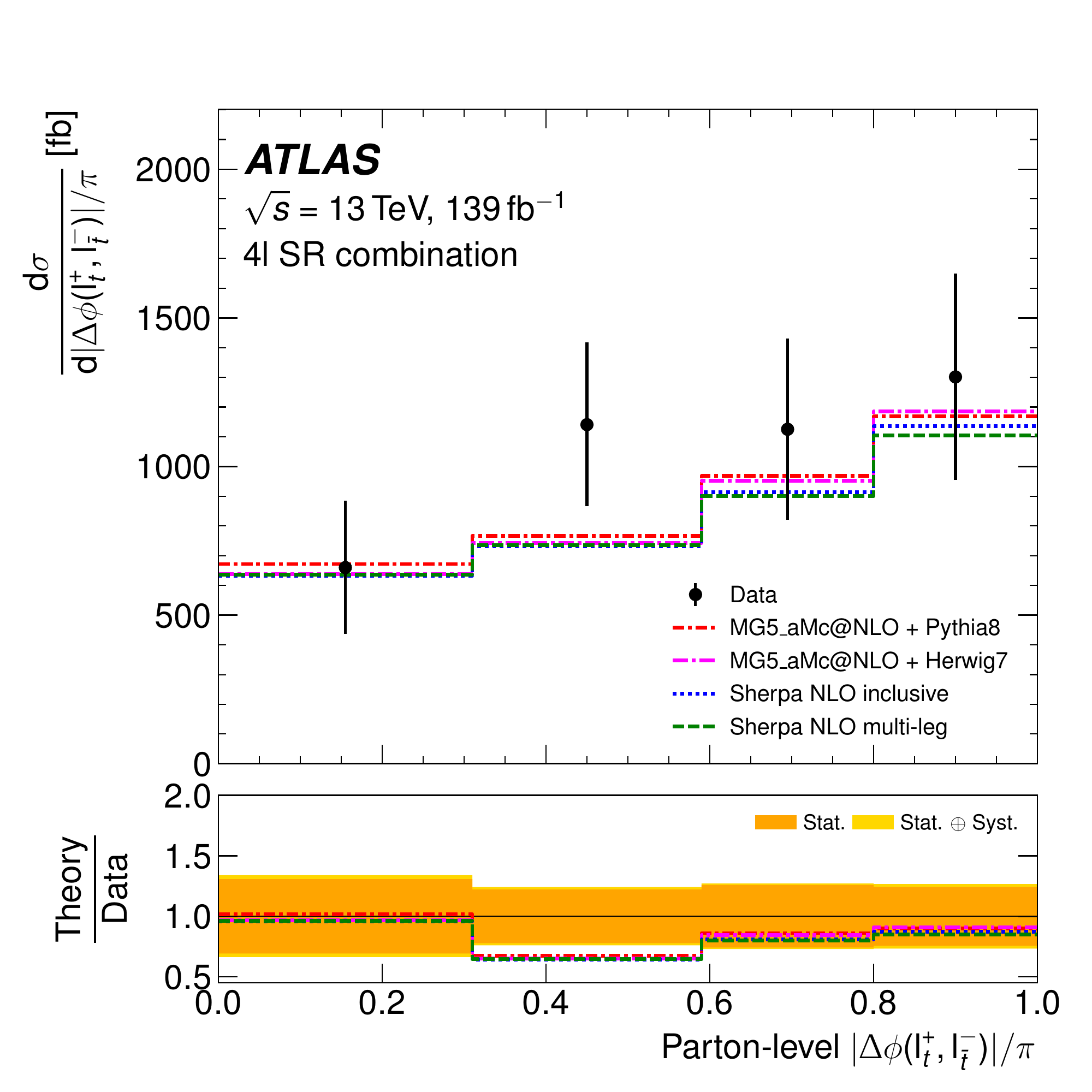} \label{fig:x-sec-results-6b}}
 
\subfloat[]{\includegraphics[width=0.49\textwidth]{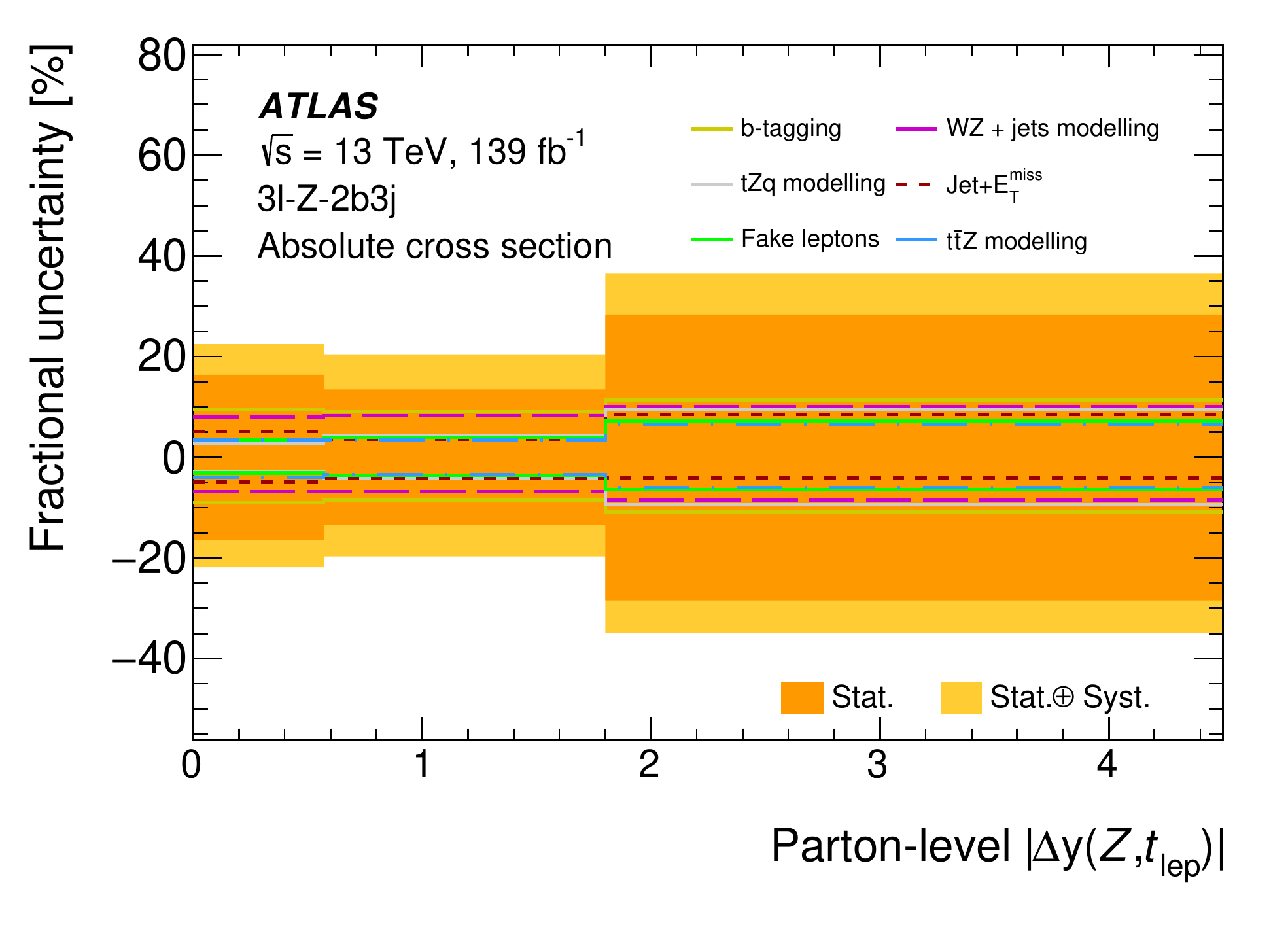}}
\subfloat[]{\includegraphics[width=0.49\textwidth]{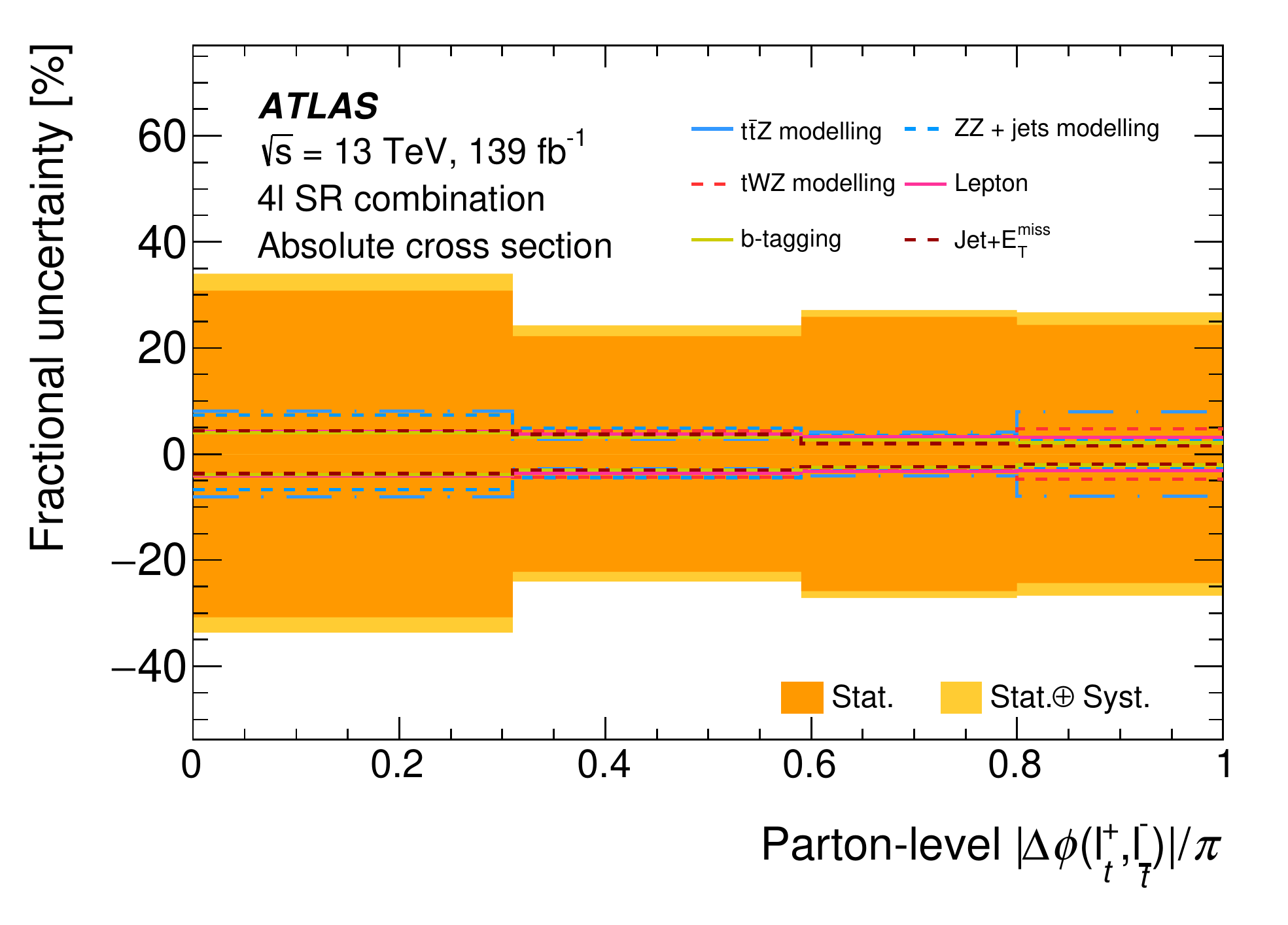}}
 
\caption{Absolute differential \ttZ cross sections measured at parton level as a function of (a)~the rapidity difference~($|\Delta y|$) between the \Zboson boson and the
reconstructed top quark and (b)~the azimuthal separation~($\Delta\phi$) between the two leptons associated with the top-quark pair. (c,d)~show the relative contributions
from different categories of systematic uncertainties per bin.}
\end{figure}

\begin{figure}[htb!]
\centering
\subfloat[]{\includegraphics[width=0.49\textwidth]{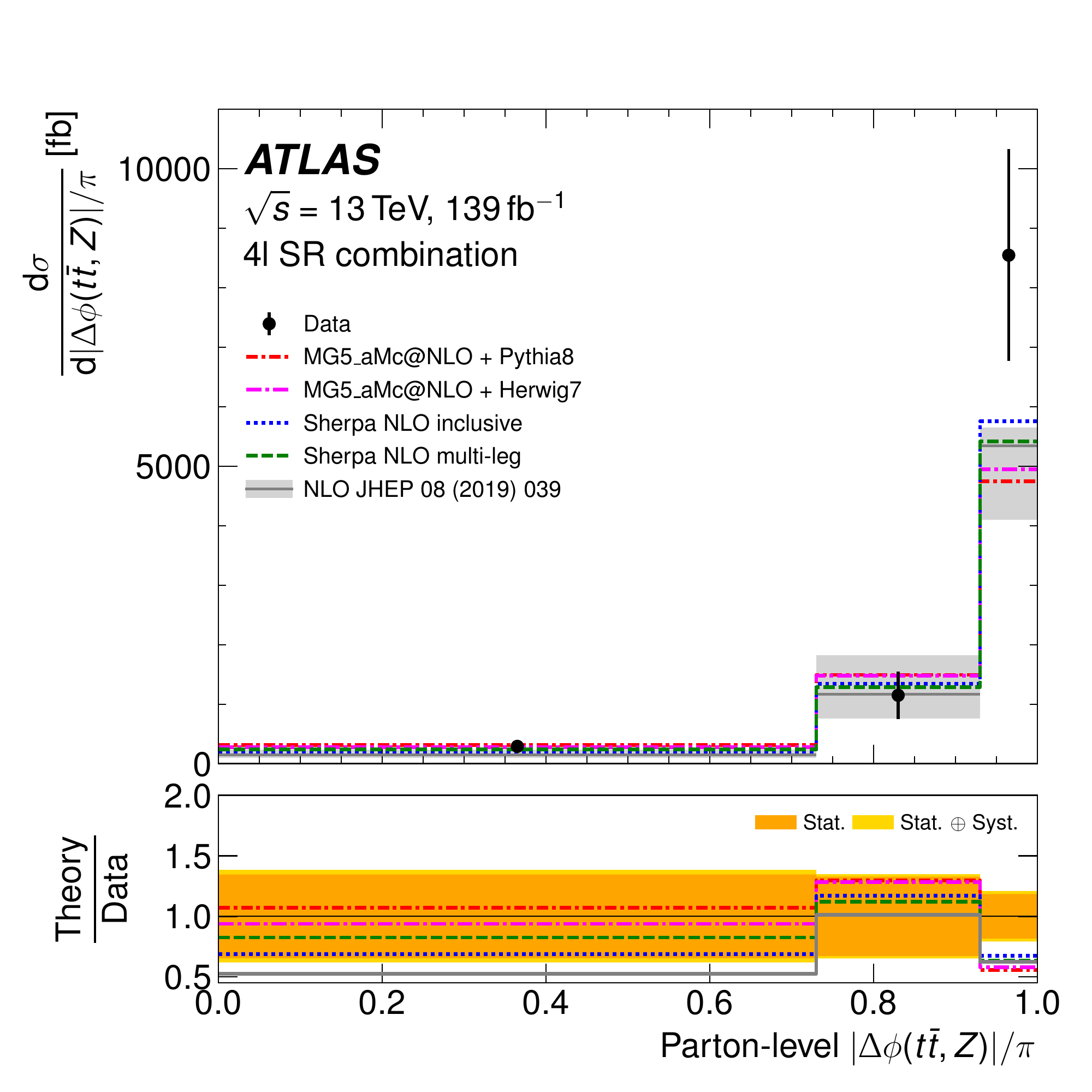} \label{fig:x-sec-results-7a}}
\subfloat[]{\includegraphics[width=0.49\textwidth]{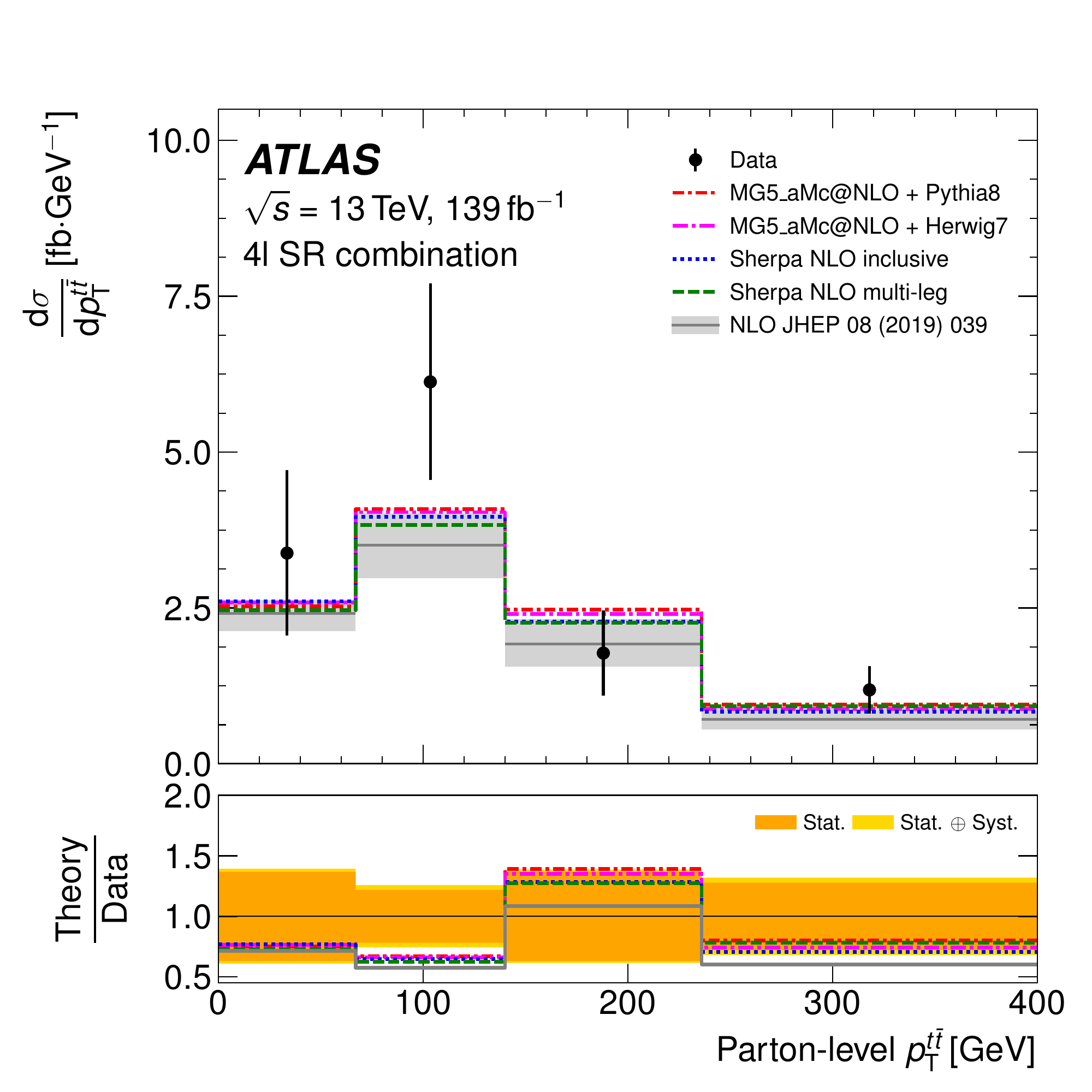} \label{fig:x-sec-results-7b}}
 
\subfloat[]{\includegraphics[width=0.49\textwidth]{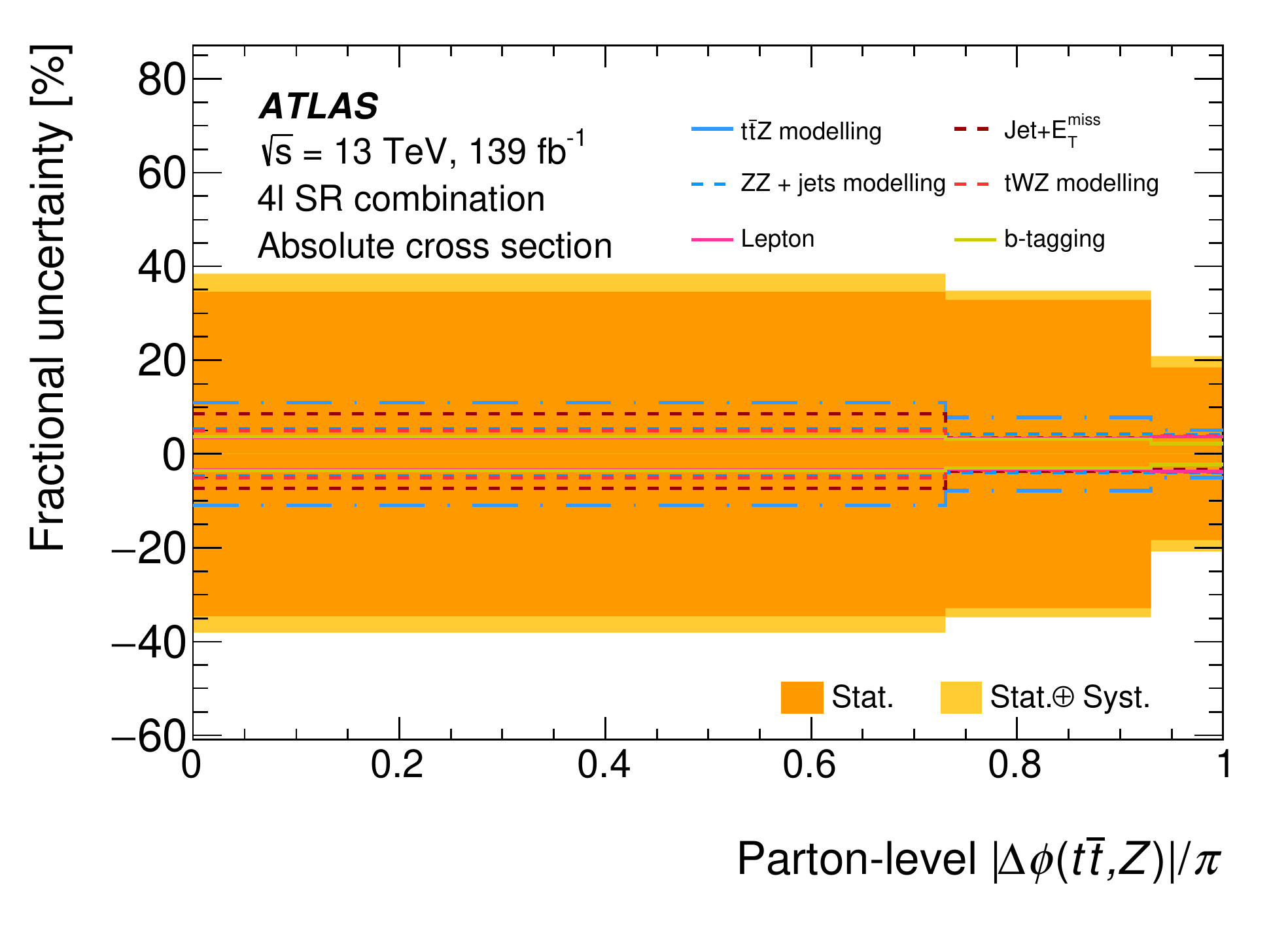}}
\subfloat[]{\includegraphics[width=0.49\textwidth]{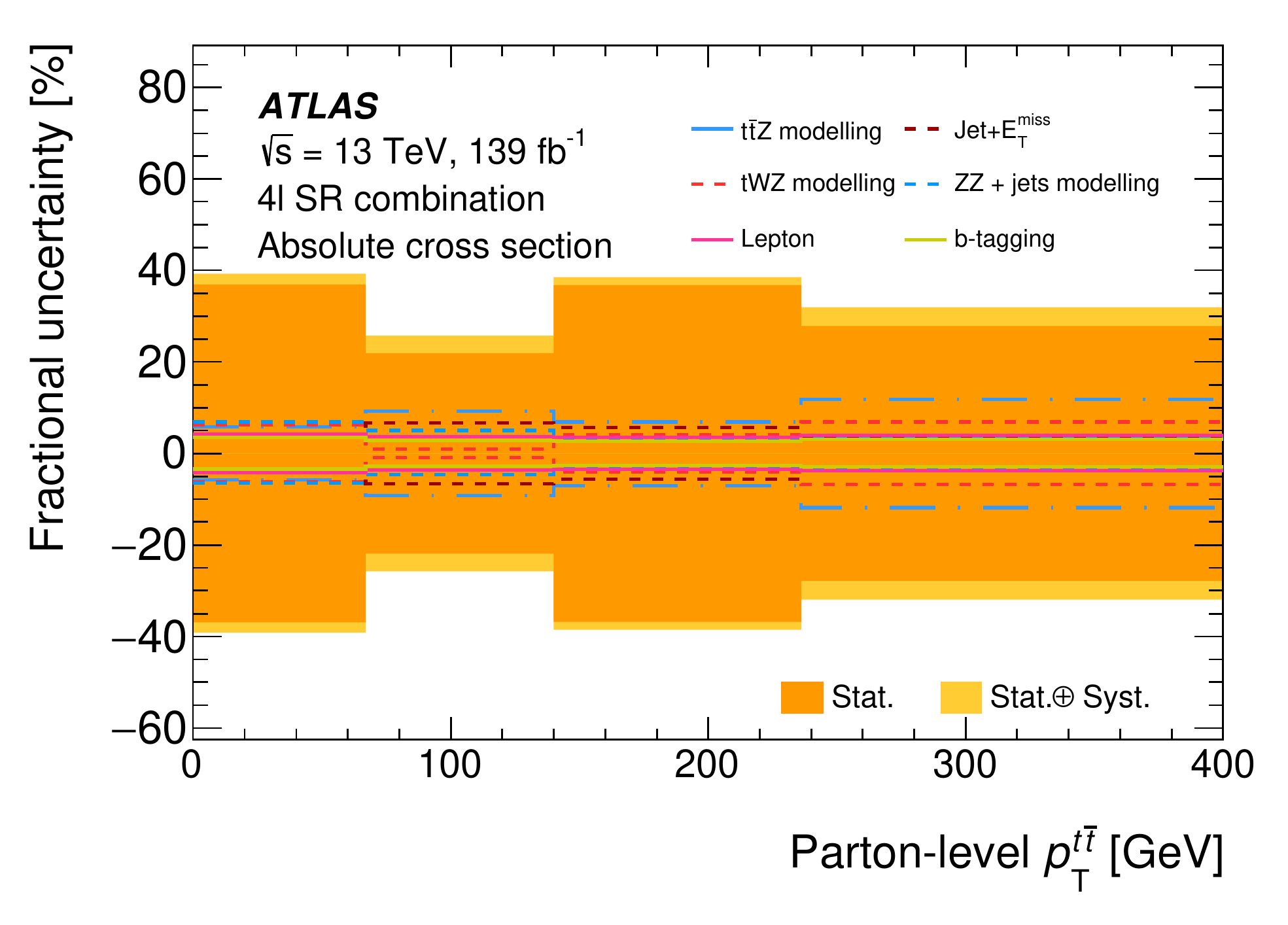}}
 
\caption{Absolute differential \ttZ cross sections measured at parton level as a function of (a)~the azimuthal separation~($\Delta\phi$) between the \Zboson
boson and the reconstructed \ttbar system and (b)~the \pt of the \ttbar system. (c,d)~show the relative contributions from different categories of systematic uncertainties per bin.}
\label{fig:x-sec-results-7}
\end{figure}

Results for the other observables described in Table~\ref{tab:variables} are presented in Figures~\ref{fig:x-sec-results-3}--\ref{fig:x-sec-results-7}. For these variables, only the
absolute parton-level differential measurements are shown, with the exception of $N_{\mathrm{jets}}$, which is unfolded only to particle level. Additional differential \ttZ predictions
at NLO, NLO+NNLL or approximate next-to-next-to-leading-order~(nNLO) precision --- including EW corrections --- are shown in grey for the parton-level results for most of the observables.
The calculations were carried out in similar fashion to that described in Ref.~\cite{Broggio:2019ewu}, but specifically performed in the context of this analysis in order to
provide predictions for the measured observables and to match the number and ranges of bins for the different variables.\footnote{The precision of the calculations
depends on the  particular variable and can be provided at either NLO, NLO+NNLL or nNLO. The orders of the calculations are indicated on the plots. Their uncertainties include the
contributions from scale variations, PDF uncertainties and MC statistics, which are added linearly to follow a conservative approach for the evaluation of the total uncertainty.}
These additional parton-level predictions are not provided for two of the observables, namely $\pT^{\ell,{\textrm{non-}}Z}$ and $|\Delta\phi(\ell_{t}^{+},\ell_{\bar{t}}^{-})|$, since the
decays of the \ttbar pair and the \Zboson boson were not included in the theoretical calculations.
 
In order to test the overall compatibility between the unfolded measurements and the various predictions, a $\chi^{2}/$ndf and corresponding $p$-value is evaluated
for each of the differential measurements, separately for the parton level and particle level as well as for the absolute and normalised cases.
 
The $\chi^{2}$ value is defined as:
\begin{equation*}
\chi^2 = \sum_{i=1}^{N_{\textrm{bins}}}\sum_{j=1}^{N_{\textrm{bins}}}\left(n_{i}-\mu_{i}\right)\left(n_{j}-\mu_{j}\right)[C^{-1}]_{ij},
\end{equation*}
where $n_{i}$ and $\mu_{i}$ correspond to the content in bin $i$ of the distributions from the unfolded data and the given prediction, respectively, and $\left[C^{-1}\right]_{ij}$
to the element in row $i$ and column $j$ of the inverse of the covariance matrix for the particular variable. The values, $n_{i}$ and $\mu_{i}$, are, therefore, notational shorthands
for $\mathrm{d}\sigma_{\ttZ} / \mathrm{d} X^{i}$, or $1/\sigma \cdot \mathrm{d} \sigma_{\ttZ} / \mathrm{d} X^{i}$ in the normalised case.
 
A given $p$-value can be interpreted as the probability of obtaining a value of $\chi^2$ greater or equal to the quoted value for a particular number of degrees of freedom,
where the latter is equal to the number of bins~($N_{\textrm{bins}}$) in the case of the absolute measurements and $N_{\textrm{bins}} - 1$ for the normalised measurements.
 
The construction of the covariance matrix is based on the approach described in Ref.~\cite{TOPQ-2018-18}, and it includes both the statistical and systematic uncertainties.
The latter include detector-related uncertainties as well as those related to the modelling of the signal and various background processes. While all sources of uncertainty
related to the measurements are incorporated in the covariance matrix elements, uncertainties in the theoretical predictions, themselves, are omitted, and their impact is, therefore,
not reflected in the quoted $\chi^2$ and corresponding $p$-values.
 
For a given variable, the elements of the covariance matrix, $C_{ij}$, are evaluated using a bootstrap technique, whereby \num{150000} Poisson-fluctuated distributions are produced,
each corresponding to a pre-unfolded distribution for a given pseudo-experiment. For the detector-related uncertainties, Gaussian-distributed shifts are added coherently to each of the
Poisson-fluctuated bin contents, with each shift corresponding to a particular uncertainty source. The shifts are applied as a multiplicative scale relative to the particular bin content,
and with the amount and direction of each shift dictated by the corresponding uncertainty source.
 
Each of the varied distributions is subsequently unfolded using the nominal acceptance and efficiency corrections, as well as the nominal migration matrix --- those derived from the
nominal \MGNLOs~+~\PYTHIA[8] signal sample. Gaussian-distributed shifts are then added coherently to the post-unfolding distributions for each of the signal- and background-modelling
uncertainty sources. These shifts are also determined and applied as relative variations for each particular source. The relative variations in this case are defined according to the
difference between the generated and the unfolded cross section of a given alternative signal or background model, using nominal corrections in the unfolding.
 
The resulting changes to the unfolded distributions are, then, used to determine the elements of the covariance matrix for that particular variable. The covariance matrices for the
normalised measurements are constructed in an analogous fashion. In this case, the distributions are normalised to unity after all effects are included.
In order to avoid performing the unfolding on distributions with negative bin contents, that can arise due to the effects of systematic uncertainties and the subtraction of backgrounds,
any such bin contents are set to zero prior to the unfolding.
 
The uncertainties reflected in the elements, $C_{ij}$, initially evaluated as relative values, are then multiplied by the differential cross-section values from the measured data in
each particular bin in order to yield absolute uncertainties.
 
In general, each of the individual sources contributing to the full covariance matrix will contribute to the off-diagonal terms, including even those from the limited data sample size
where non-diagonal contributions arise from correlations between the bins introduced during the unfolding process.
 
The correlations between the bins in the absolute differential measurements for the trilepton and combined channels are sizeable --- in many cases in the 20--55\% range --- and positive.
In the case of the tetralepton channel, where statistical uncertainties are more dominant, the correlations are generally below 20\% in absolute value. The correlations in the case of
the normalised measurements are generally negative and reach absolute values even larger than in the absolute cross-section measurements, with the most extreme case being a value of
$\rho = -0.76$ between the first and second bin of the $|\Delta y(Z,t_{\textrm{lep}})|$ variable in the trilepton channel. For the \Zboson-related variables in the combined channel,
the effect of the larger data sample is partially balanced by the increase in the number of bins, such that the correlations in the absolute measurements for $\pt^{Z}$ and $|y^{Z}|$
are also positive, but lie in the 15--45\% range. In the normalised measurements for these two variables, the correlations are also mostly negative but are smaller in magnitude
than for other variables~(strictly $|\rho| < \SI{40}{\%}$, but in most cases $|\rho| < \SI{20}{\%}$).
 
Table~\ref{tab:chi2-compatibility-main} summarises the evaluated $\chi^2/$ndf and $p$-values used to quantify the compatibility between the measured unfolded data and the various predictions.
For the parton-level measurements, the values for the additional theory predictions at NLO, NLO+NNLL or nNLO are also shown for those variables for which predictions are
available~\cite{Broggio:2019ewu}.
 
Overall, the unfolded spectra from the measured data are compatible with the various predictions for most of the variables considered. For the $\pt^{Z}$ variable in
the combined $3\ell + 4\ell$ channel, as well as for $\pT^{\ell,{\textrm{non-}}Z}$ and $|\Delta\phi(Z,t_{\textrm{lep}})|$ in the trilepton channel, slightly lower $p$-values
are obtained for several predictions, but in all cases they are found to be greater than \num{0.05}.
For the $\pt^{Z}$ variable in the combined channel, the slightly poorer agreement is driven in large part by the sixth bin ($\SI{220}{\GeV} \leq \pt^{Z} < \SI{290}{\GeV}$).
For this variable, however, the $p$-value is larger~(\num{0.17}) for the additional NLO+NNLL prediction in the absolute differential measurement. For the $|\Delta\phi(\ttbar,Z)|$ variable
in the tetralepton channel, for which the data exhibit a greater relative fraction of events with larger azimuthal separation between the \Zboson boson and the \ttbar system, a slightly
better level of agreement is observed for the \SHERPA predictions compared with those from \MGNLOs. As the statistical uncertainties of the measured data are almost always significantly
larger than the differences between the predictions, no definite conclusion about the overall compatibility for these observables can be made. The effect of the uncertainties in the
fixed-order parton-level theoretical predictions in the rightmost column of Table~\ref{tab:chi2-compatibility-main} was evaluated. This was done by adding terms of the form
$\delta\mu_i \delta\mu_j$ to the covariance matrix, where $\delta\mu_{i(j)}$ is the sum in quadrature of the uncertainties associated with the scale and PDF choice for bin~$i(j)$.
The bin contents of the theoretical predictions were, therefore, considered to be 100\% correlated. The inclusion of these uncertainties leads to a relative increase of 20--50\% in
the $p$-values for the variables $|\Delta\phi(Z,t_{\textrm{lep}})|$, $|\Delta\phi(\ttbar,Z)|$, and $\pT^{\ttbar}$ relative to those quoted in Table~\ref{tab:chi2-compatibility-main}.
For the variables $|\Delta y(Z,t_{\textrm{lep}})|$, $\pT^{Z}$, and $|y^{Z}|$, the impact is negligible.
 
\begin{table}[!htb]
\footnotesize
\caption[]{Summary of the tests of compatibility between the unfolded differential measurements and the various predictions. Quoted are the $\chi^2/$ndf and
corresponding $p$-values incorporating all bins for the given variable and based on the assumption that all sources of uncertainty are Gaussian-distributed. The values
associated with the additional theory predictions (last column) are included where applicable. These additional predictions are obtained as described in
Ref.~\cite{Broggio:2019ewu}, with their precisions depending on the particular variable: NLO for $|\Delta\phi(\ttbar,Z)|$ and $\pT^{\ttbar}$; NLO+NNLL for
$|\Delta\phi(Z,t_{\textrm{lep}})|$, $|\Delta y(Z,t_{\textrm{lep}})|$ and $\pT^{Z}$; and nNLO for $|y^{Z}|$.}
\label{tab:chi2-compatibility-main}
\def\arraystretch{1.5}
\begin{center}
\resizebox{\textwidth}{!}{
\begin{tabular}{c|c|cc|cc|c|cc|cc|cc|cc|cc}
\cline{3-17}
\cline{3-17}
\multicolumn{2}{l|}{} & \multirow{3}{*}{\rotatebox{90}{{\scriptsize Particle}}} & \multirow{3}{*}{\rotatebox{90}{{\scriptsize Parton}}} & \multirow{3}{*}{\rotatebox{90}{{\scriptsize Absolute}}} & \multirow{3}{*}{\rotatebox{90}{{\scriptsize Normalised \ }}} & \multirow{3}{*}{\rotatebox{90}{{\scriptsize Figure }}} & \multicolumn{2}{c|}{\MGNLOsV{2.3.3}} & \multicolumn{2}{c|}{\MGNLOsV{2.3.3}} & \multicolumn{2}{c|}{\SHERPA[2.2.1]} & \multicolumn{2}{c|}{\SHERPA[2.2.1]} & \multicolumn{2}{c}{Additional} \\
\multicolumn{2}{l|}{} & & & & & & \multicolumn{2}{c|}{+ \PYTHIA[8]} & \multicolumn{2}{c|}{+ \HERWIG[7]} & \multicolumn{2}{c|}{NLO multi-leg} & \multicolumn{2}{c|}{NLO inclusive} & \multicolumn{2}{c}{Theory} \\
\cline{2-2}
\cline{8-17}
\multicolumn{1}{c|}{} & Variable & & & & & & $\chi^2/$ndf & $p$-value & $\chi^2/$ndf & $p$-value & $\chi^2/$ndf & $p$-value & $\chi^2/$ndf & $p$-value & $\chi^2/$ndf & $p$-value \\ \hline \hline
\multirow{5}{*}{\rotatebox{90}{$3\ell+4\ell$}} & $\pT^{Z}$ & $\checkmark$ & & $\checkmark$ & &  \ref{fig:x-sec-results-1a} & 12.8/7 & 0.08 & 12.0/7 & 0.10 & 11.6/7 & 0.11 & 12.1/7 & 0.10 & / & / \\
& $\pT^{Z}$ & & $\checkmark$ & $\checkmark$ & & \ref{fig:x-sec-results-1b} & 12.8/7 & 0.08 & 11.7/7 & 0.11 & 11.2/7 & 0.13 & 11.3/7 & 0.13 & 10.4/7 & 0.17 \\
& $\pT^{Z}$ & $\checkmark$ & & & $\checkmark$ & \ref{fig:x-sec-results-2a} & 11.0/6 & 0.09 & 10.8/6 & 0.09 & 10.6/6 & 0.10 & 10.7/6 & 0.10 & / & / \\
& $\pT^{Z}$ & & $\checkmark$ & & $\checkmark$ & \ref{fig:x-sec-results-2b} & 11.0/6 & 0.09 & 10.8/6 & 0.10 & 10.7/6 & 0.10 & 10.6/6 & 0.10 & 10.5/6 & 0.11 \\
& $|y^{Z}|$ & & $\checkmark$ & $\checkmark$ & & \ref{fig:x-sec-results-3a} & 2.8/8 & 0.95 & 2.9/8 & 0.94 & 4.0/8 & 0.85 & 2.7/8 & 0.95 & 2.9/8 & 0.94 \\   \hline \hline
\multirow{4}{*}{\rotatebox{90}{$3\ell$}} & $N_{\textrm{jets}}$ & $\checkmark$ & & $\checkmark$ & & \ref{fig:x-sec-results-4a} & 0.8/3 & 0.85 & 0.6/3 & 0.90 & 0.3/3 & 0.95 & 0.5/3 & 0.92 & / & / \\
& $\pT^{\ell,{\textrm{non-}}Z}$ & & $\checkmark$ & $\checkmark$ & & \ref{fig:x-sec-results-5a}  & 7.6/4 & 0.11 & 8.8/4 & 0.07 & 8.3/4 & 0.08 & 8.6/4 & 0.07 & / & / \\
& $|\Delta\phi(Z,t_{\textrm{lep}})|$ & & $\checkmark$ & $\checkmark$ & &  \ref{fig:x-sec-results-5b} & 5.5/3 & 0.14 & 5.8/3 & 0.12 & 5.2/3 & 0.16 & 6.9/3 & 0.07 & 6.6/3 & 0.09 \\
& $|\Delta y(Z,t_{\textrm{lep}})|$ & & $\checkmark$ & $\checkmark$ & &  \ref{fig:x-sec-results-6a}  & 0.9/3 & 0.82 & 0.7/3 & 0.88 & 0.2/3 & 0.98 & 0.5/3 & 0.92 & 0.3/3 & 0.96 \\ \hline \hline
\multirow{4}{*}{\rotatebox{90}{$4\ell$}} & $N_{\textrm{jets}}$ & $\checkmark$ & & $\checkmark$ & & \ref{fig:x-sec-results-4b} & 1.4/4 & 0.84 & 1.7/4 & 0.79 & 2.8/4 & 0.59 & 2.8/4 & 0.59 & / & / \\
& $|\Delta\phi(\ell_{t}^{+},\ell_{\bar{t}}^{-})|$ & & $\checkmark$ & $\checkmark$ & &  \ref{fig:x-sec-results-6b} & 2.1/4 & 0.72 & 2.3/4 & 0.69 & 2.7/4 & 0.62 & 2.6/4 & 0.63 & / & / \\
& $|\Delta\phi(\ttbar,Z)|$ & & $\checkmark$ & $\checkmark$ & &  \ref{fig:x-sec-results-7a} & 5.2/3 & 0.16 & 4.7/3 & 0.19 & 3.5/3 & 0.32 & 3.4/3 & 0.33 & 4.9/3 & 0.18 \\
& $\pT^{\ttbar}$ & & $\checkmark$ & $\checkmark$ & &  \ref{fig:x-sec-results-7b} & 3.5/4 & 0.47 & 3.6/4 & 0.47 & 3.5/4 & 0.48 & 3.5/4 & 0.47 & 4.6/4 & 0.33 \\ \hline \hline
\end{tabular}
}
\end{center}
\end{table}
 
A difference between the measured inclusive cross section quoted in Section~\ref{sec:results_inclusive} and the cross section based on the integrated absolute parton-level spectra
in the combined $3\ell+4\ell$ channel is observed.  The two measurements differ both in terms of the method used and in their selection due to the use of different $b$-tagging WPs
(refer to Table~\ref{tab:sr-3l-def_main}).  Approximately 67\% of selected data events are common to both measurements.  The compatibility between the two cross-section measurements
is evaluated using pseudo-experiments taking into account the correlation between uncertainties, including all sources of statistical and systematic effects, and it is found to be
at the level of two standard deviations.
\FloatBarrier
 
\section{Conclusions}
\label{sec:conclusion}
Inclusive and differential measurements of the production cross section of a \ttbar pair in association with a \Zboson boson are presented. The full $\rts = \SI{13}{\TeV}$
\pp collision  data set collected by the ATLAS detector during \RunTwo of the LHC between 2015 and 2018, corresponding to \SI{139}{\ifb}, was used for this analysis. Only
final states with three or four isolated charged leptons~(electrons or muons) were considered for the measurements. The measured inclusive cross section of the \ttZ process
is $\sigma_{\ttZ} = 0.99 \pm 0.05$~(stat.)~$\pm\, 0.08$~(syst.)~pb, in agreement with the SM prediction. The dominant sources of systematic uncertainty in this measurement
are associated with the \ttZ parton-shower modelling, $b$-tagging, and modelling of the \tWZ background.
 
Absolute and normalised differential cross sections were measured as functions of nine different observables sensitive to the MC modelling of the \ttZ process and to potential
BSM effects. The differential cross-section measurements were performed at particle and parton levels in specific fiducial volumes. The unfolded spectra from the measured data
are compared with the predictions of different NLO QCD \ttZ MC generators and theoretical predictions at NLO, NLO+NNLL and nNLO precision, based on a $\chi^{2}/$ndf and $p$-value
compatibility test. For most of the considered observables, good agreement between data and the predictions is observed. The differences between the various predictions are
determined to be smaller than the uncertainties of the unfolded data. For the variables $\pT^{Z}$, $\pT^{\ell,\textrm{non-}Z}$, $|\Delta\phi(Z,t_{\mathrm{lep}})|$ and
$|\Delta\phi(\ttbar,Z)|$, the observed and predicted differential results show slightly poorer agreement, but $p$-values $> 0.05$ are obtained in all cases.
\clearpage

\section*{Acknowledgements}

We thank CERN for the very successful operation of the LHC, as well as the
support staff from our institutions without whom ATLAS could not be
operated efficiently.
 
We acknowledge the support of
ANPCyT, Argentina;
YerPhI, Armenia;
ARC, Australia;
BMWFW and FWF, Austria;
ANAS, Azerbaijan;
SSTC, Belarus;
CNPq and FAPESP, Brazil;
NSERC, NRC and CFI, Canada;
CERN;
ANID, Chile;
CAS, MOST and NSFC, China;
Minciencias, Colombia;
MSMT CR, MPO CR and VSC CR, Czech Republic;
DNRF and DNSRC, Denmark;
IN2P3-CNRS and CEA-DRF/IRFU, France;
SRNSFG, Georgia;
BMBF, HGF and MPG, Germany;
GSRI, Greece;
RGC and Hong Kong SAR, China;
ISF and Benoziyo Center, Israel;
INFN, Italy;
MEXT and JSPS, Japan;
CNRST, Morocco;
NWO, Netherlands;
RCN, Norway;
MNiSW and NCN, Poland;
FCT, Portugal;
MNE/IFA, Romania;
JINR;
MES of Russia and NRC KI, Russian Federation;
MESTD, Serbia;
MSSR, Slovakia;
ARRS and MIZ\v{S}, Slovenia;
DSI/NRF, South Africa;
MICINN, Spain;
SRC and Wallenberg Foundation, Sweden;
SERI, SNSF and Cantons of Bern and Geneva, Switzerland;
MOST, Taiwan;
TAEK, Turkey;
STFC, United Kingdom;
DOE and NSF, United States of America.
In addition, individual groups and members have received support from
BCKDF, CANARIE, Compute Canada and CRC, Canada;
COST, ERC, ERDF, Horizon 2020 and Marie Sk{\l}odowska-Curie Actions, European Union;
Investissements d'Avenir Labex, Investissements d'Avenir Idex and ANR, France;
DFG and AvH Foundation, Germany;
Herakleitos, Thales and Aristeia programmes co-financed by EU-ESF and the Greek NSRF, Greece;
BSF-NSF and GIF, Israel;
Norwegian Financial Mechanism 2014-2021, Norway;
La Caixa Banking Foundation, CERCA Programme Generalitat de Catalunya and PROMETEO and GenT Programmes Generalitat Valenciana, Spain;
G\"{o}ran Gustafssons Stiftelse, Sweden;
The Royal Society and Leverhulme Trust, United Kingdom.
 
The crucial computing support from all WLCG partners is acknowledged gratefully, in particular from CERN, the ATLAS Tier-1 facilities at TRIUMF (Canada), NDGF (Denmark, Norway, Sweden), CC-IN2P3 (France), KIT/GridKA (Germany), INFN-CNAF (Italy), NL-T1 (Netherlands), PIC (Spain), ASGC (Taiwan), RAL (UK) and BNL (USA), the Tier-2 facilities worldwide and large non-WLCG resource providers. Major contributors of computing resources are listed in Ref.~\cite{ATL-SOFT-PUB-2021-003}.
 
 
\clearpage

\printbibliography
 
\clearpage

\clearpage
 
\begin{flushleft}
\hypersetup{urlcolor=black}
{\Large The ATLAS Collaboration}

\bigskip

\AtlasOrcid[0000-0002-6665-4934]{G.~Aad}$^\textrm{\scriptsize 102}$,    
\AtlasOrcid[0000-0002-5888-2734]{B.~Abbott}$^\textrm{\scriptsize 128}$,    
\AtlasOrcid[0000-0002-7248-3203]{D.C.~Abbott}$^\textrm{\scriptsize 103}$,    
\AtlasOrcid[0000-0002-2788-3822]{A.~Abed~Abud}$^\textrm{\scriptsize 36}$,    
\AtlasOrcid[0000-0002-1002-1652]{K.~Abeling}$^\textrm{\scriptsize 53}$,    
\AtlasOrcid[0000-0002-2987-4006]{D.K.~Abhayasinghe}$^\textrm{\scriptsize 94}$,    
\AtlasOrcid[0000-0002-8496-9294]{S.H.~Abidi}$^\textrm{\scriptsize 167}$,    
\AtlasOrcid[0000-0002-8279-9324]{O.S.~AbouZeid}$^\textrm{\scriptsize 40}$,    
\AtlasOrcid{N.L.~Abraham}$^\textrm{\scriptsize 156}$,    
\AtlasOrcid[0000-0001-5329-6640]{H.~Abramowicz}$^\textrm{\scriptsize 161}$
% ...
\bigskip
\\

% ...
$^{36}$CERN, Geneva; Switzerland.\\
% ...
$^{40}$Niels Bohr Institute, University of Copenhagen, Copenhagen; Denmark.\\
% ...
$^{53}$II. Physikalisches Institut, Georg-August-Universit\"{a}t G\"ottingen, G\"ottingen; Germany.\\
% ...
$^{94}$Department of Physics, Royal Holloway University of London, Egham; United Kingdom.\\
% ...
$^{102}$CPPM, Aix-Marseille Universit\'e, CNRS/IN2P3, Marseille; France.\\
$^{103}$Department of Physics, University of Massachusetts, Amherst MA; United States of America.\\
% ...
$^{128}$Homer L. Dodge Department of Physics and Astronomy, University of Oklahoma, Norman OK; United States of America.\\
% ...
$^{156}$Department of Physics and Astronomy, University of Sussex, Brighton; United Kingdom.\\
% ...
$^{161}$Raymond and Beverly Sackler School of Physics and Astronomy, Tel Aviv University, Tel Aviv; Israel.\\
% ...
$^{167}$Department of Physics, University of Toronto, Toronto ON; Canada.\\
% ...

\end{flushleft}

 
\end{document}